\documentclass[aps,prd,preprint,groupedaddress,
citeautoscript,
amsmath,amssymb,longbibliography  
]{revtex4-2}
\usepackage{graphicx}
\usepackage{amsmath}
\usepackage{siunitx}
\usepackage[hidelinks]{hyperref} 
\usepackage[table]{xcolor}
\usepackage{booktabs}  
\usepackage{makecell,multirow}  
\usepackage{rotating,bm}   
\usepackage{mathptmx}  
\setcitestyle{super}  
\newcommand*{\citen}[1]{%
  \begingroup
    \romannumeral-`\x 
    \setcitestyle{numbers}%
    \cite{#1}%
  \endgroup   
}
\newcommand{\hideForFinalVersion}[1]{{#1}}  

\usepackage{setspace}  

\usepackage{pdfpages} 
\usepackage{pgffor} 

\makeatletter
\AtBeginDocument{\let\LS@rot\@undefined}
\makeatother

\def\supplementfilename{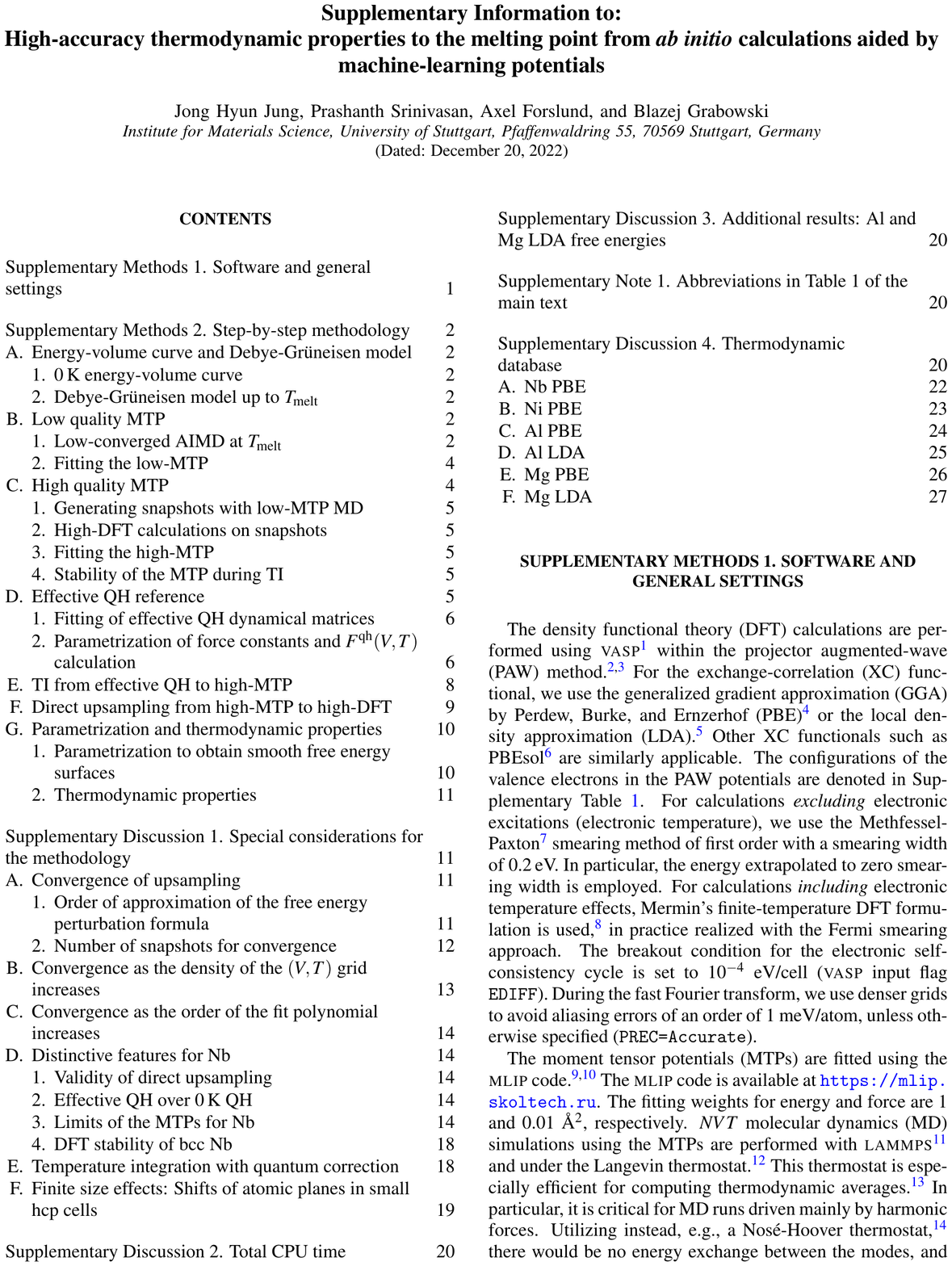}

\pdfximage{\supplementfilename}
\def\numbersupplementpages{\the\pdflastximagepages}

\newif\ifarXiv
\arXivtrue 

\begin{document}


\title{High-accuracy thermodynamic properties to the melting point from \textit{ab initio} calculations aided by machine-learning potentials}
\author{Jong Hyun Jung}
\author{Prashanth Srinivasan}
\author{Axel Forslund}
\author{Blazej Grabowski}
\email{blazej.grabowski@imw.uni-stuttgart.de}
\affiliation{Institute for Materials Science, University of Stuttgart, Pfaffenwaldring 55, 70569 Stuttgart}
\date{\today}

\begin{abstract}
Accurate prediction of thermodynamic properties requires an extremely accurate representation of the free energy surface. Requirements are twofold---first, the inclusion of the relevant finite-temperature mechanisms, and second, a dense volume-temperature grid on which the calculations are performed. A systematic workflow for such calculations requires computational efficiency and reliability, and has not been available within an \textit{ab initio} framework so far. Here, we elucidate such a framework involving \textit{direct upsampling}, thermodynamic integration and machine-learning potentials, allowing us to incorporate, in particular, the full effect of anharmonic vibrations. The improved methodology has a five-times speed-up compared to state-of-the-art methods. We calculate equilibrium thermodynamic properties up to the melting point for bcc Nb, magnetic fcc Ni, fcc Al and hcp Mg, and find remarkable agreement with experimental data. Strong impact of anharmonicity is observed specifically for Nb. The introduced procedure paves the way for the development of \textit{ab initio} thermodynamic databases.
\clearpage
\end{abstract}
\maketitle

Thermodynamic properties such as the heat capacity, the expansion coefficient, and the bulk modulus are key benchmarks in materials design. They provide insight into phase stability, phase transformations, microstructural stability and strength, thereby giving guidance to synthesis and application. The heat capacity is linked to thermodynamic potentials such as the Gibbs energy, and thus facilitates the construction of phase diagrams. By virtue of being experimentally measurable via calorimetry, the heat capacity remains---ever since the seminal work of Einstein\cite{einstein07plancksche}---among the most fundamental properties in basic research and industrial applications~\cite{zheng19understanding}. Knowledge of the thermal expansion coefficient is crucial, for example, to optimize the Invar behavior of alloys used in instrumentation applications~\cite{vanSchilfgaarde99origin}. The bulk modulus is vital in modeling strength and ductility up to high temperatures~\cite{lee20temperature}. 

Coordinated surveys of experimental data on thermodynamic properties have been carried out in multiple works, resulting in well-known series of books, e.g., Touloukian et al.\cite{touloukian75thermophysical-TPRC} and Landolt-B\"ornstein~\cite{landolt81metals}. Although these books have well-served as the basis for materials design, there is an urgent need for an efficient extension of the databases triggered by the ever-increasing demand for new and optimized materials. However, applying experimental techniques alone is time-consuming and thus inefficient. 

To complement experiments and to rapidly obtain material properties, \textit{ab initio} databases have been intensively sought after 
recently. \cite{draxl19nomad,legrain17how,kirklin15OQMD,JCA15,talirz20materials} 
These databases allow for a quick, online access to a wide range of material properties. As yet, these databases are to a large extent based on $T=0$\,K ($-273.15\,^{\circ}$C) data, and low temperature approximations to account for the effect of temperature. For example, in one such database~\cite{draxl19nomad}, from a total of 12,000,000 entries on material properties, only 1,810 thermal properties are available that are derived from harmonic or quasiharmonic approximations. Even though these approximations can provide rapid results (though not for all systems), they do not consider any explicit finite-temperature vibrations and coupling effects, rendering them unsuitable at elevated temperatures.\cite{glensk15understanding-Fah}

To predict reliable high-temperature thermodynamic properties, an accurate representation of the system's free energy is needed, including in particular explicit anharmonicity, i.e., phonon-phonon interactions and interactions of phonons with other excitation mechanisms (electronic, magnetic). Approximation schemes have been developed to account for phonon-phonon interactions. One class of approaches utilizes effective, renormalized harmonic Hamiltonians~\cite{hellman11lattice,tidholm20temperature-Nb,adams21anharmonic,souvatzis08entropy}. Another approximate route involves the use of a low temperature perturbative expansion.\cite{junkaew14ab} Though often efficient, such approximations can be inaccurate at high temperatures and for systems with strong anharmonicity.\cite{glensk15understanding-Fah,grabowski19ab}

Explicit anharmonicity can be included up to all orders using thermodynamic integration (TI), wherein one computes the free energy difference between a reference state and density-functional-theory (DFT).\cite{Vocadlo03possible} Over the last decade, several improvements have been made to TI-based methods~\cite{vocadlo02ab,pozzo19FeO,dorner18melting}. Notably, the \textit{upsampled thermodynamic integration using Langevin dynamics} (UP-TILD) (TI to low-accuracy DFT followed by `upsampling')~\cite{glensk15understanding-Fah,grabowski09ab-Al} and the \textit{two-stage upsampled thermodynamic integration using Langevin dynamics} (TU-TILD) (UP-TILD split into two stages with an intermediate potential)~\cite{duff15improved} have increased computational performance. Recent developments have been directed towards exploiting advancements from machine-learning.~\cite{ko21fourth,batzner22e3,lopanitsyna21finite} The so-called moment tensor potentials (MTPs)~\cite{shapeev16moment} have proved to be one of the most efficient machine-learning potentials.~\cite{nyshadham19machine,zuo20performance,lysogorskiy21performant}  The application of MTPs within the TU-TILD formalism has further improved the efficiency of free energy calculations~\cite{forslund22ab,grabowski19ab}.

Despite the advancements, free energy calculations including the impact of anharmonicity have remained challenging until now, even for supposedly `simple' elementary systems. A critical aspect is the accurate determination of thermodynamic properties, for which numerically converged first and second derivatives of the free energy are indispensable. The situation is illustrated in Table~\ref{table:literature-review} for four prototype systems covering the fundamental crystallographic structures of metals, i.e., bcc Nb, fcc Ni and Al, and hcp Mg. Although the first thermodynamic calculations including anharmonicity were done as early as 2002 for fcc Al,\cite{vocadlo02ab} more recent studies have either neglected anharmonicity (empty fields in the table) or have utilized approximate approaches. Only a few of the studies included explicit anharmonicity to DFT accuracy (marked in bold). The situation is worse for the other elements, for which almost no information is available on thermodynamic properties including anharmonicity at the full DFT level. A low temperature expansion as performed in Ref.~\citen{junkaew14ab} is particularly uncertain for high-melting elements with strong anharmonicity such as Nb. This explains why the data was analyzed only below 1000 K.\cite{junkaew14ab} At higher temperatures Nb experiences large anharmonicity which strongly impacts the thermodynamic properties. To account for such features, explicit anharmonicity needs to be captured to all orders and the derivatives of the free energy surface need to be stabilized by higher order parametrizations on \emph{dense} volume-temperature $(V,T)$ grids of explicitly computed free energies. 

Summarizing the current state-of-the-art in finite temperature \textit{ab initio} simulations, it has to be concluded that a holistic computational methodology to readily obtain accurate thermodynamic properties is not yet at hand. Consequently, \textit{ab initio} thermodynamic properties are only very scarcely and inconsistently available in the literature, even for elementary systems. This clearly hampers the development of \textit{ab initio} thermodynamic databases.

In the current paper, we present such a holistic computational procedure with which affordable, fully anharmonic free energy calculations become possible on a sufficiently large number of $(V,T)$ points (cf.~Table~\ref{table:literature-review}). As shown in this work, a dense sampling and appropriately chosen parametrizations of the free energy are mandatory and facilitate the computation of thermodynamic properties to highest DFT accuracy. The proposed procedure encompasses key insights and techniques distilled from several of the above mentioned studies.~\cite{glensk15understanding-Fah,duff15improved,zhang18temperature_SFEs,grabowski19ab} We lay down a detailed, complete, and pedagogical description of all the steps of the procedure (Methods section below and Supplementary Information) and discuss all relevant numerical and performance aspects.

An important ingredient is the \textit{direct upsampling} technique~\cite{zhou22thermodynamic}---a modification to TU-TILD---where the upsampling is performed directly on MTP configurations. The upsampling establishes DFT level accuracy and accounts for the impact of vibrations on electronic and magnetic free energies. Direct upsampling was introduced previously on a preliminary, purely theoretical example (multi-component alloy).\cite{zhou22thermodynamic} However, the convergence, robustness and optimization of the upsampling was not touched upon and has remained elusive. In the present work, we therefore perform a systematic and rigorous application and analysis of direct upsampling. Importantly, our analysis provides a measure of the number of configurations needed for convergence of the upsampled free energy, which is crucial in keeping the number of highly expensive DFT calculations to a minimum. Another key ingredient to our procedure is the fitting approach and the resulting robustness of the MTP interatomic potential. Here, we propose a novel two-step approach of fitting the MTP to DFT data with additional stabilization measures (by 
`harmonic'
configurations), which further reduces the number of expensive DFT calculations and ensures stable simulations.

We apply the procedure to the above-mentioned, experimentally well-assessed four prototypical systems (bcc Nb, magnetic fcc Ni, fcc Al, hcp Mg) and provide a database-like collection of highly converged equilibrium thermodynamic properties. Beyond the main thermodynamic properties of interest (heat capacity, expansion coefficient, bulk modulus) additional quantities are tabulated in the Supplementary Information, and all properties are provided online. Amongst the investigated systems, Nb offers the most numerical challenges and insights into the methodology, owing to its large anharmonicity, large impact of vibrations on electrons and long-range interactions~\cite{tidholm20temperature-Nb}. In fact, the inherent complexity even challenges the MTPs in reproducing \textit{ab initio} data for Nb. Calculations for Ni and Al are performed on much denser $(V,T)$ grids in comparison to previous results~\cite{grabowski09ab-Al,glensk15understanding-Fah} leading to an improved and systematically assessed convergence of the thermodynamic properties. Some of the thermodynamic properties of Ni (including the impact of magnetism) and Al are reported here 
at the full DFT level of accuracy (bulk modulus for Al and Ni, expansion coefficient for Ni). For both Nb and Mg, the current work is the first instance of evaluation of thermodynamic properties including explicit anharmonicity up to the melting point with DFT accuracy.

\section*{Results}

\subsection*{General overview of the methodology}

The most integral part of the procedure is an accurate representation of the free energy surface up to the melting point over the relevant volume range, from which thermodynamic properties can be derived at various pressures and temperatures. This calls not only for highly stable absolute free energy values, but also for stable first and second derivatives of the free energy surface. Essential to achieving this stability is a dense and optimally distributed set of $(V,T)$ points on which explicit free energy calculations are performed, such as to obtain a smooth and converged parametrization of the surface. This is illustrated in Fig.~\ref{fig:master-method}(f) wherein the blue dots depict a representative and converged $(V,T)$ grid and the arrows the derivatives along different directions. To this end, a fast and reliable method is needed to obtain the free energy for every single $(V,T)$ point, and on that account, we utilize the \textit{direct upsampling} method aided by the outstanding performance of MTPs (cf. Fig.~\ref{fig:master-method}(a)-(e)).

The relevant excitation mechanisms can be captured by starting with the adiabatic decomposition of the total free energy $F(V,T)$ according to the \emph{free energy} Born-Oppenheimer approximation,\cite{grabowski11formation-review} as given by
\begin{equation}
  \begin{split}
  F(V,T) & = E_{\text{0K}}(V)+F^{\text{el}}(V,T)+F^{\text{vib}}(V,T)+F^{\text{mag}}(V,T).
  \end{split}
  \label{equation:1}
\end{equation}
Here, $E_{\text{0K}}$ denotes the 0\,K total energy of the static lattice; $F^{\text{el}}$ is the electronic free energy \emph{including} coupling from atomic vibrations; $F^{\text{vib}}$ is the vibrational free energy which, in our framework, is further decomposed into an effective quasiharmonic (QH) and an anharmonic (AH) part $F^{\text{qh}}$ and $F^{\text{ah}}$ respectively; $F^{\text{mag}}$ is the magnetic free energy including the coupling from electronic and atomic vibrations. The various free energy contributions and their impact on the thermodynamic properties are exemplified for Ni in Fig.~\ref{fig:master-method}(g)-(j).

Contrary to the other excitation mechanisms, there is a lack of standard DFT methods to self-consistently calculate the magnetic excitations. The reasons for this are the complex coupled magnetic degrees of freedom~\cite{
stockem18anomalous} and the relevance of longitudinal spin fluctuations (LSFs)~\cite{ruban07temperature}. So far, one has to resort to effective Heisenberg models fitted to the experimental Curie temperature~\cite{koermann11role-Ni}, DFT-informed semi-empirical heat capacity models, and more recently, magnetic MTPs to facilitate constrained magnetic calculations to allow sampling the LSFs with MD~\cite{novikov22magnetic}. Here, for Ni, we use a thoroughly tested DFT-informed semi-empirical model \cite{zhang18temperature_SFEs}.

A key element in our methodology is the preparation of a high accuracy MTP (high-MTP) in an efficient two-step approach. First, a lower-quality MTP (low-MTP) is fitted to inexpensive low-converged DFT data. This low-MTP is used to rapidly generate MD snapshots at the melting point at various volumes. High-converged DFT (high-DFT) calculations are then performed for the snapshots and the data is used to fit the high-MTP. The high-MTP is stabilized by introducing a 
`harmonic'
configuration containing small interatomic distances into the fitting procedure. This harmonic configuration is obtained by sampling the phase space with an effective QH reference, which is fitted to low-temperature high-DFT forces. This approach ensures that the magnitude of the forces remains physical during the following TI, especially when atoms come close to each other for small $\lambda$ coupling parameter values due to the softness of the harmonic reference. The resulting high-MTP is thus accurate and stable over the relevant part of the phase space.

To obtain the free energy, we perform a $\lambda$-based TI from the effective QH reference to the high-MTP. The TI is performed on $(V,T)$ points on a dense grid in large-size supercells and over long time scales to achieve statistical convergence. To corroborate the free energy differences calculated using the TI, we also perform a temperature integration, which serves both as a cross-check for the TI and substitutes in cases where the TI becomes unstable (e.g., when there is diffusion during the TI and a fixed-lattice reference becomes inadequate). In order to achieve high-accuracy free energies, we propose a modification, specifically a quantum correction, to the conventional temperature integration, to yield the same free energy differences as the TI. A detailed description of the quantum-corrected temperature integration is provided in Supplementary 
Discussion~1\,E.

Direct upsampling on high-MTP snapshots is then utilized to efficiently obtain DFT accuracy including the electronic and magnetic contributions. These contributions implicitly include the coupling effect from thermal vibrations. Once the total free energies are calculated across all the $(V,T)$ grid points, they are parametrized to obtain highly converged free energy surfaces. A Legendre transformation on the free energy surface gives the Gibbs energy $G(p,T)=F(V,T)+pV$, from which thermodynamic properties at a given pressure are obtained. The Methods section contains further information on the involved steps and the Supplementary Information all relevant details.

\subsection*{Accuracy of the MTPs}

An accurate MTP that can reproduce \textit{ab initio} data is crucial for the efficiency of our proposed methodology. The smaller the RMSE, the fewer the number of snapshots that are needed for converging the direct upsampling, as will be discussed below.
In this regard, Fig.~\ref{fig:MTP-error-main} shows the root-mean-square error (RMSE) in the energies and forces of the high-MTP on a high-DFT test set at the respective melting point for the four systems. Comparatively, although the MTP for Nb shows a larger RMSE, the values are still sufficiently small for an efficient evaluation of thermodynamic properties, given our detailed analysis and understanding of direct upsampling. The relatively large RMSE of Nb probably arises from the large anharmonicity and a relatively complex atomic distribution for Nb, as also observed for other highly anharmonic systems in prior works~\cite{grabowski19ab,zhou22thermodynamic,ferrari20frontiers}. In Supplementary 
Methods 2\,C\,3, 
we assess the performance of MTPs with increasing levels.

\subsection*{Optimization of direct upsampling}

The most expensive stage during the free energy calculations is by far the upsampling from high-MTP snapshots to high-DFT (cf.~Table~\ref{table:CPU-time-main-text}). The efficiency of the methodology is thus significantly improved by minimising the number of high-DFT calculations during direct upsampling, while still maintaining highest accuracy in the final free energies. The number of high-DFT calculations needed to achieve a certain accuracy is correlated to the RMSE of the high-MTPs.

Figure~\ref{fig:upsample-num-samples-vs-RMSE-main} shows such a relation for Nb for different convergence criteria (symbols connected with lines), featuring a decreasing number of snapshots for a decreasing energy RMSE of the MTPs. A model MTP is used here as the target system in the upsampling (further details in Supplementary 
Discussion 1\,A).
The trend can be analytically formulated irrespective of the system solely by using the energy RMSE of the MTP and the target accuracy. By approximating the standard deviation of the upsampling as the MTP energy RMSE, assuming a normal distribution, and using the standard error within a 95\% confidence interval, the number of required snapshots is estimated as
 \begin{gather}
     n = \left(\frac{2 \,\, \mathrm{RMSE}}{c} \right)^2,\label{eq-main:snapshot-estimate}
 \end{gather}
where $\pm c$ is the target accuracy. For instance, for Nb with an MTP RMSE of 2\,meV/atom, around 40 snapshots are needed to achieve a target accuracy of 0.6\,meV/atom. This is indicated with the purple star in the figure. Besides Nb, the other systems in this work have high-MTPs fitted even to within 0.5\,meV/atom and \SI{0.05}{eV/\angstrom} accuracy in energies and forces respectively (Fig.~\ref{fig:MTP-error-main}), requiring much fewer snapshots for the convergence of the direct upsampling. For instance, Mg with an MTP RMSE of 0.16\,meV/atom would require only around 10 snapshots to reach 0.1\,meV/atom accuracy, as indicated by the red star in Fig.~\ref{fig:upsample-num-samples-vs-RMSE-main}. The RMSEs here are evaluated at the melting temperature, and provide the largest estimates of the standard deviation. At lower temperatures the standard deviation decreases and fewer snapshots are needed.

\subsection*{Significance of the \texorpdfstring{$(V,T)$}{(V,T)} grid density}

The proposed formalism enables affordable full free energy calculations on a higher number of $(V,T)$ points as compared to previous works. 
The importance of the $(V,T)$ grid density is illustrated in Figs.~\ref{fig:convergence-main} (a) and (b), which show the anharmonic free energy for Nb at $T_\mathrm{melt}$ and the resulting bulk modulus calculated using different grid densities. The anharmonic free energies in Fig.~\ref{fig:convergence-main} (a) are given with respect to the anharmonic free energy calculated using the highest-density grid $(11 V \times 13 T)$. 
As the grid density increases, the anharmonic free energy begins to converge as noticed by the closer proximity of the dashed red curve to the solid black line, in comparison to the dotted gray curve. It is observed that even a small difference in the anharmonic free energy (about 0.5~meV/atom at $T_\mathrm{melt}$) can lead to a considerable change in the high temperature bulk modulus (18$\%$ change at $T_\mathrm{melt}$ in the drop in the bulk modulus from 0\,K). Moreover, calculations on a coarse $4 V \times 4 T$ grid are also not sufficient to capture the qualitative (more quadratic) behavior of the bulk modulus at higher temperature (Fig.~\ref{fig:convergence-main} (b)). This reveals the importance of a highly converged free energy surface with respect to the grid density,
especially at higher temperatures and higher volumes. For further improvement, we additionally choose grid points above $T_\textrm{melt}$ and $V_\textrm{melt}$ to obtain converged thermodynamic properties also at the melting point. Such a study is made feasible owing to the rapidness of our methodology.

Once free energy calculations are performed on a sufficiently dense $(V,T)$ grid, it is crucial to parametrize the surface, in particular the anharmonic free energy, with a sufficiently high-order polynomial basis. This is illustrated in Figs.~\ref{fig:convergence-main} (c) and (d) which show the anharmonic free energy (with respect to the 4th order parametrization) calculated using different orders and the resulting thermal expansion coefficient. Although a second order polynomial fit (gray curve) of $F^\mathrm{ah}(V,T)$ differs from the fourth order by less than $1$\,meV/atom, it amounts to a 20 percent difference in the expansion coefficient at the melting point. As the order of the polynomial increases further to three and four, the free energy differs by less than $0.2$\,meV/atom, leading to converged thermodynamic properties (red dashed and black solid curves fall on each other).

\subsection*{Benchmarks of the methodology}

Taking both an optimized direct upsampling and a converged $(V,T)$ grid into account, Table~\ref{table:CPU-time-main-text} shows the total computational cost of obtaining thermodynamic properties to the desired accuracy using the current framework for Nb. The values are compared to the state-of-the-art TU-TILD+MTP scheme, in which MTPs were fitted directly to AIMD energies and forces, and a second TI was performed from the MTP to DFT, prior to upsampling. A 4.5-times speed-up is achieved during high-MTP fitting by using the two-stage training procedure (see the top half of Table~\ref{table:CPU-time-main-text}). Through direct upsampling from precise high-MTPs and an optimized number of snapshots, we completely do away with the second stage of TU-TILD (TI from MTP to DFT), thereby achieving a 4.8-times gain in speed during free energy calculations using a $11\times13$ $(V,T)$ grid, shown in the bottom half of the table. Although the speed-up coming from direct upsampling in comparison to TU-TILD was mentioned in Ref.~\citen{zhou22thermodynamic}, here we optimize both the number of snapshots for a single $(V,T)$ point and the total number of grid points needed for the evaluation of thermodynamic properties. 

\subsection*{Free energies for the prototype systems\label{section:result-free-energies}}

In Figure~\ref{fig:free-energy}, we highlight key insights from the calculated free energy surfaces, from which the target thermodynamic properties are derived. The first column provides the Gibbs energy $G(T)$ at ambient pressure for the four elements using the GGA-PBE approximation to the exchange-correlation (XC) functional. Gibbs energies are the fundamental input to the calculation of phase diagrams (CALPHAD\cite{dinsdale91SGTE}). On the total scale, the full DFT Gibbs energy curves (solid lines) are closely tracing the CALPHAD values. Differences can be noticed at high temperatures when the DFT curves are referenced with respect to the CALPHAD data as shown in the insets. The discrepancy is not a shortcoming of the present methodology, but instead due to the inherent limitation of the local nature of the standard XC functionals. For Al and Mg, we also demonstrate the results using the LDA XC functional. Results from both functionals can act as an `\textit{ab initio} confidence interval', as was shown previously\cite{grabowski07ab-fcc}.

In the second column in Fig.~\ref{fig:free-energy}, the free energy at the melting point is plotted as a function of volume. The $F$-$V$ curves including all excitation mechanisms (solid lines) are analogous to the conventional $E$-$V$ curves at 0\,K, but corresponding to $T_\textrm{melt}$. They contain, for example, information on the equilibrium volume and the (isothermal) bulk modulus at the melting point. In contrast to the $E$-$V$ curves at 0\,K, the $F$-$V$ curves are dominated by thermal excitations. The anharmonic contribution (calculated with the effective QH as reference) is large for Nb ($\approx+50$\,meV/atom, see also third column). This strong anharmonic behavior can be intuited by the open bcc structure of Nb that favors vibrational entropy, as compared to the close packed fcc and hcp structures for the other elements. The electronic free energies are large for Nb and Ni ($\approx-100$ and $-50$ meV/atom), which can be corroborated with the Fermi-level contribution of a smeared-out electronic density of states.\cite{zhang17accurate} For Ni, the magnetic contribution is as large as the electronic contribution, with its strength determined by the local magnetic moments on the Ni atoms. It needs to be stressed that the impact of thermal vibrations on the electronic and magnetic free energy is important, as the vibrations break the symmetric arrangement of the atoms and thereby significantly smoothen the electronic density of states.

As noted above, reaching the desired accuracy in the thermodynamic properties that require first and second derivatives of the free energy requires control over sub-meV differences in the free energies. In particular, it is of crucial importance to faithfully describe the physically relevant variations of the free energy with volume and temperature (cf.~the wavy dependence for Ni's $F^\text{ah}(V)$ in Fig.~\ref{fig:free-energy}), while at the same time avoiding any overfitting. The sufficiently dense sets of explicitly computed $F(V,T)$ points are seen in the right column in Fig.~\ref{fig:free-energy}, guaranteeing convergence with respect to the number of basis elements in the expansion of the free energy surface.

\subsection*{Thermodynamic properties for the prototype systems}

Figure~\ref{fig:properties} shows the target thermodynamic properties---the isobaric heat capacity $C_p(T)$, the linear thermal expansion coefficient $\alpha(T)$ and the adiabatic bulk modulus $B_S(T)$---calculated up to the melting point using the current $\textit{ab initio}$ framework, for Nb, Ni, Al and Mg, including the different excitation mechanisms (provided in the legend). Experimental values are shown as blue circles for comparison, and our calculations including all excitation mechanisms (solid lines) show excellent  agreement. 

The unprecedented accuracy achievable with the current framework is apparent in the results for Nb (first row in Fig.~\ref{fig:properties}). We have shown in the previous section that Nb has the largest electronic and anharmonic contribution of the four systems studied. Nb also possesses long-range interactions at 0\,K that gradually disappear as temperature increases.~\cite{tidholm20temperature-Nb} The disappearance is validated by explicit TI calculations on large-size cells from high-MTP to DFT (see Supplementary 
Discussion 1\,D). 
Additionally for Nb, the energies predicted by the high-MTP (fitted at $T_\textrm{melt}$) on low temperature configurations with de-coupled phonons become less accurate (see Supplementary 
Discussion 1\,D\,3).
However, the loss in accuracy gets fully compensated for in the directly upsampled free energy. Considering all such challenges offered by Nb, we still achieve remarkable accuracy with experiments. In particular, the calculations are able to reproduce the strong temperature dependence and the curvature of the expansion coefficient all the way to the melting point. 
This is made possible by virtue of a dense $(V,T)$ grid on which the free energy calculations are performed, leading to highly converged numerical first and second derivatives. 

The results for Nb also showcase how the different thermodynamic properties probe distinct features of the free energy surface. The strong anharmonic free energy discussed above affects significantly the expansion coefficient (yellow dotted vs red dashed line) and contributes to its curvature. In contrast, the heat capacity is much less affected by the anharmonic free energy. The situation is opposite for the electronic thermal excitations which strongly increase the computed heat capacity of Nb bringing it close to experiment, while the expansion coefficient is less affected. It is the dependence on derivatives along distinct directions on the free energy surface that brings about the different behavior of the thermodynamic properties. 

In the second row in Fig.~\ref{fig:properties}, we present the results for Ni. By virtue of its construction, the model can well predict the peak in the heat capacity at the Curie temperature arising from the second order magnetic phase transition. The validity of the current approach for calculating magnetic free energies and the negligibility of contributions beyond (e.g., LSFs) have been thoroughly proven in earlier works.\cite{zhang18temperature_SFEs,fang10origin} However, the magnetic model utilized for Ni cannot capture the small peak in the experimental expansion coefficient originating from the magnetic phase transition. Other, more elaborate magnetic models could be incorporated into the present framework to further improve the magnetic description.

The results for Al and Mg (last two rows in Fig.~\ref{fig:properties}) are provided for both the LDA and PBE XC functionals. The difference coming from the XC functionals is evident in the calculated bulk modulus, where PBE and LDA results are identified as a lower and upper bound to the experimental bulk modulus, providing an `\textit{ab initio} confidence interval'\cite{grabowski07ab-fcc} similarly as for the Gibbs energies mentioned above.
Although this has been documented for some systems and properties in literature~\cite{grabowski07ab-fcc,grabowski09ab-Al}, we report it for the first time for the bulk moduli at full DFT accuracy.

\section*{Discussion}

The procedure described in this work presents a complete and very efficient methodology to predict highly accurate \textit{ab initio} free energy surfaces and thermodynamic properties up to the melting point. The procedure has been developed and streamlined by considering key insights and findings from \textit{ab initio}-studies over the past decade, and by utilizing direct upsampling and advanced machine learning potentials (i.e., MTPs). The current proposition makes calculations on a very dense $(V,T)$ grid affordable. It also takes the relevant finite-temperature excitations into account---the electronic free energy, the magnetic contribution, anharmonicity, and coupling effects. Consequently, even sub-meV differences that affect high-temperature thermodynamics are factored in.

The proposed procedure can be combined with any \textit{ab initio} electronic structure approach, and with advanced exchange-correlation functionals, e.g., hybrid functionals \cite{krukau06influence}, meta-generalized gradient approximations\cite{sun15strongly}, or even with the random-phase-approximation and the adiabatic connection fluctuation-dissipation theorem\cite{grabowski15random}. The more computationally expensive functionals can be employed either for the $E$-$V$ curve or during upsampling in order to reach higher \textit{ab initio} accuracy. The application and efficiency of the procedure relies primarily on accurately fitted MTPs. Hence, the technique can also be employed to more complex and possibly disordered systems such as multi-component alloys, for some of which MTPs within 3\,meV/atom energy RMSE (similar to that for Nb) exist in literature~\cite{ferrari20frontiers}. The here-derived optimization and analysis (in particular for the direct upsampling, where the RMSE of the MTP determines the efficiency) can be applied to efficiently obtain well-converged free energies. Free energies of arbitrary phases can be computed, as long as it is possible to perform sufficient dynamics within the considered phase, such as to obtain statistically converged quantities. In addition to equilibrium phases, properties of meta-stable and dynamically stabilized phases are accessible. Once bulk free energies are available, the procedure can be extended to systems with various kinds of defects (e.g., vacancies, surfaces, interfaces, grain boundaries). With the calculated free energies of the ideal bulk and the defective structure, one can evaluate the formation Gibbs energy of the respective defect. In the case of thermal vacancies, their presence at elevated temperatures will mildly contribute to the thermodynamic properties of the system, providing an even more realistic comparison to experimental data. For example, in Al, thermal vacancies are known to add 0.07\,$k_B$ to the heat capacity at the melting point.~\cite{grabowski09ab-Al} Predictions of highly accurate solid free energy surfaces are also required as a reference for liquid phase calculations, e.g., in the TOR-TILD methodology.\cite{zhu17efficient} From the free energy surface, other thermodynamic properties such as the enthalpy, entropy and Grüneisen parameter can be likewise derived.

Data sets for the studied properties---Gibbs energy, enthalpy, entropy, isobaric and isochoric heat capacity, thermal expansion coefficient, isothermal and adiabatic bulk moduli---of the here investigated prototype systems are tabulated in  
Supplementary Discussion 4  
and provided online (see 
Data Availability and Code Availability). 
With the introduced, robust and efficient methodology, we are well-positioned to extend this work to other systems and develop an entirely \textit{ab initio} thermodynamic database.

\section*{METHODS\label{section:methods}}


An overview of the steps involved in our framework to calculate the relevant free energy contributions and eventually thermodynamic properties is provided here, with more details in the Supplementary Information.
\vspace{\baselineskip}

\noindent\textbf{Energy-volume curve and Debye-Gr\"uneisen model} 
(\textit{Supplementary Methods 2\,A})

We start with a conventional 0\,K energy-volume ($E$-$V$) curve calculation with very well converged DFT parameters (2$\times$\texttt{ENMAX}, i.e., twice the maximum recommended energy cut-off, and  $>60,000$ $k$-points (kp)$\times$atoms) in a reasonable volume range (typically $-8\%$ to $+12\%$) around the 0\,K equilibrium volume on a mesh of at least 11 volumes. The DFT computed values are used to fit the Vinet equation of state~\cite{vinet87compressibility} to obtain $E_{\text{0K}}(V)$ for Eq.~\eqref{equation:1}. The $E$-$V$ curve is also used to obtain the free energy within the Debye-Grüneisen model\cite{moruzzi88calculated}, from which we estimate the relevant volume range for the free energy calculations. For the melting temperature we use experimental values. 

\vspace{\baselineskip}

\noindent\textbf{Low quality MTP} 
(\textit{Supplementary Methods 2\,B}) 

In order to efficiently generate a highly accurate MTP (next point), first, a lower quality MTP is obtained to rapidly sample the vibrational phase space. For that purpose, AIMD is performed using low-accuracy DFT (low-DFT) parameters (1$\times$\texttt{ENMAX} energy cut-off, 288-432 kp$\times$atoms and without electronic temperature) on a coarse set of four volumes from the relevant volume range, at the melting point, on small-size supercells of 32-54 atoms and a timestep of 5\,fs for 1000 steps. Uncorrelated snapshots are chosen to train a lower quality MTP (low-MTP) with few basis functions. Specifically, levels of lev$_\mathrm{max}=6$-$14$ and radial basis sizes of $N_Q=8$ are used, resulting in 25-88 fitting parameters. The minimum distances ($R_\mathrm{min}$) are 1.33-2.00~\AA\ and the cutoff radii ($R_\mathrm{cut}$) are 4.96-6.24~\AA.
\vspace{\baselineskip}

\noindent\textbf{High quality MTP} 
(\textit{Supplementary Methods 2\,C})

The low-MTP is used to perform $NVT$ MD simulations in a medium-size supercell (96-128 atoms) for 8-10 volumes in the relevant volume range, at the melting point, using a timestep of 1\,fs for 9000 steps. From the trajectories, 30 snapshots are chosen from each volume, and DFT calculations with high converged (high-DFT) parameters (1.5$\times$\texttt{ENMAX} and 6100-8200 kp$\times$atoms) are performed to serve as the training set for a higher accuracy MTP (high-MTP). The high-MTP has significantly more basis functions than the low-MTP, i.e., a level lev$_\mathrm{max}=20$ and $N_Q=8$ radial basis functions, resulting in 332 fitting parameters. A `harmonic' snapshot generated with the effective QH reference (next point) is included into the fitting database to stabilize the high-MTP for small interatomic distances. The resulting $R_\mathrm{min}$ are 1.36-2.00~\AA\ and the $R_\mathrm{cut}$ are 4.97-6.10~\AA. The electronic contribution is not included in the DFT database for the high-MTP, because the current MTPs do not entail electronic degrees-of-freedom. Note that within our approach, no expensive high-DFT AIMD is required.
\vspace{\baselineskip}
\vspace{\baselineskip}

\noindent\textbf{Effective QH reference} 
(\textit{Supplementary Methods 2\,D})

We use an effective QH model as a reference for the TI to high-MTP. To obtain it, low temperature MD (e.g., at 20\,K) is run using the high-MTP on medium-size supercells at several volumes in the relevant range for 10,000 steps of 1\,fs each. From this, we choose 30 snapshots for each volume and calculate high-DFT forces. An effective dynamical matrix is then fitted to these forces and extended to larger system sizes. Each of the force constants is parametrized using a second-order polynomial in $V$. For each $V$, $F^{\text{qh}}(T)$ is calculated on a 30$\times$30$\times$30 $q$-point mesh in reciprocal space.

An effective QH reference is preferred to a 0\,K QH due to its wider applicability and stability (even for 0\,K unstable systems that become stable at elevated temperatures), efficiency for low-symmetry systems, and a fitting dataset where errors in atomic forces are averaged out. 
\vspace{\baselineskip}

\noindent\textbf{TI from effective QH to high-MTP} 
(\textit{Supplementary Methods 2\,E})

Next, we perform TI using Langevin dynamics (TILD) to obtain the free energy difference between the effective QH reference and the high-MTP. The high-MTP is used as an intermediate potential to minimize high-DFT calculations for the final free energy. During TILD, the free energy difference is given by
\begin{eqnarray}
    \Delta F^{\mathrm{qh \rightarrow MTP}}  =  \int _0 ^1 d \lambda \left< E^{\mathrm{MTP}} - E^{\mathrm{qh}} \right>_\lambda,
\end{eqnarray}
where $\lambda (=0\,...\,1)$ dictates the coupled system with energy $E_{\lambda}=(1-\lambda)E^{\mathrm{qh}}+\lambda E^{\mathrm{MTP}}$. TILD is performed on  large-size supercells (432-500 atoms) with a timestep of 1\,fs for 50,000 steps. For each $(V,T)$, a very dense set of around 20 $\lambda$ values is used. We then integrate over $\lambda$ with an analytical fit based on a tangential function to obtain $\Delta F^{\mathrm{qh \rightarrow MTP}}$.

In certain situations, the $\lambda$-based TILD calculations cannot be straightforwardly performed. For example, in systems that feature diffusion of atoms and exchange of sites during the TI, a fixed-lattice reference such as the effective QH becomes inadequate.
Then, it is possible to utilize an alternative method to calculate free energy differences, i.e., \textit{temperature integration}. In the present study, temperature integration has been used to corroborate the TILD calculations. Details and special considerations about this method can be found in Supplementary 
Discussion 1\,E.

One should keep in mind that the present step of the procedure involves no DFT calculations. Hence, we can afford to perform highly converged free energy calculations on large-size supercells which also include contributions from vibrations with long wavelengths. Moreover, finite-size effects (e.g., stacking fault formation in small-size Mg hcp cells, see Supplementary 
Discussion 1\,F) 
are avoided by utilizing large-size supercells.

Figure~\ref{fig:master-method}(a)-(c) encapsulates the just discussed three stages. Here, $\left< E^{\mathrm{MTP}} - E^{\mathrm{qh}} \right>$ is plotted against $\lambda$ for a single volume and a set of temperatures for a 500-atom Ni supercell, where $\lambda=0$ corresponds to the effective QH reference and $\lambda=1$ corresponds to a high-MTP. The high density of $\lambda$s as seen in the figure, achieves good convergence and is affordable due to the inexpensive nature of this step.
\vspace{\baselineskip}

\noindent\textbf{Direct upsampling from high-MTP to high-DFT} 
(\textit{Supplementary Methods~2\,F})

In the spirit of the direct-upsampling approach, we perform high-DFT runs on high-MTP-generated snapshots (illustrated by red dots in Fig.~\ref{fig:master-method}(d)). In addition to reaching DFT accuracy for the vibrational free energy, we also include the electronic contribution. The notion behind upsampling from MTP relies on its superior accuracy, due to which highly converged upsampled energies can be achieved within a few tens of snapshots. (As discussed, the speed of convergence depends on the accuracy of the MTP. For the full analysis see Supplementary 
Discussion 1\,A.)

Since this step involves computationally demanding calculations (cf.~Table \ref{table:CPU-time-main-text}), they are restricted to medium-size supercells. The upsampling is performed in two parts. First, the free energy difference between high-MTP and high-DFT is calculated using the free energy perturbation expression, as given by
\begin{eqnarray}
    \Delta F^{\mathrm{up}} =  -k_B T \ln \left< \exp \left( -\frac{E^{\mathrm{DFT}} - E^{\mathrm{MTP}}}{k_B T}  \right) \right>_\mathrm{MTP},
\label{fep}
\end{eqnarray}
where $E^{\mathrm{DFT}}$ and $E^{\mathrm{MTP}}$ are high-DFT (without electronic temperature) and high-MTP energies. The averaging is performed on uncorrelated high-MTP snapshots. Equation~\eqref{fep} corresponds to the full free energy perturbation formula. We note that at least the second order approximation of the
perturbation equation is vital to capture the full upsampled free energy difference, (see Supplementary 
Discussion 1\,A\,1).
Adding the upsampled free energy to $\Delta F^{\mathrm{qh \rightarrow MTP}}$ from the previous stage gives the anharmonic vibrational contribution:
\begin{align}
    F^{\text{ah}}=\Delta F^{\mathrm{qh \rightarrow MTP}}+ \Delta F^{\mathrm{up}}.
\end{align}
In the second part, we calculate the electronic free energy $F^{\text{el}}$ using the same snapshots, as given by 
  \begin{eqnarray}
    F^\textrm{el} = -k_B T \ln \left< \exp \left( -\frac{E^{\mathrm{DFT}}_{\mathrm{el}} - E^{\mathrm{DFT}}}{k_B T}  \right) \right>_\mathrm{MTP},
  \end{eqnarray}
where $E^{\mathrm{DFT}}_{\mathrm{el}}$ is the high-DFT energy including electronic temperature. Since it is performed on MD snapshots, the upsampling also accounts for the effect of atomic vibrations on the electronic free energy. 
 
For magnetic systems (Ni, in this case), we extract average magnetic moments from the high-DFT runs (including electronic temperature). Along with the experimental Curie temperature, they are used as model parameters for an empirical heat capacity formula as a function of temperature, and for the corresponding numerically integrated magnetic free energy $F^{\text{mag}}(V,T)$.
\vspace{\baselineskip}

\noindent\textbf{Parametrization and thermodynamic properties} 
(\textit{Supplementary Methods~2\,G})

A sufficiently dense $(V,T)$ grid is necessary to fit a smooth free energy surface and numerically calculate converged second derivatives in the evaluation of $C_{p}(T)$ and $B_S(T)$ all the way to $T_{\mathrm{melt}}$. For this purpose, it is also recommended to extend the grid further (by one or two points in $V$ and $T$) beyond the corresponding melting temperature and volume, i.e., to $T > T_{\mathrm{melt}}$ and $V > V_{\mathrm{melt}}$.

Smooth surfaces in $(V,T)$ are fit to each of the free energy contributions. For a dense temperature mesh (steps of 1 kelvin), $F^{\text{qh}}(T)$ is parametrized with a third order polynomial in $V$ to obtain the effective QH free energy surface $F^{\text{qh}}(V,T)$. The anharmonic free energies $F^{\mathrm{ah}}$ at every $(V,T)$ grid point are used to fit a continuous anharmonic free energy surface using renormalized effective anharmonic frequencies~\cite{grabowski09ab-Al,zhang18temperature_SFEs}. Here, an adequate polynomial basis is a requisite for the frequencies since the derived thermodynamic quantities are particularly sensitive to them. A fourth order polynomial in $V$ and $T$ is found to be a conservative and well-converged basis set. The total vibrational free energy can be obtained by summing up the effective QH and anharmonic surfaces: $F^{\text{vib}}(V,T)=F^{\text{qh}}(V,T)+F^{\text{ah}}(V,T)$. The electronic free energies $F^{\mathrm{el}}$ at every $(V,T)$ grid point are used to fit a polynomial in $(V,T)$~\cite{zhang17accurate} to obtain a continuous electronic free energy surface $F^{\mathrm{el}}(V,T)$. The average magnetic moments $m$ at every $(V,T)$ grid point are first parametrized with a polynomial in $T$, and later with a polynomial in $V$ for every 1 kelvin step, to obtain a continuous $F^{\text{mag}}(V,T)$ surface.

All contributions are discretized in 1 kelvin steps, and the total free energy is obtained by summing the contributions to the $E$-$V$ curve (cf.~Eq.~\eqref{equation:1}). Numerical first and second derivatives along different directions are performed to obtain $C_p(T)$, $\alpha(T)$ and $B_S(T)$.


\section*{Data Availability}
The necessary data are available in the DaRUS Repository and can be accessed via \textcolor{blue}{\url{https://doi.org/10.18419/darus-3239}}. The repository contains the training sets (VASP OUTCAR files), the low-MTPs and high-MTPs, the effective QH potentials, and the final thermodynamic database (properties) for the four unaries. 

\section*{Code Availability}
The scripts for performing thermodynamic integration, direct upsampling, free energy parametrizations, and the thermodynamic database calculations are available on request from the authors.

\begin{acknowledgments}
We appreciate fruitful discussions with Yuji Ikeda, Xi Zhang, Matthias vom Bruch, Konstantin Gubaev, and Nikolay Zotov. This project has received funding from the European Research Council (ERC) under the European Union’s Horizon 2020 research and innovation programme (grant agreement No 865855). The authors acknowledge support by the state of Baden-Württemberg through bwHPC and the German Research Foundation (DFG) through grant No.~INST 40/575-1 FUGG (JUSTUS 2 cluster). B.G. acknowledges the support by the Stuttgart Center for Simulation Science (SimTech). P.S. would like to thank the Alexander von Humboldt Foundation for their support through the Alexander von Humboldt Postdoctoral Fellowship Program.

\end{acknowledgments}

\vspace{0.6cm}
\section*{Author contributions}
All authors designed the project, discussed the results, and wrote the manuscript. B.G. provided scripts; J.J., P.S., and A.F. performed the calculations. 

\section*{Competing interests}
The authors declare no competing interests.

\section*{Additional information}
\textbf{Supplementary Information} accompanies.


\begin{thebibliography}{0}%
\makeatletter
\providecommand \@ifxundefined [1]{%
 \@ifx{#1\undefined}
}%
\providecommand \@ifnum [1]{%
 \ifnum #1\expandafter \@firstoftwo
 \else \expandafter \@secondoftwo
 \fi
}%
\providecommand \@ifx [1]{%
 \ifx #1\expandafter \@firstoftwo
 \else \expandafter \@secondoftwo
 \fi
}%
\providecommand \natexlab [1]{#1}%
\providecommand \enquote  [1]{``#1''}%
\providecommand \bibnamefont  [1]{#1}%
\providecommand \bibfnamefont [1]{#1}%
\providecommand \citenamefont [1]{#1}%
\providecommand \href@noop [0]{\@secondoftwo}%
\providecommand \href [0]{\begingroup \@sanitize@url \@href}%
\providecommand \@href[1]{\@@startlink{#1}\@@href}%
\providecommand \@@href[1]{\endgroup#1\@@endlink}%
\providecommand \@sanitize@url [0]{\catcode `\\12\catcode `\$12\catcode
  `\&12\catcode `\#12\catcode `\^12\catcode `\_12\catcode `\%12\relax}%
\providecommand \@@startlink[1]{}%
\providecommand \@@endlink[0]{}%
\providecommand \url  [0]{\begingroup\@sanitize@url \@url }%
\providecommand \@url [1]{\endgroup\@href {#1}{\urlprefix }}%
\providecommand \urlprefix  [0]{URL }%
\providecommand \Eprint [0]{\href }%
\providecommand \doibase [0]{https://doi.org/}%
\providecommand \selectlanguage [0]{\@gobble}%
\providecommand \bibinfo  [0]{\@secondoftwo}%
\providecommand \bibfield  [0]{\@secondoftwo}%
\providecommand \translation [1]{[#1]}%
\providecommand \BibitemOpen [0]{}%
\providecommand \bibitemStop [0]{}%
\providecommand \bibitemNoStop [0]{.\EOS\space}%
\providecommand \EOS [0]{\spacefactor3000\relax}%
\providecommand \BibitemShut  [1]{\csname bibitem#1\endcsname}%
\let\auto@bib@innerbib\@empty
\end{thebibliography}%


\begin{thebibliography}{10}
\expandafter\ifx\csname url\endcsname\relax
  \def\url#1{\texttt{#1}}\fi
\expandafter\ifx\csname urlprefix\endcsname\relax\def\urlprefix{URL }\fi
\providecommand{\bibinfo}[2]{#2}
\providecommand{\eprint}[2][]{\url{#2}}

\bibitem{einstein07plancksche}
\bibinfo{author}{Einstein, A.}
\newblock \bibinfo{title}{{Die Plancksche Theorie der Strahlung und die Theorie
  der spezifischen W\"arme}}.
\newblock \emph{\bibinfo{journal}{Ann. Phys. (Leipzig)}}
  \textbf{\bibinfo{volume}{327}}, \bibinfo{pages}{180--190}
  (\bibinfo{year}{1907}).

\bibitem{zheng19understanding}
\bibinfo{author}{Zheng, Q.} \emph{et~al.}
\newblock \bibinfo{title}{Understanding glass through differential scanning
  calorimetry}.
\newblock \emph{\bibinfo{journal}{Chem. Rev.}} \textbf{\bibinfo{volume}{119}},
  \bibinfo{pages}{7848--7939} (\bibinfo{year}{2019}).

\bibitem{vanSchilfgaarde99origin}
\bibinfo{author}{van Schilfgaarde, M.}, \bibinfo{author}{Abrikosov, I.~A.} \&
  \bibinfo{author}{Johansson, B.}
\newblock \bibinfo{title}{Origin of the {Invar} effect in iron--nickel alloys}.
\newblock \emph{\bibinfo{journal}{Nature}} \textbf{\bibinfo{volume}{400}},
  \bibinfo{pages}{46--49} (\bibinfo{year}{1999}).

\bibitem{lee20temperature}
\bibinfo{author}{Lee, C.} \emph{et~al.}
\newblock \bibinfo{title}{Temperature dependence of elastic and plastic
  deformation behavior of a refractory high-entropy alloy}.
\newblock \emph{\bibinfo{journal}{Sci. Adv.}} \textbf{\bibinfo{volume}{6}},
  \bibinfo{pages}{eaaz4748} (\bibinfo{year}{2020}).

\bibitem{touloukian75thermophysical-TPRC}
\bibinfo{author}{Touloukian, Y.~S.}, \bibinfo{author}{Kirby, R.},
  \bibinfo{author}{Taylor, R.} \& \bibinfo{author}{Desai, P.}
\newblock \emph{\bibinfo{title}{Thermal Expansion Metallic Elements and
  Alloys}} (\bibinfo{publisher}{Thermophysical Properties of Matter -
  the {TPRC} Data Series vol.~12, IFI/Plenum}, \bibinfo{year}{1975}).

\bibitem{landolt81metals}
\bibinfo{editor}{Hellwege, K.-H.} \& \bibinfo{editor}{Olsen, J.~L.} (eds.)
  \emph{\bibinfo{title}{Metals: Phonon States, Electron States and {Fermi}
  Surfaces.}} (\bibinfo{publisher}{Landolt-B\"ornstein - Group {III}
  Condensed Matter vol.~13A, Springer-Verlag}, \bibinfo{year}{1981}).

\bibitem{draxl19nomad}
\bibinfo{author}{Draxl, C.} \& \bibinfo{author}{Scheffler, M.}
\newblock \bibinfo{title}{The {NOMAD} laboratory: from data sharing to
  artificial intelligence}.
\newblock \emph{\bibinfo{journal}{J. Phys.: Mater.}}
  \textbf{\bibinfo{volume}{2}}, \bibinfo{pages}{036001} (\bibinfo{year}{2019}).

\bibitem{legrain17how}
\bibinfo{author}{Legrain, F.}, \bibinfo{author}{Carrete, J.},
  \bibinfo{author}{van Roekeghem, A.}, \bibinfo{author}{Curtarolo, S.} \&
  \bibinfo{author}{Mingo, N.}
\newblock \bibinfo{title}{How chemical composition alone can predict
  vibrational free energies and entropies of solids}.
\newblock \emph{\bibinfo{journal}{Chem. Mater.}} \textbf{\bibinfo{volume}{29}},
  \bibinfo{pages}{6220--6227} (\bibinfo{year}{2017}).

\bibitem{kirklin15OQMD}
\bibinfo{author}{Kirklin, S.} \emph{et~al.}
\newblock \bibinfo{title}{The {Open Quantum Materials Database (OQMD)}:
  assessing the accuracy of {DFT} formation energies}.
\newblock \emph{\bibinfo{journal}{npj Comput. Mater.}}
  \textbf{\bibinfo{volume}{1}}, \bibinfo{pages}{15010} (\bibinfo{year}{2015}).

\bibitem{JCA15}
\bibinfo{author}{de~Jong, M.} \emph{et~al.}
\newblock \bibinfo{title}{Charting the complete elastic properties of inorganic
  crystalline compounds}.
\newblock \emph{\bibinfo{journal}{Sci. Data}} \textbf{\bibinfo{volume}{2}},
  \bibinfo{pages}{150009} (\bibinfo{year}{2015}).

\bibitem{talirz20materials}
\bibinfo{author}{Talirz, L.} \emph{et~al.}
\newblock \bibinfo{title}{{Materials Cloud}, a platform for open computational
  science}.
\newblock \emph{\bibinfo{journal}{Sci. Data}} \textbf{\bibinfo{volume}{7}},
  \bibinfo{pages}{299} (\bibinfo{year}{2020}).

\bibitem{glensk15understanding-Fah}
\bibinfo{author}{Glensk, A.}, \bibinfo{author}{Grabowski, B.},
  \bibinfo{author}{Hickel, T.} \& \bibinfo{author}{Neugebauer, J.}
\newblock \bibinfo{title}{Understanding anharmonicity in fcc materials: From
  its origin to ab initio strategies beyond the quasiharmonic approximation}.
\newblock \emph{\bibinfo{journal}{Phys. Rev. Lett.}}
  \textbf{\bibinfo{volume}{114}}, \bibinfo{pages}{195901}
  (\bibinfo{year}{2015}).

\bibitem{hellman11lattice}
\bibinfo{author}{Hellman, O.}, \bibinfo{author}{Abrikosov, I.~A.} \&
  \bibinfo{author}{Simak, S.~I.}
\newblock \bibinfo{title}{Lattice dynamics of anharmonic solids from first
  principles}.
\newblock \emph{\bibinfo{journal}{Phys. Rev. B}} \textbf{\bibinfo{volume}{84}},
  \bibinfo{pages}{180301} (\bibinfo{year}{2011}).

\bibitem{tidholm20temperature-Nb}
\bibinfo{author}{Tidholm, J.} \emph{et~al.}
\newblock \bibinfo{title}{Temperature dependence of the {Kohn} anomaly in bcc
  {Nb} from first-principles self-consistent phonon calculations}.
\newblock \emph{\bibinfo{journal}{Phys. Rev. B}}
  \textbf{\bibinfo{volume}{101}}, \bibinfo{pages}{115119}
  (\bibinfo{year}{2020}).

\bibitem{adams21anharmonic}
\bibinfo{author}{Adams, D.~J.}, \bibinfo{author}{Wang, L.},
  \bibinfo{author}{Steinle-Neumann, G.}, \bibinfo{author}{Passerone, D.} \&
  \bibinfo{author}{Churakov, S.~V.}
\newblock \bibinfo{title}{Anharmonic effects on the dynamics of solid aluminium
  from ab initio simulations}.
\newblock \emph{\bibinfo{journal}{J. Phys.: Condens. Matter}}
  \textbf{\bibinfo{volume}{33}}, \bibinfo{pages}{175501}
  (\bibinfo{year}{2021}).

\bibitem{souvatzis08entropy}
\bibinfo{author}{Souvatzis, P.}, \bibinfo{author}{Eriksson, O.},
  \bibinfo{author}{Katsnelson, M.~I.} \& \bibinfo{author}{Rudin, S.~P.}
\newblock \bibinfo{title}{Entropy driven stabilization of energetically
  unstable crystal structures explained from first principles theory}.
\newblock \emph{\bibinfo{journal}{Phys. Rev. Lett.}}
  \textbf{\bibinfo{volume}{100}}, \bibinfo{pages}{095901}
  (\bibinfo{year}{2008}).

\bibitem{junkaew14ab}
\bibinfo{author}{Junkaew, A.}, \bibinfo{author}{Ham, B.},
  \bibinfo{author}{Zhang, X.} \& \bibinfo{author}{Arróyave, R.}
\newblock \bibinfo{title}{Ab-initio calculations of the elastic and
  finite-temperature thermodynamic properties of niobium- and magnesium
  hydrides}.
\newblock \emph{\bibinfo{journal}{Int. J. Hydrogen Energy}}
  \textbf{\bibinfo{volume}{39}}, \bibinfo{pages}{15530--15539}
  (\bibinfo{year}{2014}).

\bibitem{grabowski19ab}
\bibinfo{author}{Grabowski, B.} \emph{et~al.}
\newblock \bibinfo{title}{Ab initio vibrational free energies including
  anharmonicity for multicomponent alloys}.
\newblock \emph{\bibinfo{journal}{npj Comput. Mater.}}
  \textbf{\bibinfo{volume}{5}}, \bibinfo{pages}{80} (\bibinfo{year}{2019}).

\bibitem{Vocadlo03possible}
\bibinfo{author}{Vo{\v{c}}adlo, L.} \emph{et~al.}
\newblock \bibinfo{title}{Possible thermal and chemical stabilization of
  body-centred-cubic iron in the {Earth}'s core}.
\newblock \emph{\bibinfo{journal}{Nature}} \textbf{\bibinfo{volume}{424}},
  \bibinfo{pages}{536--539} (\bibinfo{year}{2003}).

\bibitem{vocadlo02ab}
\bibinfo{author}{Vo\ifmmode~\check{c}\else \v{c}\fi{}adlo, L.} \&
  \bibinfo{author}{Alf\`e, D.}
\newblock \bibinfo{title}{Ab initio melting curve of the fcc phase of
  aluminum}.
\newblock \emph{\bibinfo{journal}{Phys. Rev. B}} \textbf{\bibinfo{volume}{65}},
  \bibinfo{pages}{214105} (\bibinfo{year}{2002}).

\bibitem{pozzo19FeO}
\bibinfo{author}{Pozzo, M.}, \bibinfo{author}{Davies, C.},
  \bibinfo{author}{Gubbins, D.} \& \bibinfo{author}{Alf\`e, D.}
\newblock \bibinfo{title}{{FeO} content of earth's liquid core}.
\newblock \emph{\bibinfo{journal}{Phys. Rev. X}} \textbf{\bibinfo{volume}{9}},
  \bibinfo{pages}{041018} (\bibinfo{year}{2019}).

\bibitem{dorner18melting}
\bibinfo{author}{Dorner, F.}, \bibinfo{author}{Sukurma, Z.},
  \bibinfo{author}{Dellago, C.} \& \bibinfo{author}{Kresse, G.}
\newblock \bibinfo{title}{Melting {Si}: Beyond density functional theory}.
\newblock \emph{\bibinfo{journal}{Phys. Rev. Lett.}}
  \textbf{\bibinfo{volume}{121}}, \bibinfo{pages}{195701}
  (\bibinfo{year}{2018}).

\bibitem{grabowski09ab-Al}
\bibinfo{author}{Grabowski, B.}, \bibinfo{author}{Ismer, L.},
  \bibinfo{author}{Hickel, T.} \& \bibinfo{author}{Neugebauer, J.}
\newblock \bibinfo{title}{Ab initio up to the melting point: Anharmonicity and
  vacancies in aluminum}.
\newblock \emph{\bibinfo{journal}{Phys. Rev. B}} \textbf{\bibinfo{volume}{79}},
  \bibinfo{pages}{134106} (\bibinfo{year}{2009}).

\bibitem{duff15improved}
\bibinfo{author}{Duff, A.~I.} \emph{et~al.}
\newblock \bibinfo{title}{Improved method of calculating ab initio
  high-temperature thermodynamic properties with application to {ZrC}}.
\newblock \emph{\bibinfo{journal}{Phys. Rev. B}} \textbf{\bibinfo{volume}{91}},
  \bibinfo{pages}{214311} (\bibinfo{year}{2015}).

\bibitem{ko21fourth}
\bibinfo{author}{Ko, T.~W.}, \bibinfo{author}{Finkler, J.~A.},
  \bibinfo{author}{Goedecker, S.} \& \bibinfo{author}{Behler, J.}
\newblock \bibinfo{title}{A fourth-generation high-dimensional neural network
  potential with accurate electrostatics including non-local charge transfer}.
\newblock \emph{\bibinfo{journal}{Nat. Commun.}} \textbf{\bibinfo{volume}{12}},
  \bibinfo{pages}{398} (\bibinfo{year}{2021}).

\bibitem{batzner22e3}
\bibinfo{author}{Batzner, S.} \emph{et~al.}
\newblock \bibinfo{title}{{E}(3)-equivariant graph neural networks for
  data-efficient and accurate interatomic potentials}.
\newblock \emph{\bibinfo{journal}{Nat. Commun.}} \textbf{\bibinfo{volume}{13}},
  \bibinfo{pages}{2453} (\bibinfo{year}{2022}).

\bibitem{lopanitsyna21finite}
\bibinfo{author}{Lopanitsyna, N.}, \bibinfo{author}{Ben~Mahmoud, C.} \&
  \bibinfo{author}{Ceriotti, M.}
\newblock \bibinfo{title}{Finite-temperature materials modeling from the
  quantum nuclei to the hot electron regime}.
\newblock \emph{\bibinfo{journal}{Phys. Rev. Materials}}
  \textbf{\bibinfo{volume}{5}}, \bibinfo{pages}{043802} (\bibinfo{year}{2021}).

\bibitem{shapeev16moment}
\bibinfo{author}{Shapeev, A.~V.}
\newblock \bibinfo{title}{Moment tensor potentials: A class of systematically
  improvable interatomic potentials}.
\newblock \emph{\bibinfo{journal}{Multiscale Model. Simul.}}
  \textbf{\bibinfo{volume}{14}}, \bibinfo{pages}{1153--1173}
  (\bibinfo{year}{2016}).

\bibitem{nyshadham19machine}
\bibinfo{author}{Nyshadham, C.} \emph{et~al.}
\newblock \bibinfo{title}{Machine-learned multi-system surrogate models for
  materials prediction}.
\newblock \emph{\bibinfo{journal}{npj Comput. Mater.}}
  \textbf{\bibinfo{volume}{5}}, \bibinfo{pages}{51} (\bibinfo{year}{2019}).

\bibitem{zuo20performance}
\bibinfo{author}{Zuo, Y.} \emph{et~al.}
\newblock \bibinfo{title}{Performance and cost assessment of machine learning
  interatomic potentials}.
\newblock \emph{\bibinfo{journal}{J. Phys. Chem. A}}
  \textbf{\bibinfo{volume}{124}}, \bibinfo{pages}{731--745}
  (\bibinfo{year}{2020}).

\bibitem{lysogorskiy21performant}
\bibinfo{author}{Lysogorskiy, Y.} \emph{et~al.}
\newblock \bibinfo{title}{Performant implementation of the atomic cluster
  expansion {(PACE)} and application to copper and silicon}.
\newblock \emph{\bibinfo{journal}{npj Comput. Mater.}}
  \textbf{\bibinfo{volume}{7}}, \bibinfo{pages}{97} (\bibinfo{year}{2021}).

\bibitem{forslund22ab}
\bibinfo{author}{Forslund, A.} \& \bibinfo{author}{Ruban, A.}
\newblock \bibinfo{title}{Ab initio surface free energies of tungsten with full
  account of thermal excitations}.
\newblock \emph{\bibinfo{journal}{Phys. Rev. B}}
  \textbf{\bibinfo{volume}{105}}, \bibinfo{pages}{045403}
  (\bibinfo{year}{2022}).

\bibitem{zhang18temperature_SFEs}
\bibinfo{author}{Zhang, X.} \emph{et~al.}
\newblock \bibinfo{title}{Temperature dependence of the stacking-fault {Gibbs}
  energy for {Al, Cu, and Ni}}.
\newblock \emph{\bibinfo{journal}{Phys. Rev. B}} \textbf{\bibinfo{volume}{98}},
  \bibinfo{pages}{224106} (\bibinfo{year}{2018}).

\bibitem{zhou22thermodynamic}
\bibinfo{author}{Zhou, Y.} \emph{et~al.}
\newblock \bibinfo{title}{Thermodynamics up to the melting point in a {TaVCrW}
  high entropy alloy: Systematic \textit{ab initio} study aided by machine
  learning potentials}.
\newblock \emph{\bibinfo{journal}{Phys. Rev. B}}
  \textbf{\bibinfo{volume}{105}}, \bibinfo{pages}{214302}
  (\bibinfo{year}{2022}).

\bibitem{grabowski11formation-review}
\bibinfo{author}{Grabowski, B.}, \bibinfo{author}{Hickel, T.} \&
  \bibinfo{author}{Neugebauer, J.}
\newblock \bibinfo{title}{Formation energies of point defects at finite
  temperatures}.
\newblock \emph{\bibinfo{journal}{Phys. Status Solidi B}}
  \textbf{\bibinfo{volume}{248}}, \bibinfo{pages}{1295--1308}
  (\bibinfo{year}{2011}).

\bibitem{stockem18anomalous}
\bibinfo{author}{Stockem, I.} \emph{et~al.}
\newblock \bibinfo{title}{Anomalous phonon lifetime shortening in paramagnetic
  {CrN} caused by spin-lattice coupling: A combined spin and ab initio
  molecular dynamics study}.
\newblock \emph{\bibinfo{journal}{Phys. Rev. Lett.}}
  \textbf{\bibinfo{volume}{121}}, \bibinfo{pages}{125902}
  (\bibinfo{year}{2018}).

\bibitem{ruban07temperature}
\bibinfo{author}{Ruban, A.~V.}, \bibinfo{author}{Khmelevskyi, S.},
  \bibinfo{author}{Mohn, P.} \& \bibinfo{author}{Johansson, B.}
\newblock \bibinfo{title}{Temperature-induced longitudinal spin fluctuations in
  {Fe} and {Ni}}.
\newblock \emph{\bibinfo{journal}{Phys. Rev. B}} \textbf{\bibinfo{volume}{75}},
  \bibinfo{pages}{054402} (\bibinfo{year}{2007}).

\bibitem{koermann11role-Ni}
\bibinfo{author}{K\"ormann, F.}, \bibinfo{author}{Dick, A.},
  \bibinfo{author}{Hickel, T.} \& \bibinfo{author}{Neugebauer, J.}
\newblock \bibinfo{title}{Role of spin quantization in determining the
  thermodynamic properties of magnetic transition metals}.
\newblock \emph{\bibinfo{journal}{Phys. Rev. B}} \textbf{\bibinfo{volume}{83}},
  \bibinfo{pages}{165114} (\bibinfo{year}{2011}).

\bibitem{novikov22magnetic}
\bibinfo{author}{Novikov, I.}, \bibinfo{author}{Grabowski, B.},
  \bibinfo{author}{K{\"o}rmann, F.} \& \bibinfo{author}{Shapeev, A.}
\newblock \bibinfo{title}{Magnetic moment tensor potentials for collinear
  spin-polarized materials reproduce different magnetic states of bcc {Fe}}.
\newblock \emph{\bibinfo{journal}{npj Comput. Mater.}}
  \textbf{\bibinfo{volume}{8}}, \bibinfo{pages}{13} (\bibinfo{year}{2022}).

\bibitem{ferrari20frontiers}
\bibinfo{author}{Ferrari, A.} \emph{et~al.}
\newblock \bibinfo{title}{Frontiers in atomistic simulations of high entropy
  alloys}.
\newblock \emph{\bibinfo{journal}{J. Appl. Phys. (Melville, NY, U. S.)}}
  \textbf{\bibinfo{volume}{128}}, \bibinfo{pages}{150901}
  (\bibinfo{year}{2020}).

\bibitem{dinsdale91SGTE}
\bibinfo{author}{Dinsdale, A.}
\newblock \bibinfo{title}{{SGTE} data for pure elements}.
\newblock \emph{\bibinfo{journal}{Calphad}} \textbf{\bibinfo{volume}{15}},
  \bibinfo{pages}{317--425} (\bibinfo{year}{1991}).

\bibitem{grabowski07ab-fcc}
\bibinfo{author}{Grabowski, B.}, \bibinfo{author}{Hickel, T.} \&
  \bibinfo{author}{Neugebauer, J.}
\newblock \bibinfo{title}{Ab initio study of the thermodynamic properties of
  nonmagnetic elementary fcc metals: Exchange-correlation-related error bars
  and chemical trends}.
\newblock \emph{\bibinfo{journal}{Phys. Rev. B}} \textbf{\bibinfo{volume}{76}},
  \bibinfo{pages}{024309} (\bibinfo{year}{2007}).

\bibitem{zhang17accurate}
\bibinfo{author}{Zhang, X.}, \bibinfo{author}{Grabowski, B.},
  \bibinfo{author}{Körmann, F.}, \bibinfo{author}{Freysoldt, C.} \&
  \bibinfo{author}{Neugebauer, J.}
\newblock \bibinfo{title}{Accurate electronic free energies of the 3d, 4d, and
  5d transition metals at high temperatures}.
\newblock \emph{\bibinfo{journal}{Phys. Rev. B}} \textbf{\bibinfo{volume}{95}},
  \bibinfo{pages}{165126} (\bibinfo{year}{2017}).

\bibitem{fang10origin}
\bibinfo{author}{Fang, C.~M.}, \bibinfo{author}{Sluiter, M. H.~F.},
  \bibinfo{author}{van Huis, M.~A.}, \bibinfo{author}{Ande, C.~K.} \&
  \bibinfo{author}{Zandbergen, H.~W.}
\newblock \bibinfo{title}{Origin of predominance of cementite among iron
  carbides in steel at elevated temperature}.
\newblock \emph{\bibinfo{journal}{Phys. Rev. Lett.}}
  \textbf{\bibinfo{volume}{105}}, \bibinfo{pages}{055503}
  (\bibinfo{year}{2010}).

\bibitem{krukau06influence}
\bibinfo{author}{Krukau, A.~V.}, \bibinfo{author}{Vydrov, O.~A.},
  \bibinfo{author}{Izmaylov, A.~F.} \& \bibinfo{author}{Scuseria, G.~E.}
\newblock \bibinfo{title}{Influence of the exchange screening parameter on the
  performance of screened hybrid functionals}.
\newblock \emph{\bibinfo{journal}{J. Chem. Phys.}}
  \textbf{\bibinfo{volume}{125}}, \bibinfo{pages}{224106}
  (\bibinfo{year}{2006}).

\bibitem{sun15strongly}
\bibinfo{author}{Sun, J.}, \bibinfo{author}{Ruzsinszky, A.} \&
  \bibinfo{author}{Perdew, J.~P.}
\newblock \bibinfo{title}{Strongly constrained and appropriately normed
  semilocal density functional}.
\newblock \emph{\bibinfo{journal}{Phys. Rev. Lett.}}
  \textbf{\bibinfo{volume}{115}}, \bibinfo{pages}{036402}
  (\bibinfo{year}{2015}).

\bibitem{grabowski15random}
\bibinfo{author}{Grabowski, B.}, \bibinfo{author}{Wippermann, S.},
  \bibinfo{author}{Glensk, A.}, \bibinfo{author}{Hickel, T.} \&
  \bibinfo{author}{Neugebauer, J.}
\newblock \bibinfo{title}{Random phase approximation up to the melting point:
  Impact of anharmonicity and nonlocal many-body effects on the thermodynamics
  of {Au}}.
\newblock \emph{\bibinfo{journal}{Phys. Rev. B}} \textbf{\bibinfo{volume}{91}},
  \bibinfo{pages}{201103} (\bibinfo{year}{2015}).

\bibitem{zhu17efficient}
\bibinfo{author}{Zhu, L.-F.}, \bibinfo{author}{Grabowski, B.} \&
  \bibinfo{author}{Neugebauer, J.}
\newblock \bibinfo{title}{Efficient approach to compute melting properties
  fully from ab initio with application to {Cu}}.
\newblock \emph{\bibinfo{journal}{Phys. Rev. B}} \textbf{\bibinfo{volume}{96}},
  \bibinfo{pages}{224202} (\bibinfo{year}{2017}).

\bibitem{vinet87compressibility}
\bibinfo{author}{Vinet, P.}, \bibinfo{author}{Ferrante, J.},
  \bibinfo{author}{Rose, J.~H.} \& \bibinfo{author}{Smith, J.~R.}
\newblock \bibinfo{title}{Compressibility of solids}.
\newblock \emph{\bibinfo{journal}{J. Geophys. Res.: Solid Earth}}
  \textbf{\bibinfo{volume}{92}}, \bibinfo{pages}{9319--9325}
  (\bibinfo{year}{1987}).

\bibitem{moruzzi88calculated}
\bibinfo{author}{Moruzzi, V.~L.}, \bibinfo{author}{Janak, J.~F.} \&
  \bibinfo{author}{Schwarz, K.}
\newblock \bibinfo{title}{Calculated thermal properties of metals}.
\newblock \emph{\bibinfo{journal}{Phys. Rev. B}} \textbf{\bibinfo{volume}{37}},
  \bibinfo{pages}{790--799} (\bibinfo{year}{1988}).

\bibitem{abdullaev15density}
\bibinfo{author}{Abdullaev, R.~N.}, \bibinfo{author}{Kozlovskii, Y.~M.},
  \bibinfo{author}{Khairulin, R.~A.} \& \bibinfo{author}{Stankus, S.~V.}
\newblock \bibinfo{title}{Density and thermal expansion of high purity nickel
  over the temperature range from 150 {K} to 2030 {K}}.
\newblock \emph{\bibinfo{journal}{Int. J. Thermophys.}}
  \textbf{\bibinfo{volume}{36}}, \bibinfo{pages}{603--619}
  (\bibinfo{year}{2015}).

\bibitem{prikhodko03elastic}
\bibinfo{author}{Prikhodko, S.~V.} \emph{et~al.}
\newblock \bibinfo{title}{Elastic constants of face-centered cubic and {L1$_2$}
  {Ni-Si} alloys: Composition and temperature dependence}.
\newblock \emph{\bibinfo{journal}{Metall. Mater. Trans. A}}
  \textbf{\bibinfo{volume}{34}}, \bibinfo{pages}{1863--1868}
  (\bibinfo{year}{2003}).

\bibitem{arblaster17thermodynamic-Nb}
\bibinfo{author}{Arblaster, J.~W.}
\newblock \bibinfo{title}{The thermodynamic properties of niobium}.
\newblock \emph{\bibinfo{journal}{J. Phase Equilib. Diffus.}}
  \textbf{\bibinfo{volume}{38}}, \bibinfo{pages}{707--722}
  (\bibinfo{year}{2017}).

\bibitem{wang98role-Nb-alpha}
\bibinfo{author}{Wang, K.} \& \bibinfo{author}{Reeber, R.~R.}
\newblock \bibinfo{title}{The role of defects on thermophysical properties:
  Thermal expansion of {V, Nb, Ta, Mo and W}}.
\newblock \emph{\bibinfo{journal}{Mater. Sci. Eng., R}}
  \textbf{\bibinfo{volume}{23}}, \bibinfo{pages}{101--137}
  (\bibinfo{year}{1998}).

\bibitem{bujard81elastic-Nb}
\bibinfo{author}{Bujard, P.}, \bibinfo{author}{Sanjines, R.},
  \bibinfo{author}{Walker, E.}, \bibinfo{author}{Ashkenazi, J.} \&
  \bibinfo{author}{Peter, M.}
\newblock \bibinfo{title}{Elastic constants in {Nb-Mo} alloys from zero
  temperature to the melting point: experiment and theory}.
\newblock \emph{\bibinfo{journal}{J. Phys. F: Met. Phys.}}
  \textbf{\bibinfo{volume}{11}}, \bibinfo{pages}{775--786}
  (\bibinfo{year}{1981}).

\bibitem{wang00perfect-Al-alpha}
\bibinfo{author}{Wang, K.} \& \bibinfo{author}{Reeber, R.~R.}
\newblock \bibinfo{title}{The perfect crystal, thermal vacancies and the
  thermal expansion coefficient of aluminium}.
\newblock \emph{\bibinfo{journal}{Philos. Mag. A}}
  \textbf{\bibinfo{volume}{80}}, \bibinfo{pages}{1629--1643}
  (\bibinfo{year}{2000}).

\bibitem{slutsky57elastic}
\bibinfo{author}{Slutsky, L.~J.} \& \bibinfo{author}{Garland, C.~W.}
\newblock \bibinfo{title}{Elastic constants of magnesium from
  4.2\ifmmode^\circ\else\textdegree\fi{}{K} to
  300\ifmmode^\circ\else\textdegree\fi{}{K}}.
\newblock \emph{\bibinfo{journal}{Phys. Rev.}} \textbf{\bibinfo{volume}{107}},
  \bibinfo{pages}{972--976} (\bibinfo{year}{1957}).

\bibitem{mehta06ab-Mg}
\bibinfo{author}{Mehta, S.}, \bibinfo{author}{Price, G.~D.} \&
  \bibinfo{author}{Alfè, D.}
\newblock \bibinfo{title}{Ab initio thermodynamics and phase diagram of solid
  magnesium: A comparison of the {LDA} and {GGA}}.
\newblock \emph{\bibinfo{journal}{J. Chem. Phys.}}
  \textbf{\bibinfo{volume}{125}}, \bibinfo{pages}{194507}
  (\bibinfo{year}{2006}).

\bibitem{nie07ab}
\bibinfo{author}{Nie, Y.} \& \bibinfo{author}{Xie, Y.}
\newblock \bibinfo{title}{Ab initio thermodynamics of the hcp metals {Mg, Ti,
  and Zr}}.
\newblock \emph{\bibinfo{journal}{Phys. Rev. B}} \textbf{\bibinfo{volume}{75}},
  \bibinfo{pages}{174117} (\bibinfo{year}{2007}).

\bibitem{hatt10harmonic}
\bibinfo{author}{Hatt, A.~J.}, \bibinfo{author}{Melot, B.~C.} \&
  \bibinfo{author}{Narasimhan, S.}
\newblock \bibinfo{title}{Harmonic and anharmonic properties of {Fe} and {Ni}:
  Thermal expansion, exchange-correlation errors, and magnetism}.
\newblock \emph{\bibinfo{journal}{Phys. Rev. B}} \textbf{\bibinfo{volume}{82}},
  \bibinfo{pages}{134418} (\bibinfo{year}{2010}).

\bibitem{pham11finite}
\bibinfo{author}{Pham, H.~H.} \emph{et~al.}
\newblock \bibinfo{title}{Finite-temperature elasticity of fcc {Al}: Atomistic
  simulations and ultrasonic measurements}.
\newblock \emph{\bibinfo{journal}{Phys. Rev. B}} \textbf{\bibinfo{volume}{84}},
  \bibinfo{pages}{064101} (\bibinfo{year}{2011}).

\bibitem{wrobel12thermodynamic}
\bibinfo{author}{Wróbel, J.} \emph{et~al.}
\newblock \bibinfo{title}{Thermodynamic and mechanical properties of
  lanthanum–magnesium phases from density functional theory}.
\newblock \emph{\bibinfo{journal}{J. Alloys Compd.}}
  \textbf{\bibinfo{volume}{512}}, \bibinfo{pages}{296--310}
  (\bibinfo{year}{2012}).

\bibitem{metsue14contribution}
\bibinfo{author}{Metsue, A.}, \bibinfo{author}{Oudriss, A.},
  \bibinfo{author}{Bouhattate, J.} \& \bibinfo{author}{Feaugas, X.}
\newblock \bibinfo{title}{Contribution of the entropy on the thermodynamic
  equilibrium of vacancies in nickel}.
\newblock \emph{\bibinfo{journal}{J. Chem. Phys.}}
  \textbf{\bibinfo{volume}{140}}, \bibinfo{pages}{104705}
  (\bibinfo{year}{2014}).

\bibitem{togo15first}
\bibinfo{author}{Togo, A.} \& \bibinfo{author}{Tanaka, I.}
\newblock \bibinfo{title}{First principles phonon calculations in materials
  science}.
\newblock \emph{\bibinfo{journal}{Scr. Mater.}} \textbf{\bibinfo{volume}{108}},
  \bibinfo{pages}{1--5} (\bibinfo{year}{2015}).

\bibitem{minakov15melting}
\bibinfo{author}{Minakov, D.~V.} \& \bibinfo{author}{Levashov, P.~R.}
\newblock \bibinfo{title}{Melting curves of metals with excited electrons in
  the quasiharmonic approximation}.
\newblock \emph{\bibinfo{journal}{Phys. Rev. B}} \textbf{\bibinfo{volume}{92}},
  \bibinfo{pages}{224102} (\bibinfo{year}{2015}).

\bibitem{sjostrom16multiphase}
\bibinfo{author}{Sjostrom, T.}, \bibinfo{author}{Crockett, S.} \&
  \bibinfo{author}{Rudin, S.}
\newblock \bibinfo{title}{Multiphase aluminum equations of state via density
  functional theory}.
\newblock \emph{\bibinfo{journal}{Phys. Rev. B}} \textbf{\bibinfo{volume}{94}},
  \bibinfo{pages}{144101} (\bibinfo{year}{2016}).

\bibitem{gupta17low}
\bibinfo{author}{Gupta, A.} \emph{et~al.}
\newblock \bibinfo{title}{Low-temperature features in the heat capacity of
  unary metals and intermetallics for the example of bulk aluminum and
  {${\mathrm{Al}}_{3}\mathrm{Sc}$}}.
\newblock \emph{\bibinfo{journal}{Phys. Rev. B}} \textbf{\bibinfo{volume}{95}},
  \bibinfo{pages}{094307} (\bibinfo{year}{2017}).

\bibitem{wang21DFTTK}
\bibinfo{author}{Wang, Y.} \emph{et~al.}
\newblock \bibinfo{title}{{DFTTK}: Density functional theory toolkit for
  high-throughput lattice dynamics calculations}.
\newblock \emph{\bibinfo{journal}{CALPHAD: Comput. Coupling Phase Diagrams
  Thermochem.}} \textbf{\bibinfo{volume}{75}}, \bibinfo{pages}{102355}
  (\bibinfo{year}{2021}).

\bibitem{zhang21first}
\bibinfo{author}{Zhang, J.}, \bibinfo{author}{Korzhavyi, P.~A.} \&
  \bibinfo{author}{He, J.}
\newblock \bibinfo{title}{First-principles modeling of solute effects on
  thermal properties of nickel alloys}.
\newblock \emph{\bibinfo{journal}{Mater. Today Commun.}}
  \textbf{\bibinfo{volume}{28}}, \bibinfo{pages}{102551}
  (\bibinfo{year}{2021}).

\end{thebibliography}


 \newcommand{\noop}[1]{}

\clearpage


\vspace{6cm}

\begin{table*}[!h]
\section*{Figures}
\end{table*}

\begin{figure*}[p]  
    \hideForFinalVersion{\includegraphics[width=0.95\textwidth]{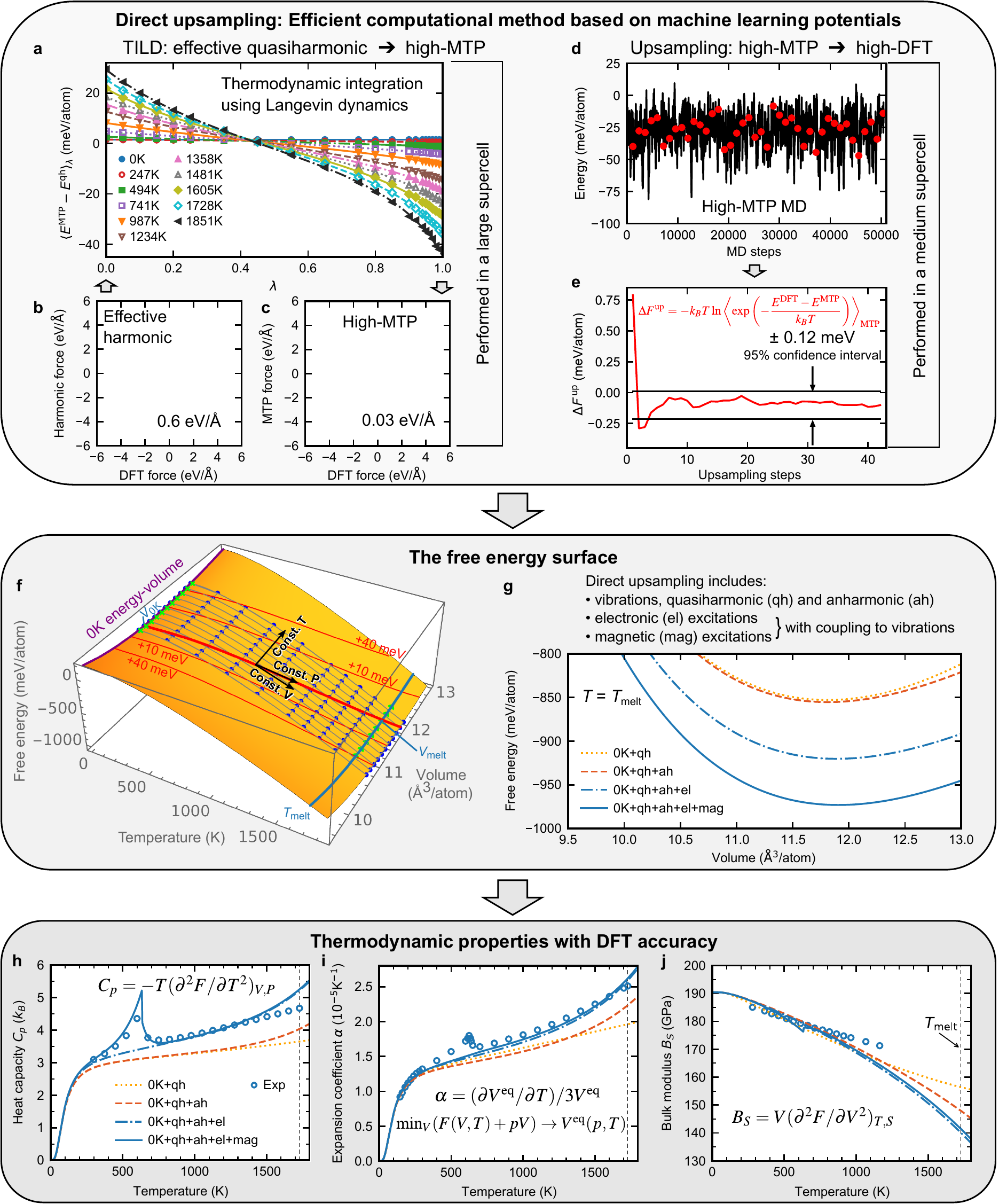}}  
\end{figure*}

\begin{figure*}[p]  
    \caption{Schematic description of the entire workflow using fcc Ni as an example. The upper box illustrates different stages of direct upsampling: (a) TILD at different temperatures from effective QH to MTP (whose force RMSEs are shown in (b) and (c)), and (d) and (e) the upsampling to high-DFT. The center box is the crux of the workflow. (f) is a representation of the free energy surface, with the $(V,T)$ mesh on which free-energy calculations are performed represented by blue dots, volumes at $T_\textrm{melt}$ chosen for MTP training set represented by green dots, volumes for the low-temperature effective QH fitting represented by green crosses, the 0\,K $E$-$V$ curve in purple and different derivatives represented by black arrows. (g) is the free energy as a function of volume at the melting point calculated while including different excitations. The lower box shows the numerically computed (h) isobaric heat capacity $C_p$, (i) linear thermal expansion coefficient $\alpha$ and (j) adiabatic bulk modulus $B_S$ along with comparison to experimentally calculated values~\cite{dinsdale91SGTE,abdullaev15density,prikhodko03elastic}. ($V^\textrm{eq}=V^\textrm{eq}(T)$ denotes the equilibrium volume at $T$ and a given pressure, $V_\textrm{0K}=V^\textrm{eq}(\textrm{0\,K})$ and $V_\textrm{melt}=V^\textrm{eq}(T_\textrm{melt})$, where $T_\textrm{melt}$ is the experimental melting temperature at ambient pressure; $S$ is the entropy.)
    }
    \label{fig:master-method}
\end{figure*}

\begin{figure}[p]  
   \centering
       \hideForFinalVersion{ \includegraphics[width=9cm]{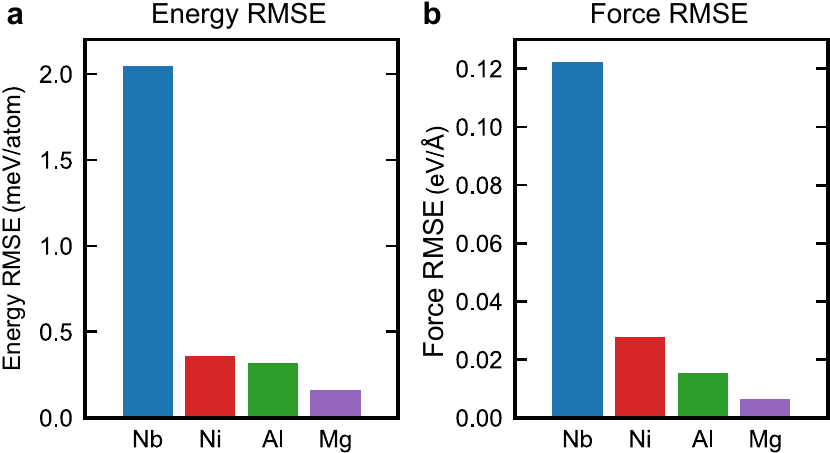}   }
   \caption{The test set root-mean-square errors (RMSEs) of the high-MTPs (level 20) in (a) energy (per atom) and (b) forces in medium-size supercells.
   }
   \label{fig:MTP-error-main}
\end{figure}


\begin{figure}[p]  
   \centering
   \hideForFinalVersion{ \includegraphics[width=9cm]{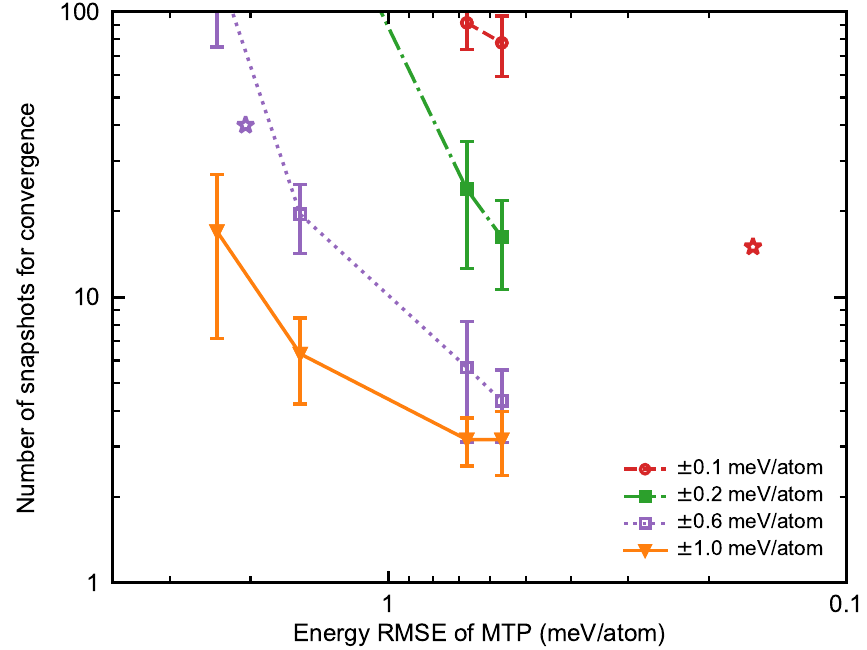} }
   \caption{Number of snapshots required for convergence of the direct upsampling as a function of the energy RMSE of the MTP. The colors represent different target accuracy as indicated in the legend. The symbols connected by lines denote upsampling to an Nb MTP model system. The purple and the red star show upsampling from MTP to DFT at $T_\mathrm{melt}$ for Nb within $\pm$ 0.6 meV/atom and for Mg within $\pm$ 0.1 meV/atom, respectively. The error bars denote a 95\% confidence interval derived from independent sets of calculations.
   }
   \label{fig:upsample-num-samples-vs-RMSE-main}
\end{figure}


\begin{figure*}[p]  
   \centering
       \hideForFinalVersion{ \includegraphics{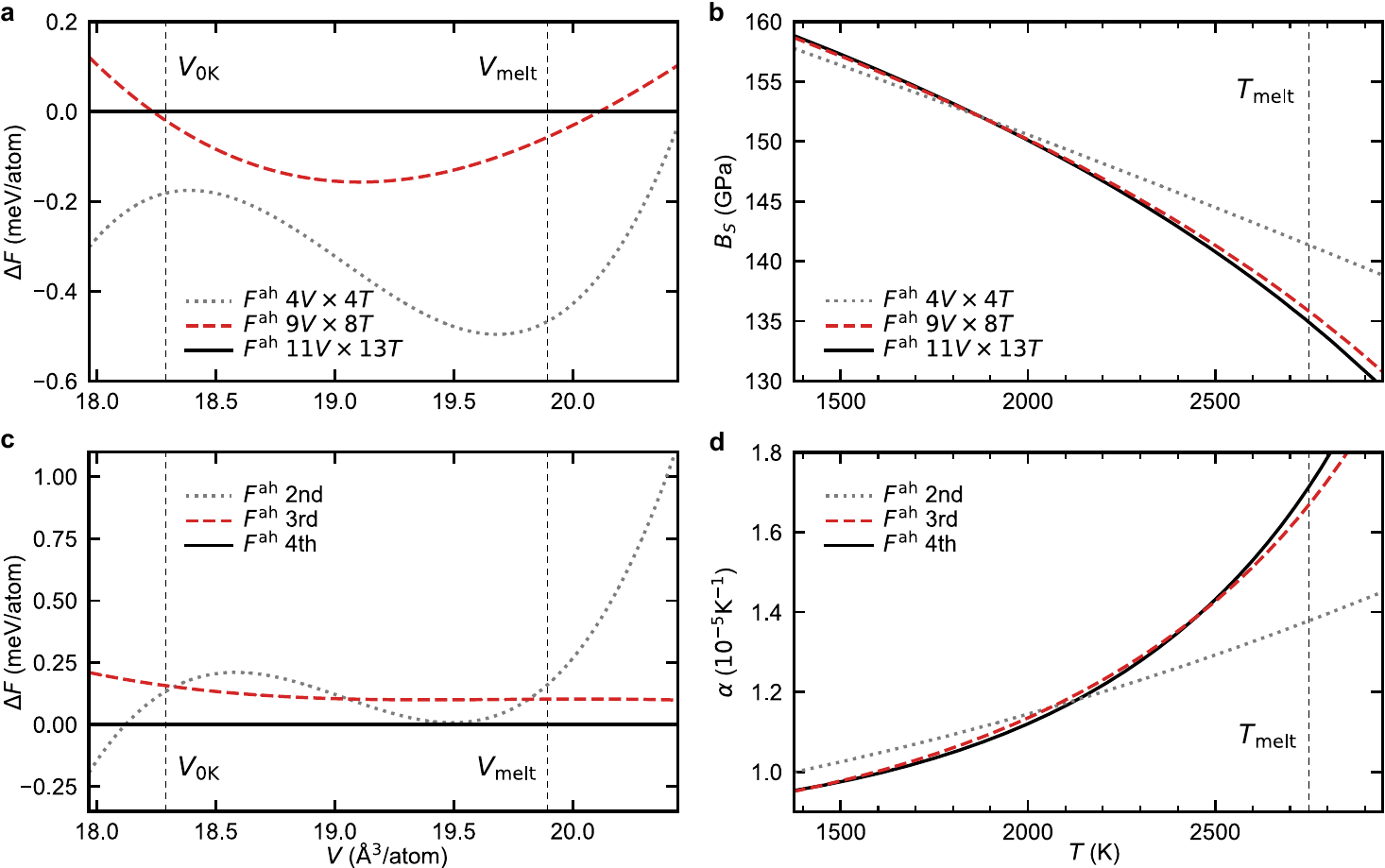}   }
   \caption{Convergence of thermodynamic properties. (a) Anharmonic free energy as a function of volume for Nb at $T_\mathrm{melt}$ calculated by using different grid densities. The free energy is plotted with respect to the values calculated using the largest grid ($11 V \times 13 T$). (b) The corresponding adiabatic bulk modulus for the three grids. (c) Anharmonic free energy for Nb at $T_\mathrm{melt}$ calculated using a $11 V \times 13 T$ grid but parametrized with polynomials of different order. The free energy is plotted with respect to the values calculated using the 4th order parametrization. (d) Corresponding thermal expansion coefficient for the three parametrizations.
   }
   \label{fig:convergence-main}
\end{figure*}

\begin{figure*}[p]  
   \centering
       \hideForFinalVersion{ \includegraphics[width=\textwidth]{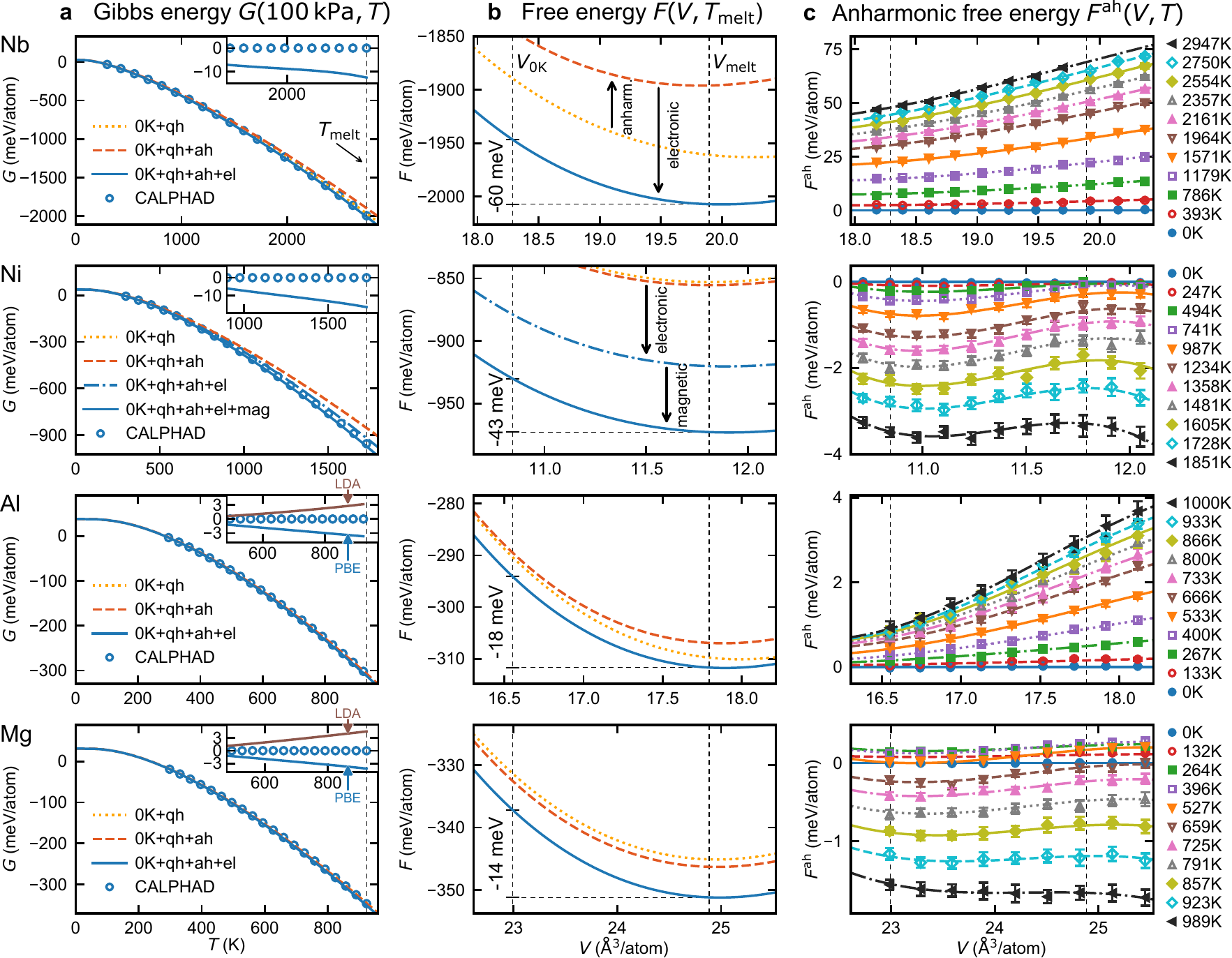}   }
   \caption{\textit{Ab initio} calculated (a) Gibbs energy $G(T)$ at 100\,kPa, (b) free energy $F(V)$ at $T_\mathrm{melt}$, and (c) 
   anharmonic free energy $F^\mathrm{ah}(V,T)$, for the four elements using the PBE XC functional. The $G(T)$ values are referenced to the minimum energy of the static lattice at 0\,K. Calculations using the CALPHAD method~\cite{dinsdale91SGTE} (aligned to the \textit{ab initio} values at room temperature) are shown in blue dots for comparison. The insets contain the full \textit{ab initio} Gibbs energy at high temperatures with respect to the CALPHAD values, with the results for LDA added for Al and Mg. For $G(T)$ and $F(V,T_\text{melt})$, curves including different excitation mechanisms are shown. The melting temperatures $T_\textrm{melt}$ correspond to experimental values. 
   The error bars denote a 95\% confidence interval.
   }
   \label{fig:free-energy}
\end{figure*}

\begin{figure*}[p]  
    \centering
         \includegraphics[width=0.09\textwidth]{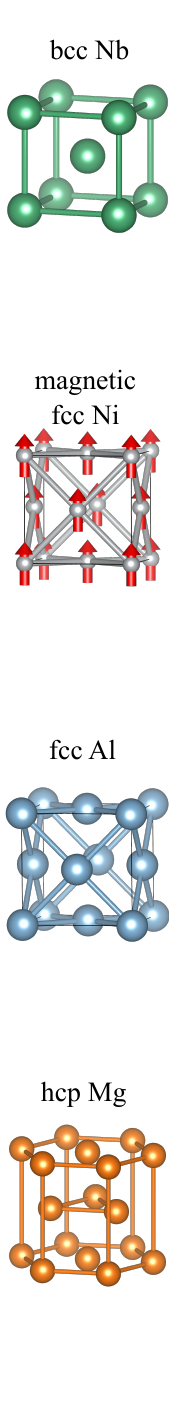}
        \includegraphics[width=0.90\textwidth]{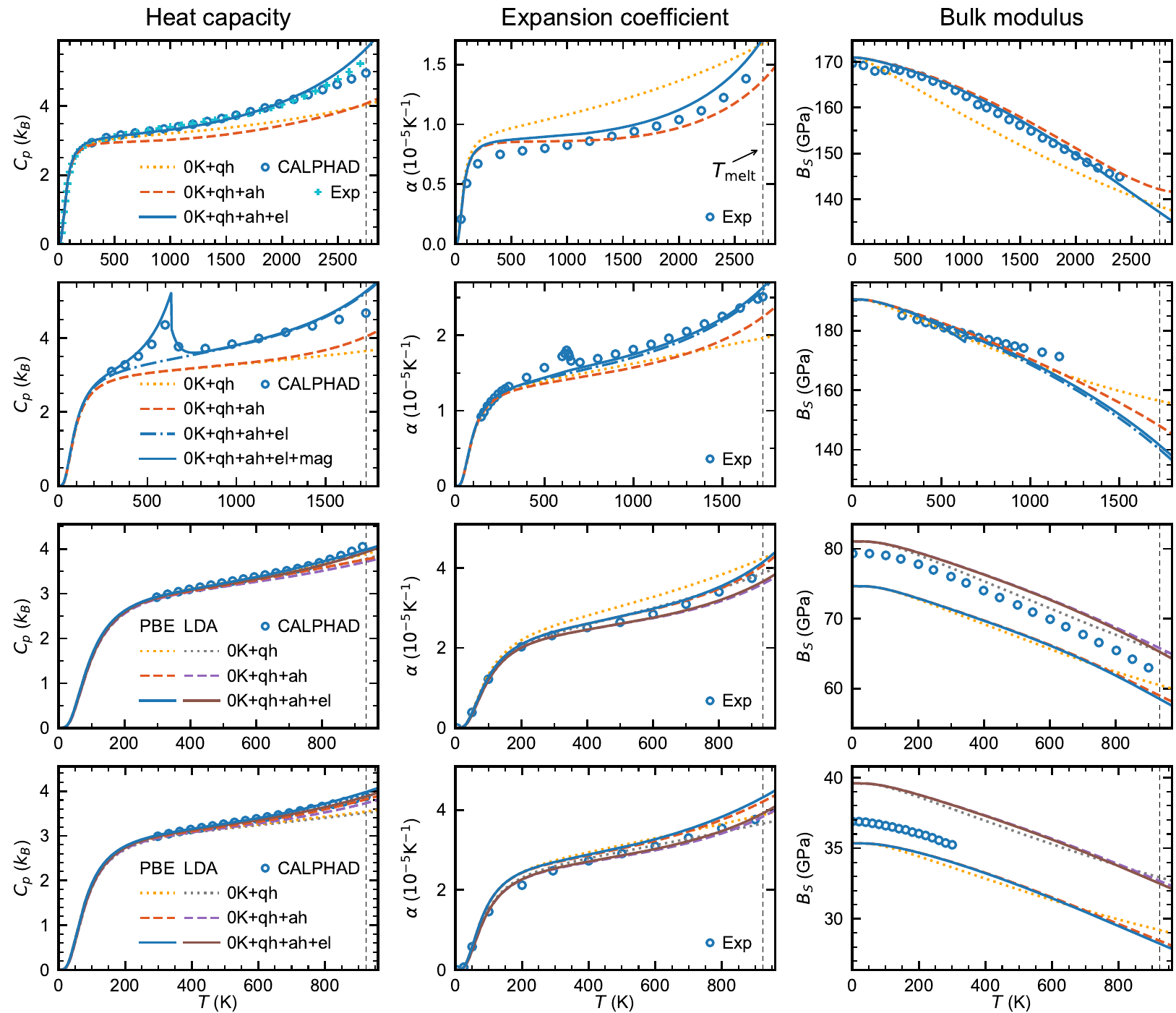}
    \caption{\textit{Ab initio} calculated $C_p(T)$, $\alpha(T)$, and $B_S(T)$ up to the melting point for Nb, Ni, Al and Mg. Calculations are compared to experimental results shown in blue circles. Results considering different excitation mechanisms (effective QH (qh), anharmonic (ah), electronic (el) and magnetic (mag)) are shown. Nb and Ni results are for PBE, while for Al and Mg LDA results are additionally shown. The following experimental values are used for comparison: Nb~\cite{dinsdale91SGTE,arblaster17thermodynamic-Nb,wang98role-Nb-alpha,bujard81elastic-Nb}, Ni~\cite{dinsdale91SGTE,abdullaev15density,prikhodko03elastic}, Al~\cite{dinsdale91SGTE,touloukian75thermophysical-TPRC,wang00perfect-Al-alpha}, and Mg~\cite{dinsdale91SGTE,touloukian75thermophysical-TPRC,slutsky57elastic}.
    }
    \label{fig:properties}
\end{figure*}

\clearpage

\vspace{6cm}

{\setstretch{1.0}  

\begin{table*}[!h]
\section*{Tables}
\end{table*}

\begin{sidewaystable*}[tbp]  
\caption{\textit{Ab initio} studies of the isobaric heat capacity ($C_p$), the thermal expansion coefficient ($\alpha$) and the bulk modulus ($B$) for bcc Nb, fcc Ni, fcc Al, and hcp Mg. The corresponding exchange-correlation (XC) functionals and the potentials representing the core electrons are listed. The columns under `Contributions to $F$' indicate which terms (0\,K static energy `$E_{\mathrm{0K}}$', electronic `el', quasiharmonic `qh' and anharmonic `ah') were included to the total free energy. For works that included the anharmonic contribution, the number of volume $V$ (or pressure $P$) and temperature $T$ data points is mentioned. Studies including the explicit anharmonic contribution up to all orders and to the accuracy of DFT are printed in boldface. All of the used abbreviations are expanded in the Supplementary Information.}
\label{table:literature-review}
\centering
\small
\begin{ruledtabular}
\begin{tabular}{ccccccccccccc}
     Year & Ref.                             & Elements & \multicolumn{2}{c}{DFT methodology}           & \multicolumn{5}{c}{Contributions to $F$} & $C_p^\mathrm{ah}$ & $\alpha^\mathrm{ah}$ & $B^\mathrm{ah}$ \\ \cline{4-5} \cline{6-10}
          &                                  &          & XC       & Potential  & $E_{\mathrm{0K}}$ & el         & qh          & ah      &    Grid for ah         \\
    \hline
     1988 & \citen{moruzzi88calculated}       & Al, Nb   & LDA      & ASW        & x        &            & x  &      &           \\
     2002 & \citen{vocadlo02ab}              & Al       & PW91 & USP         & x        &  & x & \textbf{TI} & 7 $V,T$ points           \\
     2006 & \citen{mehta06ab-Mg}              & Mg       & LDA, PW91 & PAW         & x        & x & x &            \\
     2007 & \citen{grabowski07ab-fcc}         & Al       & LDA, PBE & PAW, AE       & x        &            & x &            \\
     2007 & \citen{nie07ab}                   & Mg       & LDA      & NC       & x        & x & x &            \\
     2009 & \citen{grabowski09ab-Al}                   & Al       & LDA, PBE   & NC       & x        & x & x &  \textbf{UP-TILD}  & $\approx40$ $V,T$ points & x & x \\
     2010 & \citen{hatt10harmonic}            & Ni       & LDA, PBE   & USP       & x        &           & x &          &         \\
     2011 & \citen{koermann11role-Ni}         & Ni       & PBE      & PAW  & x        &            & x &            \\
     2011 & \citen{pham11finite}              & Al       & PW91      & PAW       & x        &  x  & x &  tight-binding MD &  $1P \times 10T$   & & & x     \\
     2012 & \citen{wrobel12thermodynamic}     & Mg       & PBE       & PAW       & x        &  x  & x &                      &                          \\
     2014 & \citen{metsue14contribution}      & Ni       & LDA, PBE & PAW        & x        & x & x &            \\
     2014 & \citen{junkaew14ab}               & Nb, Mg     & PBE       & PAW            & x        & x & x &  low $T$ expansion  & analytic, $T$$<$1000 K  & x & x & x  \\
     2015 & \citen{glensk15understanding-Fah} & Al, Ni   & LDA, PBE & PAW        & x        & x & x & \textbf{UP-TILD}& $4V \times 4T$ & x \\
     2015 & \citen{togo15first}               & Al       & PBE      & PAW        & x        &                  & x &             \\
     2015 & \citen{minakov15melting}          & Al, Ni    & LDA, PBE, GW  & PAW  & x        &  x     & x &             \\
     2016 & \citen{sjostrom16multiphase}      & Al        & LDA, PBE     & PAW  & x  &   x       & x &  MD $\rightarrow$ phonon  &  several $V, T$ points      & x & x & x \\
     2017 & \citen{gupta17low}                & Al       & LDA, PBE & PAW       & x        & x & x   &          \\
     2021 & \citen{adams21anharmonic}       & Al       & PBE   & USP           & x        &  x   & x & SCAILD, DAMA  & $3V \times 3T$, $1P \times 3T$       & x \\
     2021 & \citen{wang21DFTTK}       & Al, Ni    & PBE   & PAW           & x        &  x    & x &          &               \\
     2021 & \citen{zhang21first}               & Ni        & PBE   & EMTO          & x        &  x    & x &          &               \\[.1cm]
     \hline\\[-.3cm]
  2022 & this work & Nb, Ni, Al, Mg & LDA, PBE & PAW &  x        &  x    & x &  \textbf{direct upsampling} & $\ge 9V \times 10T$  &  x        &  x    & x   \\
\end{tabular}
\end{ruledtabular}
\end{sidewaystable*}

\clearpage

\begin{table}[p]  
\centering
\caption{CPU-time estimate (in core hours) for different stages of the free energy calculation for Nb (PBE, 11 valence electrons) using the current framework and using the previous TU-TILD methodology. The ${\Delta}F$ and upsampling values are for a $11\times13$ $(V,T)$ grid. The speed-up is with respect to TU-TILD.
}
\label{table:CPU-time-main-text}
{\small
\begin{ruledtabular}
\begin{tabular}{clcc}
     & Stage  & Current framework & \textnormal{TU-TILD} \\
    \hline
    \settowidth\rotheadsize{{\small MTP fitting\hspace{0.5mm}}}
    \multirow{4}{*}{\rothead{{\small MTP fitting\hspace{0.5mm}}}} & Low-MTP fitting\rule{0pt}{3ex} & \phantom{000,}340  & \phantom{000,}-- \\
     & High-MTP fitting & \phantom{00,}8500 &  \phantom{0,,0}40,000 \rule[-2ex]{0pt}{0pt}\\ \cline{2-4} 
     & Total fitting time\rule{0pt}{3ex} & \phantom{00,}8800 & \phantom{0,0}40,000 \\
     & Speed-up (fitting) & \phantom{0\,0}$\times$4.5 & \phantom{0,000,0}$\times$1 \rule[-2ex]{0pt}{0pt}\\
    \hline
    \hline  
    \settowidth\rotheadsize{{\small Free energy calc.\hspace{0.5mm}}}
    \multirow{6}{*}{\rothead{{\small Free energy calc.\hspace{0.5mm}}}} & Effective QH fitting\rule{0pt}{3ex}    & \phantom{0}11,000 &  \phantom{0\,0}13,000 \\
     & $\Delta F^{\mathrm{qh}\rightarrow\mathrm{MTP}}$ & \phantom{00,}7700 &  \phantom{0,00,}7700\\
     & $\Delta F^{\mathrm{MTP}\rightarrow\mathrm{DFT,low}}$ & \phantom{00,}-- & 1,630,000 \\
     & (Direct) upsampling & 410,000 & \phantom{00}410,000 \rule[-2ex]{0pt}{0pt}\\ \cline{2-4} 
     & Total free energy calc.~time\rule{0pt}{3ex}                  & 430,000 & 2,100,000\\
     & Speed-up (free energy calc.) & \phantom{0\,0}$\times$4.8 & \phantom{0\,000\,0}$\times$1 \rule[-2ex]{0pt}{0pt} \\ \hline
     \hline
     & Total calculation time for Nb \rule[2.5ex]{0pt}{0pt} & 440,000 & 2,140,000
\end{tabular}
\end{ruledtabular}
}
\end{table}

}




\clearpage

\ifarXiv
    

\section*{Supplementary material}

**Supplementary Information to:**

**High-accuracy thermodynamic properties to the melting point from *ab initio* calculations aided by machine-learning potentials**

Jong Hyun Jung, Prashanth Srinivasan, Axel Forslund, and Blazej Grabowski

*Institute for Materials Science, University of Stuttgart, Pfaffenwaldring 55, 70569 Stuttgart, Germany*

(Dated: December 20, 2022)

## CONTENTS

## SUPPLEMENTARY METHODS 1. SOFTWARE AND GENERAL SETTINGS

The density functional theory (DFT) calculations are performed using VASP[1] within the projector augmented-wave (PAW) method.[2,3] For the exchange-correlation (XC) functional, we use the generalized gradient approximation (GGA) by Perdew, Burke, and Ernzerhof (PBE)[4] or the local density approximation (LDA).[5] Other XC functionals such as PBEsol[6] are similarly applicable. The configurations of the valence electrons in the PAW potentials are denoted in Supplementary Table 1. For calculations *excluding* electronic excitations (electronic temperature), we use the Methfessel-Paxton[7] smearing method of first order with a smearing width of 0.2 eV. In particular, the energy extrapolated to zero smearing width is employed. For calculations *including* electronic temperature effects, Mermin's finite-temperature DFT formulation is used,[8] in practice realized with the Fermi smearing approach. The breakout condition for the electronic self-consistency cycle is set to $10^{-4}$ eV/cell (VASP input flag EDIFF). During the fast Fourier transform, we use denser grids to avoid aliasing errors of an order of 1 meV/atom, unless otherwise specified (PREC=Accurate).

The moment tensor potentials (MTPs) are fitted using the MLIP code.[9,10] The MLIP code is available at https://mlip.skoltech.ru. The fitting weights for energy and force are 1 and 0.01 Å$^2$, respectively. $NVT$ molecular dynamics (MD) simulations using the MTPs are performed with LAMMPS[11] and under the Langevin thermostat.[12] This thermostat is especially efficient for computing thermodynamic properties.[13] In particular, it is critical for MD runs driven mainly by harmonic forces. Utilizing instead, e.g., a Nosé-Hoover thermostat,[14] there would be no energy exchange between the modes, and

therefore no convergence. This problem is averted using the Langevin thermostat. For the friction parameter $\gamma_{LD}$, we use a value of $0.01\,\mathrm{fs}^{-1}$, which is well-suited for thermodynamic averaging.[15] Note that such a value is too large to describe rare events, e.g., diffusion of vacancies or migration of dislocations.

The effective harmonic potentials[16] are fitted using HIPHIVE,[17] and the phonon energies are evaluated using PHONOPY.[18] The thermodynamic integration from the effective harmonic reference to MTPs is performed using an in-house code. The structures are visualized using VESTA.[19]

## SUPPLEMENTARY METHODS 2. STEP-BY-STEP METHODOLOGY

A detailed description of the methodology conceptualized in the main paper (Methods section) is provided in the following. A graphical overview is given in Supplementary Figure 1.

Our primary goal is the acquisition of the different contributions to the total free energy $F(V,T)$ in Eq. (1) (main text), i.e.,

$$F(V,T) = \underbrace{E_{0K}(V)}_{\substack{\text{energy} \\ \text{static} \\ \text{lattice}}} + \underbrace{F^{el}(V,T)}_{\substack{\text{electronic} \\ \text{free energy} \\ \text{incl. impact} \\ \text{of vibrations}}} + \underbrace{F^{vib}(V,T)}_{\substack{\text{full (qh+ah)} \\ \text{vibrational} \\ \text{free energy}}} + \underbrace{F^{mag}(V,T)}_{\substack{\text{magnetic} \\ \text{free energy} \\ \text{incl. impact} \\ \text{of vibrations}}},$$

$$(1)$$

with

$$F^{vib}(V,T) = \underbrace{F^{qh}(V,T)}_{\substack{\text{effective} \\ \text{quasiharmonic} \\ \text{free energy} \\ \text{(quantum)}}} + \underbrace{F^{ah}(V,T)}_{\substack{\text{explicitly} \\ \text{anharmonic} \\ \text{free energy} \\ \text{(classical)}}}.$$

$$(2)$$

The methodology is detailed for the four prototype systems—bcc Nb, magnetic fcc Ni, fcc Al and fcc Mg. In Supplementary Table 2, an overview for these prototype systems is given in terms of the supercell sizes used in the different stages, and the corresponding number of atoms.

### A. Energy-volume curve and Debye-Grüneisen model

#### 1. 0 K energy-volume curve

To obtain the 0 K energy-volume ($E$-$V$) curve $E_{0K}(V)$, a dense volume mesh of 11-15 points is used over a range of $-8\%$ to $+12\%$ around the equilibrium volume (Supplementary Table 1). This volume range should enclose the volume at the melting point obtained at later stages (possibly by adapting the initial range). Once the energies at all volumes are calculated, the $E$-$V$ curve is parametrized with the equation of state by Vinet.[20,21] The DFT parameters for these energies are a $k$-point density of 63,000-66,000 kp·atom and an energy cutoff of twice the recommended value ENMAX. To minimize

the effect of cell-shape dependent $k$-point sampling on the energy, a unit cell that is consistent with later-used supercells is employed here. For example, a conventional four-atom unit cell is used in the case of fcc, if a cubic supercell is used later.

For hcp structures, the $c/a$ ratio is optimized (Supplementary Table 3). Firstly, the volume is relaxed at 0 K for a set of fixed $c/a$'s by fitting the Vinet equation of state. Secondly, by using the minimum energies for each $c/a$, the equilibrium $c/a$ value is obtained from a third order polynomial fit to the energies. For the present case of hcp Mg, the relaxed $c/a$ at 0 K is then used throughout for finite temperatures, assuming that the $c/a$ has a negligible effect on the energy compared to volume (as observed, e.g., in Ref. [22]). The temperature dependence of $c/a$ or $a$ and $c$ axis-dependent thermodynamic properties[23] are, however, in general accessible within our framework.

#### 2. Debye-Grüneisen model up to $T_{melt}$

The 0 K $E$-$V$ curve is used to calculate an approximate free energy surface based on the Debye-Grüneisen model.[24,25] In particular, the equilibrium energy, volume, bulk modulus and its pressure derivative as obtained from the $E$-$V$ curve are used as input parameters. The free energies are calculated up to the experimental melting point $T_{melt}$.[26] We then estimate the thermal expansion up to $T_{melt}$, thereby obtaining the relevant volume range for the full free energy calculations. Supplementary Table 4 compiles the relevant data for the studied systems. The Debye temperatures (corresponding to the 0 K volume, $V_{0K}$) compare well to previously reported values.[24] For systems where experimental melting points are unavailable, one can use CALPHAD-based estimates[27,28] or Vegard's law for alloys.

### B. Low quality MTP

To efficiently generate a representative training set for an accurate moment-tensor potential (MTP),[9,10] a temporary low-accuracy MTP (low-MTP) is trained to short, low-converged (low-DFT) *ab initio* MD (AIMD) runs. This low-MTP is then used to simulate classical MD for a longer time to inexpensively sample the vibrational phase space.

#### 1. Low-converged AIMD at $T_{melt}$

For the purpose of training the low-MTP, low-DFT AIMD simulations are performed in small-size supercells, with 32-54 atoms (Supplementary Table 5). Since these AIMD runs are computationally lightweight, a training set for the low-MTP can quickly be generated. The AIMD is run using a Langevin thermostat with a friction parameter of $0.01\,\mathrm{fs}^{-1}$ and a time step of 5 fs for 1000 steps at $T_{melt}$. Four $V$'s are chosen that enclose the corresponding $V_{0K}$ and $V_{melt}$ estimated by the Debye-Grüneisen model. To improve the stability of the

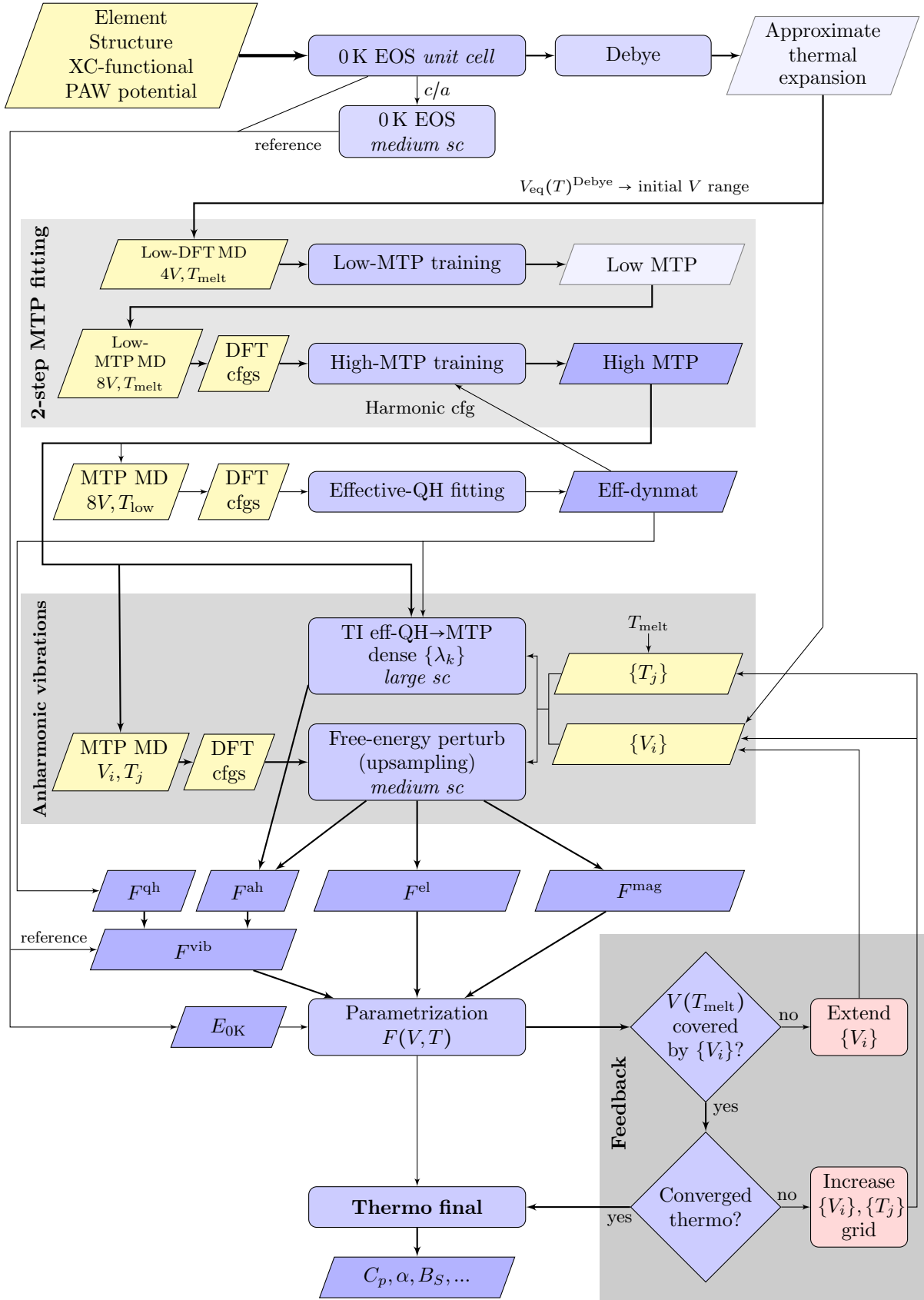

Supplementary Figure 1. Flowchart describing the entire direct upsampling framework. The main steps of the procedure are seen in the centered blue rectangular boxes. Inputs are denoted by yellow parallelograms, outputs by blue parallelograms with major output marked by a stronger blue color. The diamonds represent decisions and the red boxes represent the corresponding measures taken.

Supplementary Table 1. Input parameters and results for the 0 K $E$-$V$ curve.

| PAW potential[a] | XC | Atoms | Cutoff[b] (eV) | $k$ points | kp · atom | Volumes $V$ (Å$^3$/atom) | $a_{eq}$ (Å) | $V_{eq}$ (Å$^3$/atom) | $B_{eq}$ (GPa) | $B'_{eq}$ | $|\Delta E|_{max}$[c] (meV/atom) | $|\Delta B|_{max}$[d] (GPa) |
|---|---|---|---|---|---|---|---|---|---|---|---|---|
| | | | | | Input | | | | Output | | | |
| Nb, $4p^65s^14d^4$ bcc | PBE | 2 | 417 | $32 \times 32 \times 32$ | 66,000 | 16.54 ... 20.18, 12 $V$'s | 3.32 | 18.3 | 173 | 3.9 | 0.03 | 0.4 |
| Ni, $3d^94p^1$ fcc | PBE | 4 | 539 | $25 \times 25 \times 25$ | 63,000 | 10.00 ... 12.26, 13 $V$'s | 3.52 | 10.9 | 196 | 4.9 | 0.01 | 0.5 |
| Al, $3s^24s^1$ fcc | PBE | 4 | 480 | $25 \times 25 \times 25$ | 63,000 | 15.06 ... 18.52, 15 $V$'s | 4.04 | 16.5 | 78 | 4.6 | 0.01 | 0.4 |
| Al, $3s^24s^1$ fcc | LDA | 4 | 482 | $25 \times 25 \times 25$ | 63,000 | 15.06 ... 18.52, 15 $V$'s | 3.98 | 15.8 | 85 | 4.7 | 0.01 | 0.2 |
| Mg, $3s^2$ hcp | PBE | 2 | 400 | $37 \times 37 \times 23$ | 63,000 | 21.37 ... 25.75, 11 $V$'s | 3.19 | 22.9 | 37 | 4.1 | < 0.001 | < 0.1 |
| Mg, $3s^2$ hcp | LDA | 2 | 400 | $37 \times 37 \times 23$ | 63,000 | 20.35 ... 24.59, 11 $V$'s | 3.12 | 21.4 | 41 | 4.1 | < 0.001 | < 0.1 |

[a] Valence electron configuration corresponding to the PAW potential.
[b] The energy cutoff is $2 \times$ ENMAX.
[c] $|\Delta E|_{max}$: maximum deviation of fitted Vinet EOS in the volume range between $V_{0K}$ and $V_{melt}$.
[d] $|\Delta B|_{max}$: absolute maximum contribution to $B$ from $\Delta E$.

Supplementary Table 2. Supercells and reciprocal mesh used.

| System | Small-size | | Medium-size | | Large-size | | Reciprocal space |
|---|---|---|---|---|---|---|---|
| | Supercell | Atoms | Supercell | Atoms | Supercell | Atoms | |
| Nb–PBE | $3 \times 3 \times 3$ | 54 | $4 \times 4 \times 4$ | 128 | $6 \times 6 \times 6$ | 432 | $30 \times 30 \times 30$ $q$ mesh for all systems; corresponds to a supercell with 27,000 primitive cells |
| Ni–PBE | $2 \times 2 \times 2$ | 32 | $3 \times 3 \times 3$ | 108 | $5 \times 5 \times 5$ | 500 | |
| Al–PBE | $2 \times 2 \times 2$ | 32 | $3 \times 3 \times 3$ | 108 | $5 \times 5 \times 5$ | 500 | |
| Al–LDA | $2 \times 2 \times 2$ | 32 | $3 \times 3 \times 3$ | 108 | $5 \times 5 \times 5$ | 500 | |
| Mg–PBE | $3 \times 3 \times 2$ | 36 | $4 \times 4 \times 3$ | 96 | $7 \times 7 \times 5$ | 490 | |
| Mg–LDA | $3 \times 3 \times 2$ | 36 | $4 \times 4 \times 3$ | 96 | $7 \times 7 \times 5$ | 490 | |
| Used for: | Low-MTP fitting | | High-MTP fitting direct upsampling $\Delta F^{up}$ | | Thermodynamic integration $F^{qh \to MTP}$ | | Analytical QH reference $F^{qh}$ |

Supplementary Table 3. The $c/a$ ratios for hcp Mg.

| System | Input $c/a$ ratios | Output $(c/a)_{eq}$ |
|---|---|---|
| Mg–PBE | 1.56 ... 1.70, 0.02 | 1.625 |
| Mg–LDA | 1.56 ... 1.70, 0.02 | 1.624 |

Supplementary Table 4. Results from the Debye-Grüneisen model and the experimental melting temperature.

| System | Debye $T$ (K) | $a_{eq}(T_{melt})$ (Å) | $V_{eq}(T_{melt})$ (Å$^3$/atom) | $T_{melt}$ (K) |
|---|---|---|---|---|
| Nb–PBE | 316 | 3.39 | 19.5 | 2750 |
| Ni–PBE | 388 | 3.61 | 11.8 | 1728 |
| Al–PBE | 387 | 4.13 | 17.6 | 933 |
| Al–LDA | 400 | 4.07 | 16.8 | 933 |
| Mg–PBE | 295 | 3.29 | 25.1 | 923 |
| Mg–LDA | 308 | 3.21 | 23.3 | 923 |

high-MTP, a small margin is added to the volume range to account for deviations in experimental values.[24] (For example, we choose $[a_{0K} - 0.1(a_{melt} - a_{0K}), a_{melt} + 0.25(a_{melt} - a_{0K})]$ for Mg PBE.) A $k$-point density of 256-433 kp·atoms and an energy cutoff equal to ENMAX are used.

### 2. Fitting the low-MTP

The initial low-MTPs are fitted to the low-DFT AIMD, with a recommended root-mean-square error (RMSE) in energy between 2 and 10 meV/atom (Supplementary Table 8). In our present calculations, this is accounted for by using MTPs with a level of 6-14 and a radial basis size of 8. The higher the level of an MTP, the larger the number of basis functions. Hence, a higher-level MTP might overfit the relatively small DFT training set. As a first step of the fitting procedure, we select a subset of the configurations from the low-DFT AIMD. For each $V$, 10% of the total configurations (i.e., 100 configurations) are taken, each separated by 50 fs in the AIMD. A low-MTP is fitted to these configurations, following which more configurations are added from the low-DFT AIMD runs using the maximum volume criterion.[10] The low-MTP is then re-trained on this new training set and the retraining process is repeated multiple times (typically three to four). During fitting, the minimum distance parameter $R_{min}$ is set to a value smaller than the smallest atomic distance in the training set.

### C. High quality MTP

In this step, very accurate MTPs (high-MTPs) are trained to reproduce highly converged DFT (high-DFT) energies and

forces of configurations generated by the low-MTP.

### 1. Generating snapshots with low-MTP MD

MD simulations at $T_{melt}$ are run using the low-MTP for a denser set of at least eight volumes in the medium-size supercells (Supplementary Table 6). We use (nearly) equidistant volumes ($V$'s; the lattice constants are equidistant) across the same minimum and maximum value as used in the low-DFT MD runs (Supplementary Methods 2 B 1) (except for Nb, for which we tested a narrower volume range to improve the accuracy of the MTP). We now discard the small-size supercells used for the low-DFT MD, and subsequently use medium-size or large-size supercells for the further steps in the workflow. MD is run with the Langevin thermostat with a time step of 1 fs and a damping parameter of $\gamma_{LD} = 0.01\,fs^{-1}$. Following a pre-equilibration of 600 fs (or $6/\gamma_{LD}$), MD is run for 9000 fs from which 30 uncorrelated snapshots are chosen at intervals of 300 fs (or $3/\gamma_{LD}$).

### 2. High-DFT calculations on snapshots

Energies and forces of the previously chosen snapshots are calculated using high-DFT parameters (Supplementary Table 7). We use a dense $k$-point mesh (6100-8200 kp · atom), an energy cutoff of $1.5 \times$ ENMAX, a convergence criterion for the electronic self-consistency loop (EDIFF) of $10^{-4}$ eV/cell, and an optimal number of bands (NBANDS). We additionally use the extra grid for the evaluation of augmentation charges (ADDGRID). To optimize the high-DFT calculations, we apply a pre-convergence step, i.e., an initial faster calculation with lower parameters (PREC = Normal, ENMAX, LREAL = Auto and without ADDGRID). With the pre-converged wave function, a new calculation is performed using the high-DFT parameters. For instance, such a procedure reduces the computing time by 45% for a 128-atom Nb PBE supercell.

The electronic excitations (electronic temperature) are excluded at the current step by employing the Methfessel-Paxton smearing of first order, smearing width of 0.2 eV, and the 0 K extrapolated energy. The reason is that the presently available MTPs cannot explicitly describe the electronic degree-of-freedom. The electronic contribution is added at the upsampling stage of our procedure (Supplementary Methods 2 F).

### 3. Fitting the high-MTP

The high-MTP is then fitted to the high-DFT energies and forces in a similar way to that of the low-MTP, with a radial basis size of 8 and the other parameters as given in Supplementary Table 9. The utilized cutoff radius $R_{cut}$ of the high-MTP is large enough to include interactions up to the third neighbor shell. This is relevant because force constants up to the third neighbor shell usually approximate the phonon dispersion well and reproduce most vibrational modes. For hcp Mg, the 4th neighbor shell interactions are also included, since

distances to the 1st and 2nd neighbors are similar and would be equal for the ideal $c/a$ ratio. We fit a preliminary MTP to every 10th configuration in the initial training set, iteratively add selected configurations and re-train typically three or four times[10] to obtain the final high-MTP.

Supplementary Figure 2 shows the RMSEs in energies and forces of different level MTPs fitted to high-DFT data. As the level increases, the number of basis functions increases exponentially[10] and the error decreases approximately algebraically. This behavior has been observed before.[9] With a smaller RMSE, fewer high-DFT calculations are required at the later upsampling stage, described in Supplementary Methods 2 F, to obtain highly converged free energy differences. For a deeper analysis on the convergence of the direct upsampling, see Supplementary Discussion 1 A.

### 4. Stability of the MTP during TI

During the TI from effective QH to high-MTP (Supplementary Methods 2 E), the MTP needs to be stable even on configurations coming from the effective QH potential. Soft repulsion in the effective QH model results in small interatomic distances which might not be included in the high-MTP training set. This can result in the high-MTPs predicting unphysical forces. Such an observation was made for Nb, and is illustrated in Supplementary Figure 3 (a). A high-MTP fitted only to the initial high-DFT training set predicts inaccurate energies (and forces) of configurations with shorter bonds, shown by the blue curve. In such a scenario, the unphysical forces arise from the radial basis functions of the MTPs that are built on Chebyshev polynomials. One of the polynomials is represented in Supplementary Figure 3 (b). Extrapolation of the polynomial below $R_{min}$ causes the inaccurate energy and force predictions.

A systematic approach to stabilise the high-MTP is to include one configuration from the effective QH potential and re-train the MTP. Chronologically, the re-training can be done after the effective QH potential is prepared (described in the next step in Supplementary Methods 2 D). From a dynamics run using the effective QH at the largest $V$ and $T$ in the mesh, the snapshot with the smallest minimum distance is extracted, its energies and forces are calculated using high-DFT parameters and are added to the training set to re-train the high-MTP. In Supplementary Figure 3 (a), the green curve shows the energy prediction of the stabilized MTP. Although this stage is not a necessity for all systems, stabilizing the MTP using such a technique improves reliability and robustness.

## D. Effective QH reference

In order to obtain the free energy of a system using TI, a reference whose free energy can be analytically calculated is essential. To this end, we suggest an effective quasiharmonic (QH) reference.[29] Prior TI works have used a 0 K QH reference where the dynamical matrix is obtained using the finite displacement method.[21,30] However, a 0 K dy-

Supplementary Table 5. Low-converged DFT (low-DFT) parameters for AIMD at $T_{melt}$.

| System | Input | | | | | | | | Output |
|--------|-----------|-------|--------------|---------|-----------------------------------|-----------------|-------|--------------------|----------------------|
| | Supercell | Atoms | Cutoff (eV) | $k$ points | Volumes $V$[b] (Å³/atom) | Time step (fs) | Steps | $T_{melt}$ (K) | Min. distance[c] (Å) |
| Nb–PBE | $3 \times 3 \times 3$ | 54 | 157 | $2 \times 2 \times 2$ | 17.94 … 20.50, 4 $V$'s | 5 | 1000 | 2750 | 2.10 |
| Ni–PBE | $2 \times 2 \times 2$ | 32 | 270[a] | $2 \times 2 \times 2$ | 10.87 … 12.12, 4 $V$'s | 5 | 1000 | 1728 | 1.83 |
| Al–PBE | $2 \times 2 \times 2$ | 32 | 181 | $2 \times 2 \times 2$ | 16.22 … 18.50, 4 $V$'s | 5 | 1000 | 933 | 2.22 |
| Al–LDA | $2 \times 2 \times 2$ | 32 | 181 | $2 \times 2 \times 2$ | 15.49 … 17.70, 4 $V$'s | 5 | 1000 | 933 | 2.19 |
| Mg–PBE | $3 \times 3 \times 2$ | 36 | 200[a] | $2 \times 2 \times 2$ | 22.91 … 25.66, 4 $V$'s | 5 | 1000 | 923 | 2.36 |
| Mg–LDA | $3 \times 3 \times 2$ | 36 | 200[a] | $2 \times 2 \times 2$ | 21.38 … 23.81, 4 $V$'s | 5 | 1000 | 923 | 2.33 |

[a] The energy cutoff corresponds to ENMAX.
[b] The volumes enclose $V_{0K}$ and $V_{melt}$ from the Debye-Grüneisen model.
[c] 'Min. distance' is the minimum distance between any two atoms during the MD.

Supplementary Table 6. Snapshot-generation parameters with low-MTP MD in medium-size supercells at $T_{melt}$.

| System | Input | | | | | Output |
|--------|-----------|-------|-------------------------|----------------|---------------------|----------------------|
| | Supercell | Atoms | Volumes $V$ (Å³/atom) | $T_{melt}$ (K) | Number of snapshots | Min. distance (Å) |
| Nb–PBE | $4 \times 4 \times 4$ | 128 | 18.26 … 19.95, 8 $V$'s | 2750 | 240 | 2.05 |
| Ni–PBE | $3 \times 3 \times 3$ | 108 | 10.87 … 12.12, 8 $V$'s | 1728 | 240 | 1.86 |
| Al–PBE | $3 \times 3 \times 3$ | 108 | 16.22 … 18.50, 10 $V$'s | 933 | 300 | 2.22 |
| Al–LDA | $3 \times 3 \times 3$ | 108 | 15.49 … 17.70, 10 $V$'s | 933 | 300 | 2.21 |
| Mg–PBE | $4 \times 4 \times 3$ | 96 | 22.91 … 25.66, 8 $V$'s | 923 | 240 | 2.45 |
| Mg–LDA | $4 \times 4 \times 3$ | 96 | 21.38 … 23.81, 8 $V$'s | 923 | 240 | 2.36 |

Supplementary Table 7. High-DFT parameters in medium-size supercells.

| System | Supercell | Atoms | Cutoff[a] (eV) | $k$ points | kp · atoms |
|--------|-----------|-------|----------------|------------|------------|
| Nb–PBE | $4 \times 4 \times 4$ | 128 | 313 | $4 \times 4 \times 4$ | 8200 |
| Ni–PBE | $3 \times 3 \times 3$ | 108 | 404 | $4 \times 4 \times 4$ | 6900 |
| Al–PBE | $3 \times 3 \times 3$ | 108 | 360 | $4 \times 4 \times 4$ | 6900 |
| Al–LDA | $3 \times 3 \times 3$ | 108 | 360 | $4 \times 4 \times 4$ | 6900 |
| Mg–PBE | $4 \times 4 \times 3$ | 96 | 300 | $4 \times 4 \times 4$ | 6100 |
| Mg–LDA | $4 \times 4 \times 3$ | 96 | 300 | $4 \times 4 \times 4$ | 6100 |

[a] The energy cutoff corresponds to $1.5 \times$ ENMAX.

namical matrix cannot be used for dynamically unstable systems. Additionally, systems with long-range interactions require comparably large supercells (e.g., Nb), and, further, obtaining a 0 K QH matrix for disordered multicomponent systems is highly expensive owing to the large number of displacements.[31] Hence, we resort to an effective QH reference where the dynamical matrix is obtained from finite-temperature DFT forces and the free energy is calculated analytically using the quantum-mechanical formula. In order for the effective QH reference to capture low-temperature quantum-mechanical phonon behaviour as accurately as possible, it is fitted to low-temperature forces.

*1. Fitting of effective QH dynamical matrices*

Dynamical matrices are fitted to high-DFT forces of configurations generated from low-temperature high-MTP MD for 9-11 $V$'s (Supplementary Table 10). The snapshots are generated in medium-size supercells at 20 K. This temperature can be adjusted for dynamically unstable systems that are stabilized only at elevated temperatures. After a pre-equilibration stage, 30 uncorrelated snapshots separated by 300 fs ($3/\gamma_{LD}$) are generated at each volume and recalculated with high-DFT parameters. One dynamical matrix is fitted (independently) for each volume. A cutoff radius including the 3rd neighbor shell is used to restrict the interactions. Once the effective dynamical matrices are fitted for the medium-size supercells, they are extended to large-size supercells (432-500 atoms) by adding zero force interactions beyond the cutoff radius (using ForceConstantPotential of HIPHIVE). The extended dynamical matrices are required as a reference for the QH to high-MTP TI calculations in the large-size supercells.

*2. Parametrization of force constants and $F^{qh}(V,T)$ calculation*

The 9-11 extended dynamical matrices at the different volumes are used to obtain a parametrization of the force constants as a function of volume. Specifically, each force constant in the dynamical matrix is parametrized using a second order polynomial in $V$, using the explicitly computed force constants as input to the fits. Parametrizing the force constants provides a smooth dependence of $F^{qh}$ on $V$ and thereby also

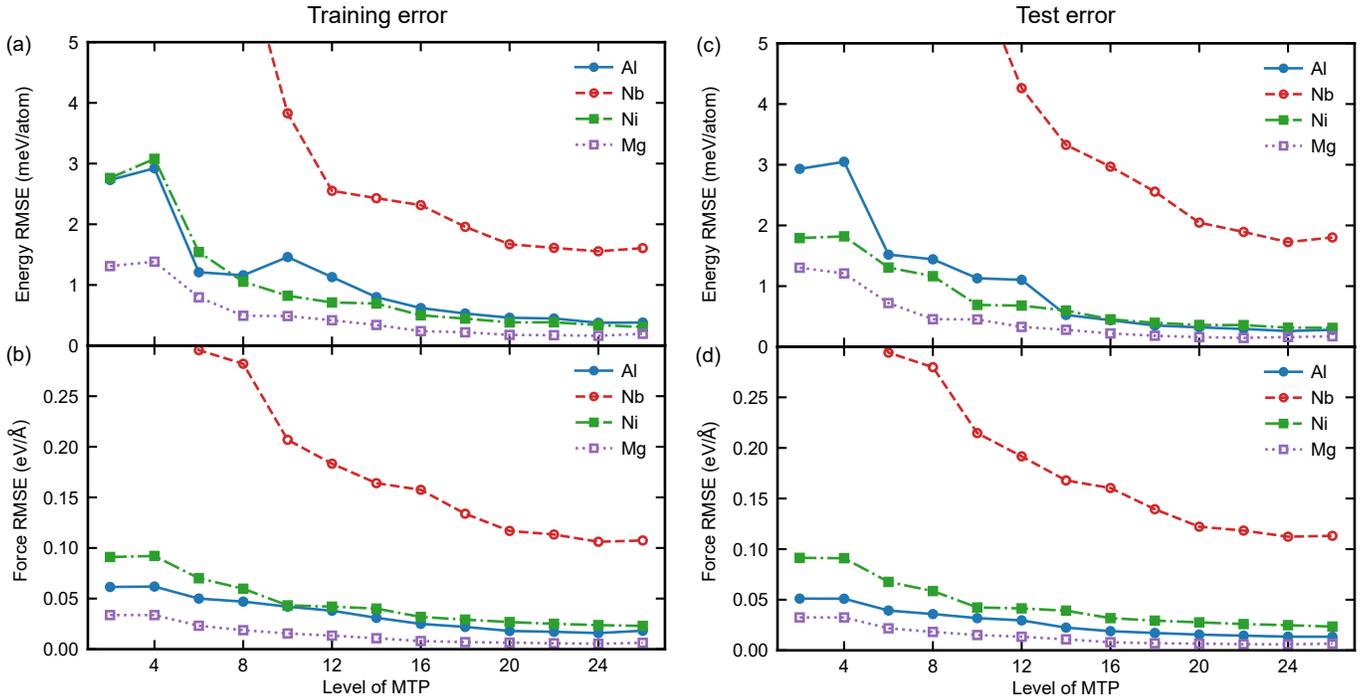

Supplementary Figure 2. Training and test errors of high-MTPs. The training root-mean-square errors (RMSEs) of the (a) energy per atom and (b) forces in medium-size supercells. The MTPs are trained to PBE DFT energies and forces of configurations at $T_{melt}$ in the relevant $V$ range ($V_0$...$V_{melt}$), and tested to the configurations of upsampling at $T_{melt}$ and near $V_{melt}$. The test RMSEs of the (c) energy per atom and (d) forces.

### Supplementary Table 8. Low-MTP fitting parameters and results.

| System | Input | | | | Output | | |
|---|---|---|---|---|---|---|---|
| | $R_{min}$ (Å) | $R_{cut}$ (Å) | Level | No. of MTP parameters | Energy RMSE (meV/atom) | Force RMSE (eV/Å) | $\Delta F^{up,\,a}$ (meV/atom) |
| Nb–PBE | 2.00 | 5.00 | 10 | 44 | 8.3 | 0.203 | - |
| Ni–PBE | 1.33 | 4.96 | 14 | 88 | 1.1 | 0.040 | - |
| Al–PBE | 2.00 | 5.00 | 8 | 29 | 2.6 | 0.091 | - |
| Al–LDA | 2.00 | 5.00 | 8 | 29 | 2.4 | 0.087 | - |
| Mg–PBE | 1.86 | 6.24 | 6 | 25 | 1.0 | 0.026 | - |
| Mg–LDA | 1.83 | 6.24 | 6 | 25 | 0.9 | 0.025 | $4.80^{b} \pm 0.23$ |

[a] $\Delta F^{up}$ is evaluated at the $(V,T)$ grid point closest to $(V_{melt}, T_{melt})$. The $\pm$ error is given for 95% confidence interval.
[b] The energy is larger than the energy RMSE because this MTP is fitted to low-DFT.

### Supplementary Table 9. High-MTP fitting parameters and results.

| System | Input | | | | Output | | |
|---|---|---|---|---|---|---|---|
| | $R_{min}$ (Å) | $R_{cut}^{\,a}$ (Å) | Level | No. of MTP parameters | Energy RMSE (meV/atom) | Force RMSE (eV/Å) | $\Delta F^{up,\,b}$ (meV/atom) |
| Nb–PBE | 1.55 | 5.33 | 20 | 332 | 2.05 | 0.122 | $-0.08 \pm 0.65$ |
| Ni–PBE | 1.36 | 4.97 | 20 | 332 | 0.36 | 0.028 | $-0.06 \pm 0.11$ |
| Al–PBE | 2.00 | 5.50 | 20 | 332 | 0.32 | 0.016 | $0.05 \pm 0.10$ |
| Al–LDA | 1.71 | 5.57 | 20 | 332 | 0.31 | 0.012 | $0.16 \pm 0.09$ |
| Mg–PBE | 1.95 | 5.89 | 20 | 332 | 0.16 | 0.007 | $-0.01 \pm 0.05$ |
| Mg–LDA | 1.83 | 6.10 | 20 | 332 | 0.14 | 0.006 | $0.02 \pm 0.04$ |

[a] The cutoff radius $R_{cut}$ corresponds to the 3rd nearest neighbor distance (NND) (for largest $V$) +0.5 Å for Al LDA, Nb, Ni. For hcp Mg, it is instead the 4th NND since the 1st and 2nd NNDs are similar, or equal in case of the ideal hcp structure.
[b] $\Delta F^{up}$ is evaluated at the $(V,T)$ grid point closest to $(V_{melt}, T_{melt})$. The $\pm$ error is given for a 95% confidence interval.

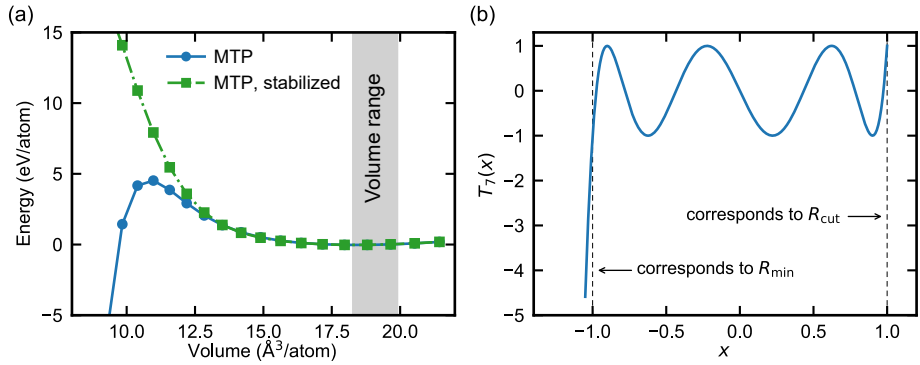

Supplementary Figure 3. (a) 0 K $E$-$V$ curves for bcc Nb using a high-MTP (MTP), and a high-MTP stabilized by a harmonic snapshot (MTP, stabilized). The gray shaded area indicates the volume range between 0 K and $T_{melt}$. In the MTPs, Chebyshev polynomials are used as a part of the radial functions. (b) shows one such Chebyshev polynomial, specifically $T_7(x)$.

Supplementary Table 10. Effective QH potential parameters for fitting to high-DFT forces from snapshots generated using high-MTP MD at 20 K.

| | Input | | | | | | | | Output | |
|---|---|---|---|---|---|---|---|---|---|---|
| | Cell for fitting | | for extended dyn. matrix | | | | | | | |
| System | Supercell | Atoms | Supercell | Atoms | Volumes[a] $V$ (Å³/atom) | $R_{cut}$[b] (Å) | $N_{param}$[c] | | $R^{2,d}$ | Force RMSE[d] (eV/Å) |
| Nb–PBE | $4 \times 4 \times 4$ | 128 | $6 \times 6 \times 6$ | 432 | 18.18 … 20.37, 11 $V$'s | 4.9 | 7 | | 0.933 | 0.020 |
| Ni–PBE | $3 \times 3 \times 3$ | 108 | $5 \times 5 \times 5$ | 500 | 10.71 … 12.05, 11 $V$'s | 4.5 | 9 | | 0.995 | 0.007 |
| Al–PBE | $3 \times 3 \times 3$ | 108 | $5 \times 5 \times 5$ | 500 | 16.55 … 18.11,  9 $V$'s | 5.3 | 9 | | 0.988 | 0.008 |
| Al–LDA | $3 \times 3 \times 3$ | 108 | $5 \times 5 \times 5$ | 500 | 15.87 … 17.28, 10 $V$'s | 5.3 | 9 | | 0.989 | 0.008 |
| Mg–PBE | $4 \times 4 \times 3$ | 96 | $7 \times 7 \times 5$ | 490 | 22.99 … 25.45,  9 $V$'s | 4.7 | 12 | | 0.992 | 0.005 |
| Mg–LDA | $4 \times 4 \times 3$ | 96 | $7 \times 7 \times 5$ | 490 | 21.46 … 23.61,  9 $V$'s | 4.7 | 12 | | 0.992 | 0.005 |

[a] The volumes should enclose $V_{0K}$ and $V_{melt}$ from the final total free energy. Two volumes greater than $V_{melt}$ are used.
[b] The cutoff radius $R_{cut}$ includes third nearest neighbors in the static structure for all $V$. The written $R_{cut}$ is for the smallest $V$ and as $V$ increases, cutoff is scaled.
[c] Here, $N_{param}$ is the number of fitting parameters.
[d] The $R^2$ value and force RMSE for the largest $V$ are written.

of $F^{ah}(V)$ in the later stage of the procedure. The smoothened force constants are Fourier transformed and the resulting reciprocal dynamical matrices are diagonalized to obtain frequencies on a dense $30 \times 30 \times 30$ wave vector mesh (in the reciprocal space of the primitive cell), and at the required volumes (i.e., the ones used for the anharmonic calculations, cf. Supplementary Table 11 below). This is important in order to interpolate long-wavelength phonons and to include their contribution to the vibrational free energy. For every $V$, the free energy $F^{qh}(V,T)$ is then calculated with the quantum-mechanical harmonic formula

$$F^{qh}(V,T) = \sum_{\mathbf{q}j} \left\{ \frac{\hbar\omega_{\mathbf{q}j}}{2} + k_B T\,\omega_{\mathbf{q}j} \ln\left[1 - \exp\left(-\frac{\hbar\omega_{\mathbf{q}j}}{k_B T}\right)\right] \right\},$$ (3)

where $\omega_{\mathbf{q}j} = \omega_{\mathbf{q}j}(V)$ is the frequency of the phonon for wavevector $\mathbf{q}$ and band $j$ at $V$. Note that a consistent normalization (e.g., per atom) is required for $F^{qh}$ as well as all other contributions entering Eqs. (1) and (2).

It is worth to stress that the described parametrization procedure works well for the studied prototype systems and likewise for the complex multicomponent alloy studied in Ref. [29]. A possible concern may have been that the independently fitted dynamical matrices at the different volumes are optimized into distinct local minima of the objective function. Such distinct minima may then lead to discontinuous jumps in the force constants as a function of volume impeding the parametrization procedure. It could possibly be that the restriction of the interactions to within a limited cutoff and the linearity of the fitting problem facilitate the fitting and thereby counteract any such difficulties.

### E. TI from effective QH to high-MTP

To obtain $\Delta F^{qh \to MTP}(V,T)$ (main contribution to $F^{ah}$), we perform thermodynamic integration using Langevin dynamics from the effective QH reference to the high-MTP on a dense $(V,T)$ grid. Specifically, $\Delta F^{qh \to MTP}(V,T)$ is given by[13,32]

$$\Delta F^{qh \to MTP} = \int_0^1 d\lambda \left\langle E^{MTP} - E^{qh} \right\rangle_\lambda.$$ (4)

Here, $\langle \ldots \rangle_\lambda$ denotes the ensemble average for a given coupling parameter $\lambda$, and $E^{MTP}$ and $E^{qh}$ are the classical vibra-

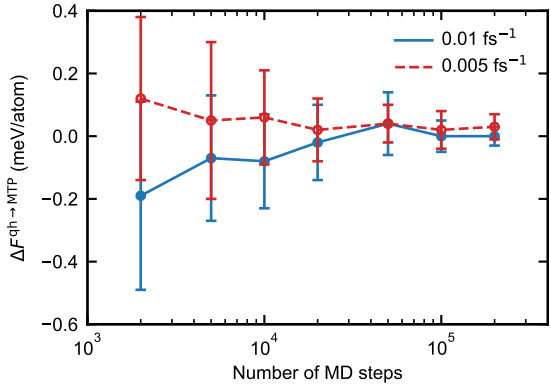

Supplementary Figure 4. Convergence of $\Delta F^{\text{qh}\rightarrow\text{MTP}}$ from TI with respect to the number of MD steps for two different friction parameters $\gamma_{\text{LD}}$.

tional energies of the high-MTP and the effective QH reference. Owing to the inexpensive nature of the MTPs used in the TI, large-size supercells and well-converged, dense $(V,T)$ meshes can be used at this stage (Supplementary Table 11). At each $(V,T)$, we use at least 20 $\lambda$ values, with a much denser spacing near $\lambda = 1$, where the dependence on $\lambda$ is stronger. (See Fig. 1 (a) in the main text, $\lambda = \{0, 0.05, 0.10, 0.15, 0.20, 0.25, 0.35, 0.45, 0.55, 0.65, 0.75, 0.80, 0.85, 0.90, 0.92, 0.94, 0.95, 0.96, 0.97, 0.98, 0.99, 1.00\}$.) In the TI calculations, we use a time step of 1 fs and a friction parameter $\gamma_{\text{LD}} = 0.01\,\text{fs}^{-1}$. After a pre-equilibration for 500 steps, TI is run for 50,000 steps resulting in statistically very well converged average energies. One can also check for the convergence of the average temperature to fall within ~10 K of the given $T$. Supplementary Figure 4 shows a convergence test of time-steps and friction parameter for Mg where $F^{\text{ah}}$ from TI converges to within 0.1 meV/atom.

In the calculation of $\Delta F^{\text{qh}\rightarrow\text{MTP}}(V,T)$ in Eq. (4), we account only for the contribution of vibrations, without the energy of the static lattice. That is,

$$E^{\text{MTP}} - E^{\text{qh}} = \frac{\left(E_{N,\text{tot}}^{\text{MTP}} - E_{0K,N}^{\text{DFT}} - E_N^{\text{qh}}\right)}{N-1},\qquad(5)$$

where $E_{N,\text{tot}}^{\text{MTP}}$ is the total energy using the high-MTP of a supercell with $N$ atoms, $E_{0K,N}^{\text{DFT}}$ is the DFT static 0 K energy, and $E_N^{\text{qh}}$ is the effective QH energy calculated using the classical formula (i.e., it is 0 at 0 K). In order for the $E_{0K,N}^{\text{DFT}}$ reference to be consistent with the DFT calculations in the upsampling step, it should be calculated using the high-DFT settings, i.e, the same cell shape, the same energy cutoff, and the same $k$-point mesh (cf. Supplementary Methods 2 C 2). Otherwise, an order of 1 meV/atom difference could occur.

After $\langle E^{\text{MTP}} - E^{\text{qh}}\rangle$ is obtained for all $\lambda$'s, an integration over $\lambda$ is done using an analytical fit based on the tangent function (Fig. 3 in Ref. [33]). The fit is given by

$$\langle\cdots\rangle_{NVT,\lambda} = a_0\cot\left[\pi\left(a_1\lambda^2 + a_2\lambda + a_3\right)\right] + a_4,\qquad(6)$$

involving five fitting parameters. Prior to integration, the $\lambda$ dependence needs to be checked carefully using visual inspec-

tion to ensure the validity of the analytical fit. To validate the integral, we also compare the value from the fit with integration by cubic splines and the Simpson rule. For example, at $T = 2947$ K and $V = 19.7\,\text{Å}^3$/atom for Nb, the difference between the integrations is 0.05 meV/atom, well within the 95% confidence interval.

### F. Direct upsampling from high-MTP to high-DFT

Unlike the conventional *two-stage upsampled thermodynamic integration using Langevin dynamics* (TU-TILD) method, here we perform upsampling directly on configurations from the high-MTP (direct upsampling). This is possible owing to the accuracy of the high-MTP in reproducing DFT energies and forces. The direct upsampling is performed in medium-size supercells and on the same dense $(V,T)$ mesh as the TI from QH to high-MTP (cf. Supplementary Table 11), to be able to add the two contributions point wise together.

In a first step, we obtain the free energy difference between the high-MTP and the high-DFT using the free energy perturbation expression[34,35] as given by

$$\Delta F^{\text{up}} = -k_B T \ln\left\langle\exp\left(-\frac{E^{\text{DFT}} - E^{\text{MTP}}}{k_B T}\right)\right\rangle_{\text{MTP}},\qquad(7)$$

where $E^{\text{DFT}}$ and $E^{\text{MTP}}$ are high-DFT (without electronic temperature) and high-MTP energies,[36] and $\langle\ldots\rangle_{\text{MTP}}$ denotes the ensemble average. The speed of the convergence depends on the RMSE of the MTP, see Supplementary Discussion 1 A. In this work, the same high number of uncorrelated snapshots is used for all elements in order to test for convergence. The free-energy difference, $\Delta F^{\text{up}}$, together with $\Delta F^{\text{qh}\rightarrow\text{MTP}}$ from Eq. (4), gives the anharmonic vibrational contribution

$$F^{\text{ah}}(V,T) = \Delta F^{\text{qh}\rightarrow\text{MTP}}(V,T) + \Delta F^{\text{up}}(V,T).\qquad(8)$$

Since both are calculated on the same $(V,T)$ mesh, we are able to sum them point wise and parametrize them together. For the upsampling to high-DFT without electronic temperature, the smearing method by Methfessel and Paxton is used. This is essential because if Fermi smearing was used with a large smearing width of $\sigma = k_B T$ at high $T$, the extrapolated 0 K energy would be inaccurate.[7,37] Then errors around 10 meV/atom would be possible (for Nb).

In the second step of the direct upsampling, we calculate the electronic free energy using the same snapshots as in the first step. Here, we use also the thermodynamic perturbation, which is now applied between DFT with and without electronic temperature, as given by

$$F^{\text{el}} = -k_B T \ln\left\langle\exp\left(-\frac{E_{\text{el}}^{\text{DFT}} - E^{\text{DFT}}}{k_B T}\right)\right\rangle_{\text{MTP}},\qquad(9)$$

where $E_{\text{el}}^{\text{DFT}}$ is the high-DFT energy including thermal electronic excitations obtained by the Fermi smearing corresponding to the relevant temperature. (Note that the number of bands NBANDS might need to be increased for Fermi smearing calculations with respect to the Methfessel-Paxton calculations.) The term $F^{\text{el}}$ accounts for both the thermal electronic

Supplementary Table 11. Parameters for the anharmonic calculations; TI from QH to high-MTP and direct upsampling from high-MTP to high-DFT.

| System | TI QH to high-MTP | | Direct upsampling | | Volumes $V$ (Å³/atom) | Temperature mesh (K) |
|---|---|---|---|---|---|---|
| | Supercell | Atoms | Supercell | Atoms | | |
| Nb–PBE | $6 \times 6 \times 6$ | 432 | $4 \times 4 \times 4$ | 128 | 18.18 … 20.37, 11 V's | $2750/7 \times \{0, 0.33, 0.67, 1, 1.5, 2, 3, 4, 5, 5.5, 6, 6.5, 7, 7.5\}$ |
| Ni–PBE | $5 \times 5 \times 5$ | 500 | $3 \times 3 \times 3$ | 108 | 10.71 … 12.05, 11 V's | $1728/7 \times \{0, 1, 2, 3, 4, 5, 5.5, 6, 6.5, 7, 7.5\}$ |
| Al–PBE | $5 \times 5 \times 5$ | 500 | $3 \times 3 \times 3$ | 108 | 16.55 … 18.11, 9 V's | $933/7 \times \{0, 1, 2, 3, 4, 5, 5.5, 6, 6.5, 7, 7.5\}$ |
| Al–LDA | $5 \times 5 \times 5$ | 500 | $3 \times 3 \times 3$ | 108 | 15.87 … 17.28, 10 V's | $933/7 \times \{0, 1, 2, 3, 4, 5, 5.5, 6, 6.5, 7, 7.5\}$ |
| Mg–PBE | $7 \times 7 \times 5$ | 490 | $4 \times 4 \times 3$ | 96 | 22.99 … 25.45, 9 V's | $923/7 \times \{0, 1, 2, 3, 4, 5, 5.5, 6, 6.5, 7, 7.5\}$ |
| Mg–LDA | $7 \times 7 \times 5$ | 490 | $4 \times 4 \times 3$ | 96 | 21.46 … 23.61, 9 V's | $923/7 \times \{0, 1, 2, 3, 4, 5, 5.5, 6, 6.5, 7, 7.5\}$ |

free energy of the static lattice and its coupling to atomic vibrations.

Additionally, for magnetic systems (here, Ni), we also extract the magnetic moments from the high-DFT runs in the second step. The magnetic moments $m$ for each $(V, T)$ are parametrized and, along with the Curie temperature, used as model parameters to calculate the magnetic free energy $F^{\mathrm{mag}}(V, T)$ as discussed in the next section.

### G. Parametrization and thermodynamic properties

#### 1. Parametrization to obtain smooth free energy surfaces

The free energy contributions $F^{\mathrm{qh}}$, $F^{\mathrm{ah}}$, and $F^{\mathrm{el}}$, and the average magnetic moment $m$ (which is used to calculate $F^{\mathrm{mag}}$), explicitly available at specific $V$ and $T$ points from the previous step are parametrized in $V$ and $T$. This smoothens the free energy surface and allows for numerically converged first and second derivatives. Each contribution is parametrized separately with the basis sets shown in Supplementary Table 12 and as discussed next.

Since $F^{\mathrm{qh}}$ is available as an analytical function in $T$ for every $V$ point in the grid (Eq. (3)), we need to parametrize it only in $V$. For that purpose, we use a 3rd order polynomial in $V$ (basis set in first column of Supplementary Table 12). The parametrization is done on a very fine temperature mesh with 1 kelvin steps.

The anharmonic free energies $F^{\mathrm{ah}}$ (Eq. (8)) calculated previously on a dense set of $(V, T)$ grid points are used to obtain a smooth anharmonic free energy surface $F^{\mathrm{ah}}(V, T)$. The fitting is done in terms of a renormalized anharmonic frequency $\bar{\omega}^{\mathrm{ah}}(V, T)$, which is expanded as a polynomial in $V, T$. The basis set for the expansion is given in the second column of Supplementary Table 12. The free energy is computed as

$$F^{\mathrm{ah}}(V, T) = F_{\mathrm{eff}}[T; \bar{\omega}(V) + \bar{\omega}^{\mathrm{ah}}(V, T)] - F_{\mathrm{eff}}[T; \bar{\omega}(V)], \quad (10)$$

where

$$F_{\mathrm{eff}}[T; \bar{\omega}] = k_B T \ln\left[1 - \exp\left(-\frac{\hbar \bar{\omega}}{k_B T}\right)\right]. \quad (11)$$

Here, $\bar{\omega}(V)$ is an effective QH frequency computed as the average frequency of the effective QH reference; $\bar{\omega}^{\mathrm{ah}}(V, T)$ is the renormalization shift due to anharmonicity. To simplify

the choice of the initial guess for the parameters during fitting, $V$ is rescaled to the range $[-1, 1]$ and $T$ to the range $(0, 1]$.

The electronic free energy $F^{\mathrm{el}}$ (Eq. (9)) is parametrized using a fourth order polynomial in $V$ and $T$. Such a fitting of the $F^{\mathrm{el}}$ values is analogous to the parametrization described earlier in Ref. [33].[38] The basis set of the polynomial used here is provided in the third column in Supplementary Table 12. We restrict the $T$ order of the polynomial to $T^2$ or higher, which is in line with the parametrization in Ref. [33] and the Sommerfeld approximation of $F^{\mathrm{el}}$ at low $T$ (Eq. (12) in Ref. [39]).

The parametrization of the magnetic free energy contribution $F^{\mathrm{mag}}(V, T)$ is done as follows.[21] For a given $V$, the average magnetic moment $m(T)$ is parametrized using the temperature basis shown in Supplementary Table 12. Using $m(T)$ and the experimental Curie temperature $T_{\mathrm{C}} = 633$ K for Ni, the empirical magnetic heat capacity $C^{\mathrm{mag}}$ is calculated as (Eq. (9) in Ref. [40])

$$C^{\mathrm{mag}}(T) = \begin{cases} k_f \dfrac{T}{T_{\mathrm{C}}} \exp\left[-4\left(1 - \dfrac{T}{T_{\mathrm{C}}}\right)\right] & T < T_{\mathrm{C}} \\ k_p \dfrac{T}{T_{\mathrm{C}}} \exp\left[8p\left(1 - \dfrac{T}{T_{\mathrm{C}}}\right)\right] & T > T_{\mathrm{C}}. \end{cases} \quad (12)$$

The prefactors are $k_f = 4(1 - f_s)S^{\mathrm{mag}}(T)/[1 - \exp(-4)]$ and $k_p = 8p f_s S^{\mathrm{mag}}(T)$; $f_s = 0.105$ and $p = 2$ for fcc, and the magnetic entropy is given as

$$S^{\mathrm{mag}}(T) = k_B \ln[m(T) + 1]. \quad (13)$$

Following this, the free energy as a function of $T$ is obtained from $C^{\mathrm{mag}}(T)$ by numerical integration, given as

$$F^{\mathrm{mag}}(T) = \int_0^T C^{\mathrm{mag}}(T) dT - T \int_0^T C^{\mathrm{mag}}(T)/T dT. \quad (14)$$

Based on this procedure, $F^{\mathrm{mag}}(T)$ is calculated for every $V$ in the grid. Then we parametrize the volume dependence of the magnetic free energy on the same very fine temperature mesh of 1 kelvin steps as for $F^{\mathrm{qh}}$. For this purpose, the free energy is written as

$$F^{\mathrm{mag}}(V, T) = -k_B T \ln[1 + m_{\mathrm{eff}}(V, T)], \quad (15)$$

where $m_{\mathrm{eff}}$ is an effective magnetic moment. At every 1 K, the free energy is smoothened by parametrizing $m_{\mathrm{eff}}(V)$ using the volume basis set given in the last column of Supplementary Table 12.

Supplementary Table 12. Overview of basis sets used for the parametrization of the free energy contributions. Here, $\tilde{\omega}^{\mathrm{ah}}(V,T)$ is fitted by a 4th order polynomial of $V$ and $T$.

| $F^{\mathrm{qh}}(V)$ | $F^{\mathrm{ah}}$: $\tilde{\omega}^{\mathrm{ah}}(V,T)$ | $F^{\mathrm{el}}(V,T)$ | $F^{\mathrm{mag}}(V,T)$ |
|---|---|---|---|
| $\{1, V, V^2, V^3\}$ | $\{1, V, T, V^2, VT, T^2,$ $V^3, V^2T, VT^2, T^3,$ $V^4, V^3T, V^2T^2, VT^3, T^4\}$ | $\{T^2, VT^2, T^3,$ $V^2T^2, VT^3, T^4\}$ | $m(T) : \{1, T^2, T^3\}$ $m_{\mathrm{eff}}(V) : \{1, V, V^2\}$ |

### 2. Thermodynamic properties

From the above step (Supplementary Methods 2 G 1), both $F^{\mathrm{qh}}(V,T)$ and $F^{\mathrm{mag}}(V,T)$ are available as parametrizations in volume for a very fine temperature mesh (at every 1 K). The other two contributions $F^{\mathrm{ah}}(V,T)$ and $F^{\mathrm{el}}(V,T)$ are discretized in temperature and converted to a similar parametrization in volume for every 1 K. The full free energy surface $F(V,T)$ is obtained over the relevant volume and temperature range by summing up all the just-described, separately parametrized contributions, together with the 0 K $E$-$V$ curve (Eqs. (1) and (2)).

Thermodynamic properties could be obtained directly from $F(V,T)$, however, for constant pressure properties the Gibbs energy surface $G(p,T)$ is more convenient. Thus, a Legendre transformation on the free-energy surface is performed as

$$G(p,T) = F(V,T) + pV. \qquad (16)$$

Any relevant pressure can be used that is properly captured by the volume range of the explicitly computed $(V,T)$ points. Here (Appendix F), we provide ambient pressure properties. The tabulated properties are calculated numerically according to the following equations:

- enthalpy

$$H(p,T) = G(p,T) + T\,S(p,T), \qquad (17)$$

- entropy

$$S(p,T) = -\left(\frac{\partial G(p,T)}{\partial T}\right)_p, \qquad (18)$$

- isobaric heat capacity

$$C_p(p,T) = T\left(\frac{\partial S(p,T)}{\partial T}\right)_p \qquad (19)$$

- isochoric heat capacity

$$C_V(p,T) = C_p(p,T) - T\frac{B_T(p,T)}{V^{\mathrm{eq}}(p,T)}\left(\frac{\partial V^{\mathrm{eq}}(p,T)}{\partial T}\right)^2, \qquad (20)$$

- equilibrium volume

$$V^{\mathrm{eq}}(p,T) = \left(\frac{\partial G(p,T)}{\partial p}\right)_T, \qquad (21)$$

- linear expansion coefficient

$$\alpha(p,T) = \frac{1}{3V^{\mathrm{eq}}(p,T)}\left(\frac{\partial V^{\mathrm{eq}}(p,T)}{\partial T}\right), \qquad (22)$$

- isothermal bulk modulus (via the compressibility $\kappa_T$)

$$B_T(p,T) = \frac{1}{\kappa_T}, \quad \kappa_T = -\frac{1}{V^{\mathrm{eq}}(p,T)}\left(\frac{\partial V^{\mathrm{eq}}(p,T)}{\partial p}\right)_T, \qquad (23)$$

- adiabatic bulk modulus

$$B_S(p,T) = \begin{cases} B_T(p,T)\dfrac{C_p}{C_V} & C_V(p,T) > 0.01\,k_B, \\[2mm] B_T(p,T) & C_V(p,T) < 0.01\,k_B. \end{cases} \qquad (24)$$

## SUPPLEMENTARY DISCUSSION 1. SPECIAL CONSIDERATIONS FOR THE METHODOLOGY

### A. Convergence of upsampling

In this section (Supplementary Discussion 1 A), the validity of the direct upsampling—as performed within free energy perturbation theory[34,35,41]—is thoroughly investigated. We focus on Nb at $T_{\mathrm{melt}}$ and $V_{\mathrm{melt}}$, which is the most challenging of all elements studied in this work, in terms of RMSEs and convergence. In particular, we analyze the order of approximation used for the perturbation formula (Eq. (7)) and the number of configurations needed to converge the directly upsampled free energy.

### 1. Order of approximation of the free energy perturbation formula

To test for convergence, we compare the upsampled free energies to values from an explicit TI, which is considered as the gold standard. To decrease the computational burden at this stage, we substitute the DFT system with a level-24 'model-MTP'. This model-MTP is consequently used as the target system to which the TI and upsampling is performed from different lower-level MTPs.

Supplementary Figure 5 (a) shows the upsampled free energy from a level-20 MTP to the model-MTP as a function of the number of uncorrelated snapshots, using different levels of approximation to the perturbation formula, Eq. (7). It is observed that the statistical convergence of the free energy difference is similar for all levels of approximation. The first order approximation (in green) converges statistically slightly

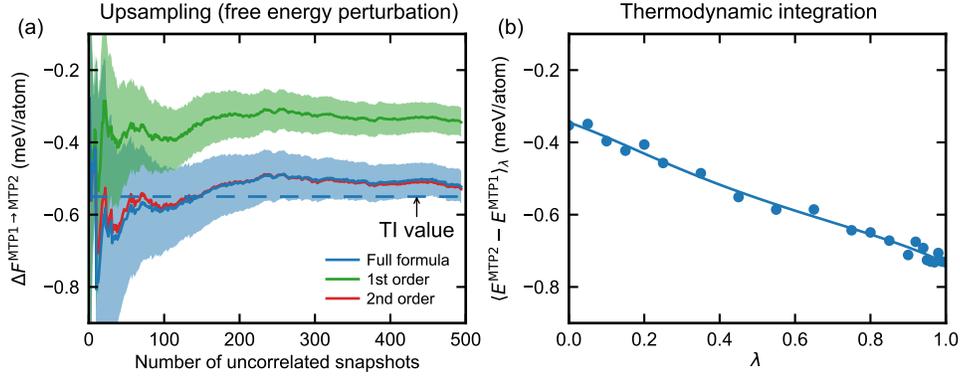

Supplementary Figure 5. Convergence of upsampling for Nb at $T_{\text{melt}}$ and near $V_{\text{melt}}$ in a large-size cell. (a) Upsampled free energy difference between a level-20 MTP (MTP1) and the level-24 model-MTP (MTP2) as a function of the number of snapshots used for the upsampling with the full free energy perturbation formula, and approximated up to the first and second-order. The shaded region indicates the estimated error bar of convergence. The 'TI value' is the free energy difference computed as the area under the curve during TI as shown in (b). (b) Energy difference versus the coupling parameter $\lambda$ between MTP1 and MTP2 during TI. The blue curve is a fit to the data points. The area under this curve is the 'TI value' in (a).

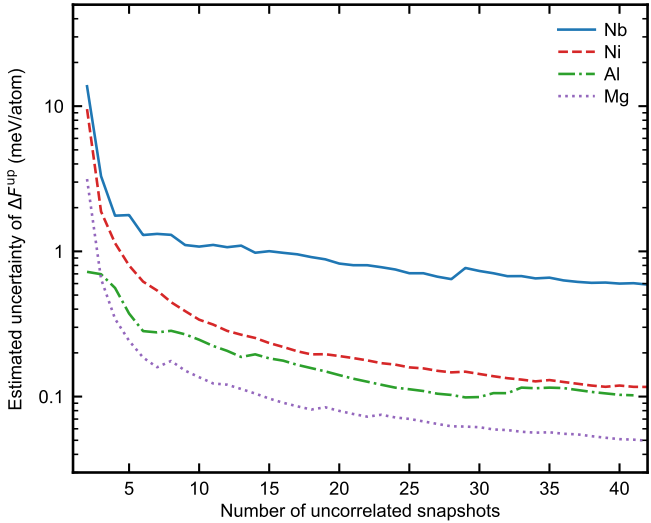

Supplementary Figure 6. Convergence of the estimated error during direct upsampling from high-MTP to DFT as the number of snapshots increase.

faster, but it gives an offset in comparison to the value calculated using the formula approximated to second order (in red). The calculation up to the second order is already converged in the final value compared to that using the full formula (in blue). The blue curve and the red curve fall on top of each other, owing to the small RMSE of the level-20 MTP (0.67 meV/atom with respect to the level-24 model-MTP). The shaded region around the mean denotes the estimated error bar.

The converged value of the upsampled free energy clearly falls very close to that obtained from TI, which is indicated by the dashed blue line denoted as the 'TI value'. The TI value of $-0.55 \pm 0.01$ meV/atom in Supplementary Figure 5 (a) is obtained from the area under the curve shown in Supplementary Figure 5 (b), which is the average energy difference

as function of $\lambda$ during the TI from the level-20 MTP to the model-MTP. Note that the converged value of the first order approximation of the perturbation formula in Supplementary Figure 5 (a) ($-0.34 \pm 0.04$ meV/atom) corresponds exactly to the value for $\lambda = 0$ in Supplementary Figure 5 (b). Since the $\lambda$-dependence is almost linear, the second order approximated value and the full perturbation formula value in Supplementary Figure 5 (a) approximately equal the value for $\lambda = 0.5$. The slope of the $\lambda$-dependence is always negative, and correspondingly the converged full perturbation will always give a lower free energy difference than the first order approximated value. This is also reflected in the second order term, which is always negative.

From the above analysis we conclude that at least the second order approximation of the perturbation equation is vital to capture the full upsampled free energy difference. Since the statistical convergence behaviour is similar between the 2nd order and the full formula, we rely in general on the full perturbation equation within our procedure and also for the further convergence analysis in the following subsection.

### 2. Number of snapshots for convergence

In addition to the order of approximation of the perturbation formula, we also investigate the number of uncorrelated snapshots needed for convergence. This is analyzed by comparing the estimated error in the upsampled free energy (from a level-20 MTP to DFT) as a function of the number of snapshots as shown in Supplementary Figure 6. It is seen that Nb has by far the largest $\Delta F^{\text{up}}$ error and is thereby also the most challenging to converge. Nonetheless, it converges to within 0.6 meV/atom with around 40 snapshots. Note that if correlated snapshots were used, sampling errors could be estimated by the block averaging procedure.[41–43] Since the snapshots used here are separated by a large enough time interval, and the systems studied here do not possess rare events, the snap-

shots can be assumed to be uncorrelated.

As noticeable from Supplementary Figure 6, the number of snapshots required for convergence depends on the RMSE of the MTP, and a certain target accuracy. Indeed, the requisite number of uncorrelated snapshots can be estimated using the first order approximation of Eq. (7), $\Delta F^{\mathrm{up}} \approx \left\langle E^{\mathrm{DFT}} - E^{\mathrm{MTP}} \right\rangle_{\mathrm{MTP}}$. As seen in Supplementary Figure 5 (a), the first order approximation shows a similar convergence dependence with the number of snapshots as the full formula. The estimated 95% confidence interval is approximately $\pm 2s/\sqrt{n}$ (which holds true for large $n$ ($> 30$), and in fact fairly well also for smaller $n$), where $s$ is the sample standard deviation of $E^{\mathrm{DFT}} - E^{\mathrm{MTP}}$ and $n$ the number of snapshots. The sample standard deviation $s$ can be approximated as the energy RMSE of the MTP (at $T < T_{\mathrm{melt}}$ the RMSE—evaluated at $T_{\mathrm{melt}}$—generally provides an overestimate of $s$). Thereby, the estimated confidence interval is $\pm 2\,\mathrm{RMSE}/\sqrt{n}$ and the corresponding required number of snapshots is given as

$$ n = \left( \frac{2\,\mathrm{RMSE}}{c} \right)^2, \qquad (25) $$

where $\pm c$ is the target accuracy. For instance, for Nb with an MTP RMSE of $2\,\mathrm{meV/atom}$, and a target accuracy of $0.6\,\mathrm{meV/atom}$, around 40 snapshots are needed. For Mg with an MTP RMSE of $0.16\,\mathrm{meV/atom}$, and a target accuracy of $0.1\,\mathrm{meV/atom}$, around 10 snapshots are needed.

In order to validate these estimates, and clarify the trends, we return to the level-24 model-MTP from the previous subsection (Supplementary Discussion 1 A 1) and perform upsampling to this system from MTPs of different levels. We then plot the number of snapshots required for convergence, against the RMSE of the MTPs (with respect to the model-MTP). Such curves are shown in Supplementary Figure 7 for different target accuracies. From the figure, it is clear that Eq. (25) holds both qualitatively with respect to the trend, and to a good extent quantitatively (within the error bars of the upsampling to the model-MTP, which are based on an average over several independent upsampling procedures).

In Supplementary Figure 7 there are, additionally, two stars marked. The purple star corresponds to the upsampling to DFT for Nb with an MTP energy RMSE of $2\,\mathrm{meV/atom}$ and a target accuracy of $0.6\,\mathrm{meV/atom}$. This falls within the error bars of the purple curve corresponding to the same target accuracy. The red star corresponds similarly to the upsampling to DFT for Mg with an MTP energy RMSE of $0.16\,\mathrm{meV/atom}$ and a target accuracy of $0.1\,\mathrm{meV/atom}$ from the above example. We see again that, if the red curve corresponding to the target of $0.1\,\mathrm{meV/atom}$ is extrapolated, it would pass closely through the red star.

It is hence clarified that, given a certain MTP with a particular RMSE and a target accuracy of the directly upsampled energy, we can *a priori* estimate the number of snapshots needed for a converged upsampled free energy value, irrespective of the system.

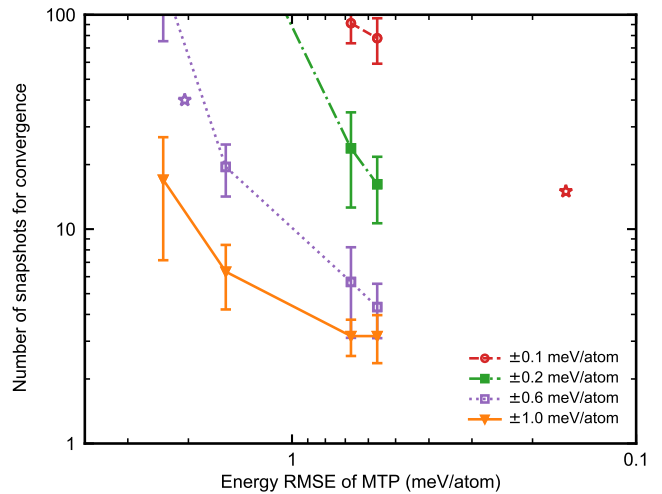

Supplementary Figure 7. Number of snapshots required for convergence of the upsampling at $T_{\mathrm{melt}}$ using the full perturbation formula as a function of the energy RMSE of the MTP. The markers connected by lines denote upsampling to the model-MTP for Nb, for RMSEs of the various MTPs with respect to the model-MTP. The purple star shows upsampling from MTP to DFT in Nb within $\pm 0.6$ meV/atom, and the red star shows upsampling from MTP to DFT in Mg within $\pm 0.1$ meV/atom. For these points the RMSE is with respect to DFT.

### B. Convergence as the density of the $(V, T)$ grid increases

A dense $(V, T)$ grid is crucial to obtain a smooth and converged free energy surface and its first and second derivatives. This requirement transfers to the Gibbs energy surface and its derivatives. In Supplementary Figure 8, we illustrate the convergence of the anharmonic free energy and the resulting thermodynamic properties calculated with different grid densities. The first column is the anharmonic free energy at the melting point with respect to the fully converged grid (cf. Supplementary Table 11). Third order polynomials are used to avoid overfitting for the coarse grid. Since the first column contains the free energy as a function of volume, we demonstrate in the second and third column thermodynamic properties that depend significantly on corresponding changes, i.e., the thermal expansion coefficient and bulk modulus. The experimental values are also provided as a visual aid to gauge the extent of convergence coming from the different grid densities. As the grid density increases, the anharmonic free energies begin to converge as noticed by the proximity of the curves. The thermodynamic properties from the maximum $V, T$ grid are similar to those from a $9V \times 8T$ grid indicating grid convergence. For the final, tabulated properties the maximum grid is used.

Even small differences in the anharmonic free energy (e.g., about 0.05 meV/atom for Mg PBE and 0.5 meV/atom for Nb at $T_{\mathrm{melt}}$) cause noticeable changes (of several percent) in the thermal expansion coefficient and bulk modulus, especially near the end of the grid points or at high $T$. This is particularly visible by noticing the difference in the gray dotted and black solid curves, for example, in the bulk modulus of Nb and Mg

PBE (about 6% difference). This shows the importance of an extremely-accurate convergence of the free energy surface by increasing the grid density especially at higher temperatures and higher volumes. We use an optimized mesh instead of an equi-spaced mesh, in order to stabilize the highly-curved nature of certain thermodynamic properties. In addition, choosing grid points above $T_{melt}$ and $V_{melt}$ is also done to obtain converged thermodynamic properties at the melting point.

### C. Convergence as the order of the fit polynomial increases

Similarly as the grid density, the polynomial basis used in parametrizing the free energy surfaces, in particular the anharmonic surface, is crucial as well. This is illustrated in Supplementary Figure 9 which shows the anharmonic free energy (with respect to the 4th order parametrization) calculated using different orders and the resulting thermal expansion coefficient and bulk modulus. A second order polynomial fit (gray dotted curve) of $F^{ah}(V,T)$ differs from the fourth order only by less than 1 meV/atom for Nb and Ni and even less than 0.1 meV/atom for Al and Mg. However, its impact on the expansion coefficient and bulk modulus is significant especially near the melting point (3% to 10% difference in both properties depending on the system). As the order of the polynomial for parametrizing $F^{ah}(V,T)$ increases further to three and four, the free energy differs by $< 0.05$ meV/atom. Thus we achieve a high convergence in the thermodynamic properties (red dashed and black solid curves fall on each other). At the same time, one should keep in mind that using much higher-order polynomials ($> 4$) can result in overfitting which can be checked by visually inspecting the graphs.

### D. Distinctive features for Nb

The outstanding performance of our methodology is clearly suggested by the remarkable agreement of the calculated properties with experimental data. Specifically, the predictions for Nb—the strong temperature dependence and curvature of the expansion coefficient—are reassuring, since Nb has a strong electronic and anharmonic contribution to the free energy at higher temperatures. Yet, Nb possesses long-range interactions at 0 K that gradually disappear as temperature increases.[51,52] This calls for a more thorough investigation of the methodology at lower temperatures as described next.

#### 1. Validity of direct upsampling

The long-range interactions in Nb at 0 K are, to some extent, also present at low temperatures.[51] MTPs, by nature of their construction, consider only local interactions (up to $R_{cut}$). Direct upsampling from the high-MTP to high-DFT is performed in a medium-size supercell (128 atoms) in which the long-range interactions may not be fully captured. The validity of direct upsampling should therefore be explicitly veri-

fied. In order to do so, we have performed explicit TI calculations from the high-MTP to DFT at 500 K (18% of $T_{melt}$) on two differently-sized, but commensurate (in terms of $k$-sampling) cells—a 54-atom and a 432-atom Nb system. By doing such a TI on a large supercell, we include long-range interactions in the DFT energies and forces, i.e., a potential impact of the long-range interactions on the phase space distribution is included. The results of the TI reveal that the free energy difference between the high-MTP and DFT changes by $< 1$ meV/atom when the size increases from 54 to 432 atoms. This shows that the interactions at this temperature are already to a good degree captured even by the small-size cell. Further, the anharmonic free energy differs by only 1 meV/atom between the value obtained from direct upsampling and the TI in the large-size supercell. These results justify the applicability of the methodology for high-temperature property predictions (from at least 500 K to $T_{melt}$), even for systems such as Nb.

#### 2. Effective QH over 0 K QH

One of the benefits of using the effective QH as a reference for TI is evident while considering Nb. The long-range interactions at 0 K in Nb warrant the need for large supercells to obtain the 0 K dynamical matrix. Otherwise, artificial imaginary phonons could appear in the 0 K dynamical matrix when extended from small-size to large-size supercells (as a consequence of the Fourier interpolation) thereby limiting its role as a reference in TI. Averaging over many configurations in a finite-temperature MD run smears out long-range effects and makes the interactions more short-ranged as discussed in Supplementary Discussion 1 D 1. Hence, by using an effective QH reference (with a three neighbor-shell cutoff) fitted to MD forces at 20 K, we achieve dynamical stability even in small supercells. The analytical formula of the effective QH also serves as a good approximation of the low temperature quantum-mechanical free energy.

#### 3. Limits of the MTPs for Nb

Since the high-MTP is fitted to $T_{melt}$ and interactions are restricted by $R_{cut}$, its predictability decreases as temperature nears 0 K with long-range interactions contributing to the true energy. One needs to keep in mind that this loss of predictability is compensated to a large extent during the upsampling stage. To test whether the predictability of the MTP can be improved at temperatures near 0 K, we have added a 0 K structure to the training set and re-trained the high-MTP. The re-trained MTP exhibits an imaginary mode near the Γ point as shown by the orange-dashed curve in Supplementary Figure 10 (a). This is corroborated in Supplementary Figure 10 (b) which displays the potential energy as atoms are increasingly moved along the eigenvectors of the dynamical matrix (corresponding to $q$ points along Γ to N). However, the high-MTP fitted only to $T_{melt}$, shown by the blue curve in Supplementary Figure 10 (a), exhibits only real frequencies and is dynamically stable.

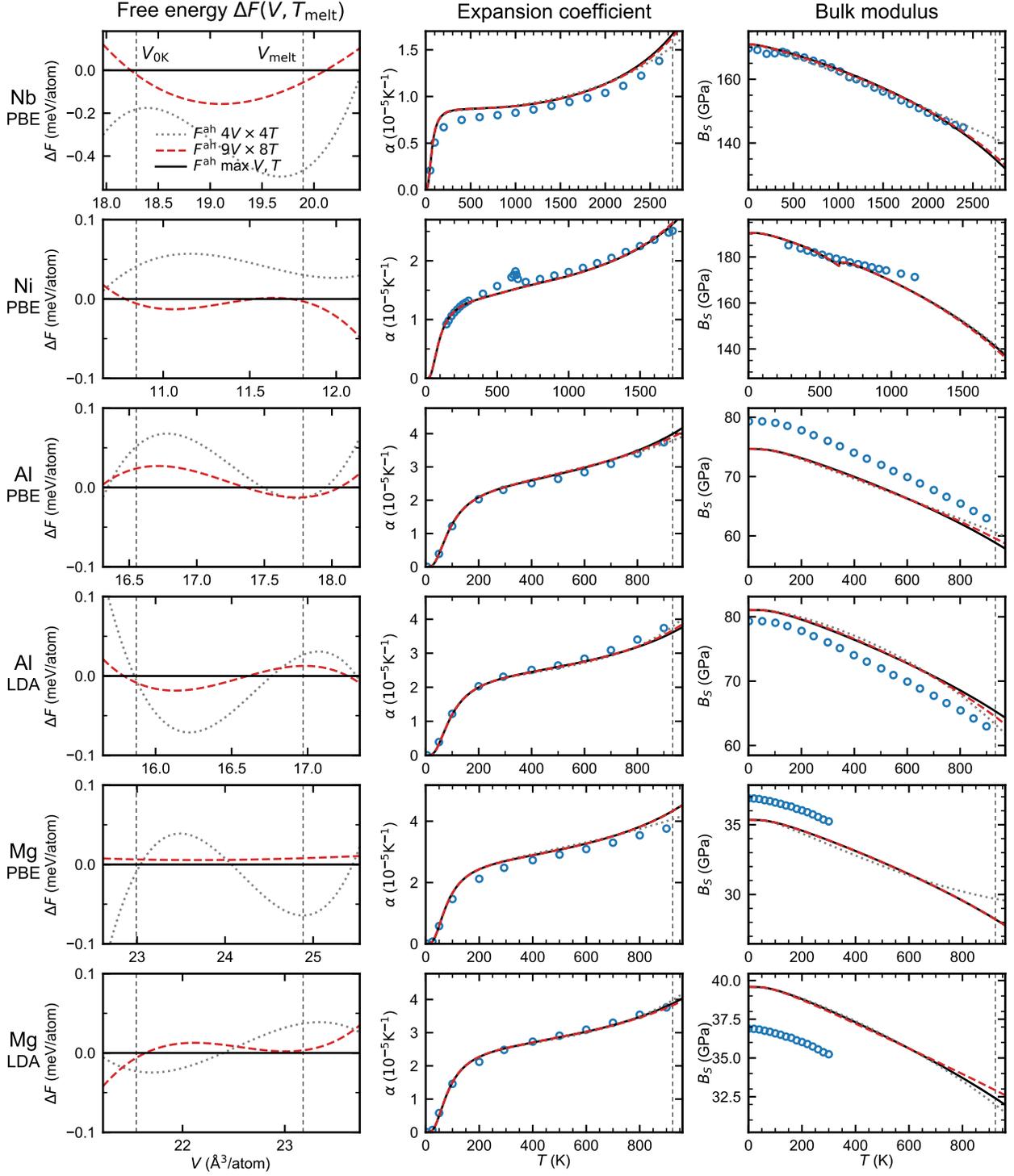

Supplementary Figure 8. Convergence of thermodynamic properties as the density of the $(V, T)$ mesh for $F^{ah}(V, T)$ increases. Here, $\Delta F$ denotes the difference in total free energy from the mesh with the highest density. In addition, the thermal expansion coefficient, and the adiabatic bulk modulus are shown for the different elements and functionals. The circles in the second and third column denote experimental data: Nb,[44,45] Ni,[46,47] Al,[48,49] and Mg.[48,50] The 3rd order polynomials are used for the parametrization of $F^{ah}$, and the total free energy includes the contribution from $F^{el}$.

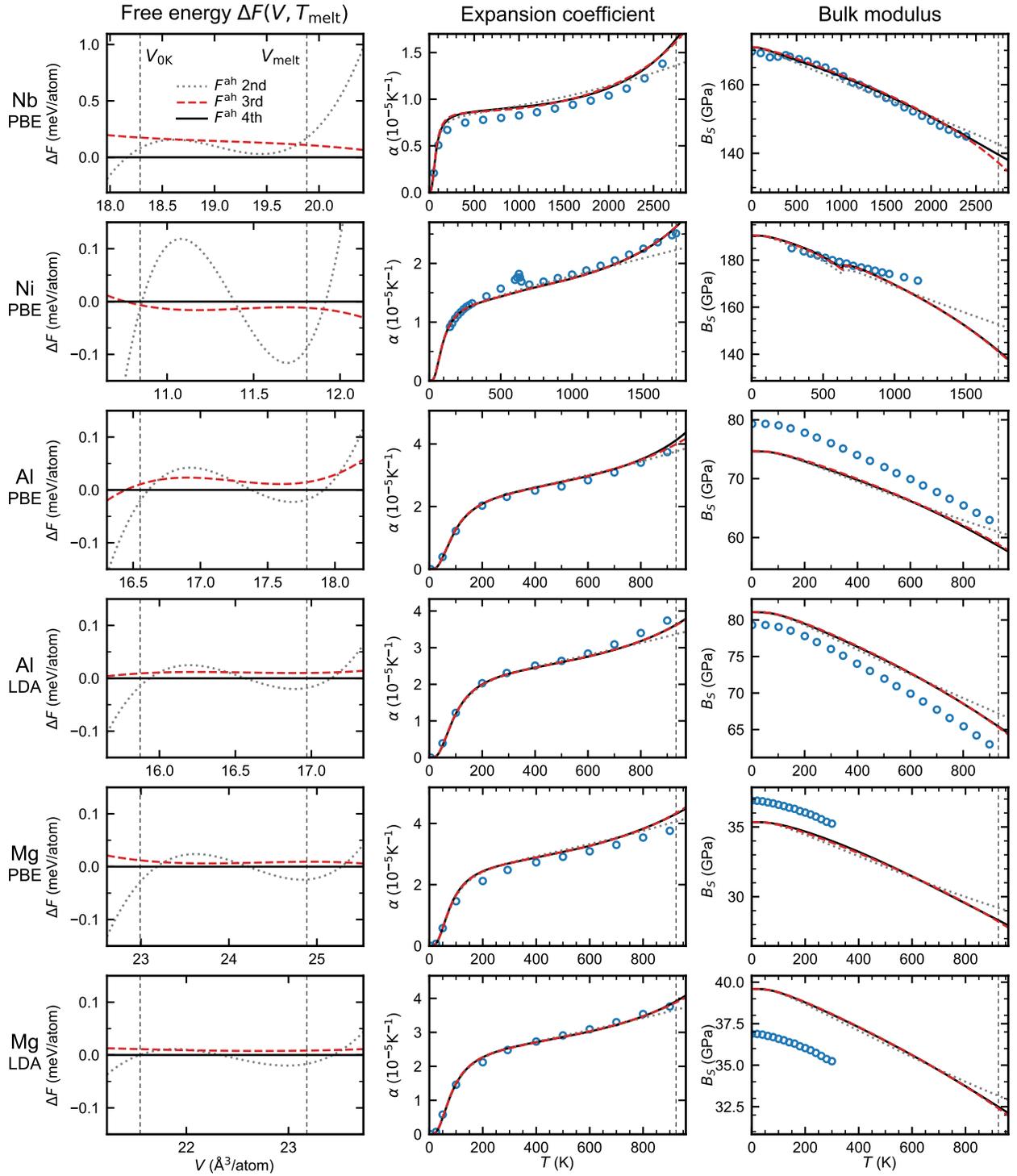

Supplementary Figure 9. Convergence of thermodynamic properties as the order of the polynomial for $\tilde{\omega}^{\mathrm{ah}}(V,T)$ increases, which is used in the parametrization of $F^{\mathrm{ah}}(V,T)$. Here, $\Delta F$ denotes the difference in total free energy from the polynomial with the highest order of 4. In addition, the thermal expansion coefficient and the adiabatic bulk modulus are shown for the different elements and functionals. The circles denote experimental data: Nb,[44,45] Ni,[46,47] Al,[48,49] and Mg.[48,50] The total free energy includes the contribution from $F^{\mathrm{el}}$.

高

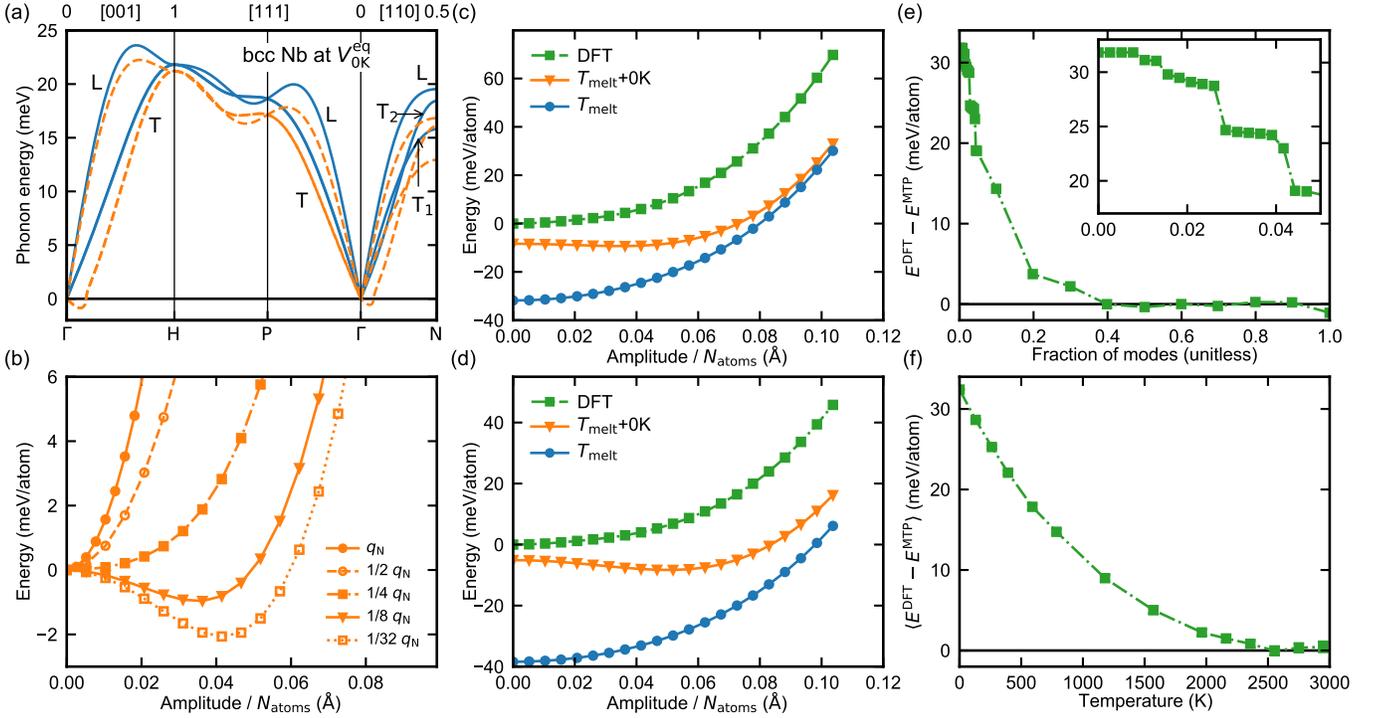

Supplementary Figure 10. (a) Phonon dispersions of effective QH potentials fitted to MTP forces at 20 K in a medium-size supercell. The blue solid line is from an MTP fitted to snapshots at $T_{melt}$. The orange dashed line is from an MTP that has an additional 0 K structure in the training set. (b) Potential energy calculated using the MTP including the 0 K structure as a function of atom displacement along different eigenvectors of the dynamical matrix of Nb. The $q$ points of the eigenvectors lie along the Γ to N direction (labelled by $q_N$) and the used wavelengths are indicated in the legend. (c, d) Comparison of the DFT and MTP energies in a 16-atom supercell at (c) $V_{0K}$ and (d) $\approx V_{melt}$, when the atoms are displaced along the wavevector $(1/8)q_N$. The $x$-axes indicate the displacement per atom. The label $T_{melt}$ represents the MTP fitted to snapshots at $T_{melt}$, and $T_{melt}$+0K corresponds to the MTP with an additional 0 K snapshot in the training set. (e) Energy difference between DFT and the MTP fitted at $T_{melt}$ for configurations obtained by exciting an increasing fraction of modes. (f) Averaged energy difference $\langle E^{DFT} - E^{MTP} \rangle$ between DFT and the MTP fitted at $T_{melt}$ at $\approx V_{0K}$ for finite-temperature snapshots (Eq. (7)).

In Supplementary Figure 10 (c) and (d), we show the potential energy as atoms are increasingly displaced along the specific wavevector $(1/8)q_N$ at $V_{0K}$ and $V_{melt}$. Energies from the two MTPs (fitted with and without the 0 K structure shown in orange and blue respectively) are compared to DFT energies. The energy difference between the MTP and DFT is much reduced when the 0 K structure is added to the training set but this adds an unwanted negative curvature (dynamical instability) at small amplitudes.

We also notice in Supplementary Figure 10 (c) and (d) that the high-MTP fitted only to $T_{melt}$ (blue curves) has an almost constant difference with DFT even with increasing amplitude. The reason for this is, as alluded to earlier, that the displacement of atoms is along a single mode, while the high-MTP is fitted to $T_{melt}$, which contains all modes interacting with one another. To further verify this, in Supplementary Figure 10 (e), we plot the difference in DFT and MTP as a function of the fraction of modes. To obtain the configurations on which we performed these calculations, we did the following. From a snapshot at $T_{melt}$ and $V_{melt}$, phonon modes were subtracted by projection, starting from the mode with the highest frequency to that with zero frequency. Thereby, we created different structures with decreasing fraction of modes. From

Supplementary Figure 10 (e), it is observed that, as the fraction of modes increases, the energy difference between DFT and MTP drops down to the RMSE in energy in the high-MTP. This is a direct consequence of the high-MTP being able to accurately predict a large fraction of the modes interacting with each other accurately.

The interacting modes are also activated as temperature increases. In Supplementary Figure 10 (f) we plot the energy difference between the high-MTP and high-DFT (similar to the upsampled energy) as a function of temperature. As expected, the predictability of the high-MTP increases with increasing temperature. Note that for the other elements, the predictability of the high-MTP is very good even at low $T$. Specifically, the energy difference at 0 K is $-1.01$ meV/atom for Ni–PBE, 0.70 meV/atom for Al–PBE, and $-0.41$ meV/atom for Mg–PBE.

To sum up, in Nb, the MTP fitted to $T_{melt}$ (where there are highly interacting phonon modes) is unable to predict the DFT energies at lower temperatures (with de-coupled phonons and long-range interactions). Trying to add a 0 K structure to the training set introduces an incorrect instability in the system. Hence, the complex interactions in Nb governing its properties over the full temperature including in particular 0 K data

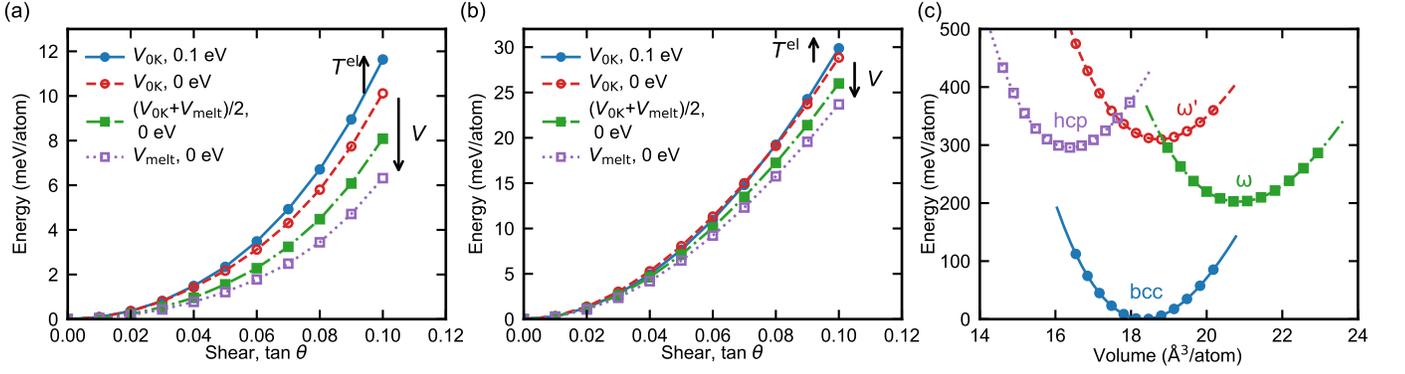

Supplementary Figure 11. (a, b) DFT energy of an Nb cell when sheared according to the long-wavelength limit of the Γ-N direction, in the (a) T2 transversal direction and (b) T1 transversal direction as a function of $\tan\theta$, where $\theta$ is the change of the angle due to the shear. The legend shows the corresponding volumes and electronic temperatures in units of eV. (c) $E$-$V$ curves of Nb in bcc, $\omega$ phase with $c/a = \sqrt{3/8}$ corresponding to bcc ($\omega$'), $\omega$ phase with relaxed $c/a$, and hcp with relaxed $c/a$.

are a challenge to the flexibility and robustness of the MTP. Further investigations are required to exploit the full capabilities of MTPs for systems like Nb.

### 4. DFT stability of bcc Nb

Here, we ensure that the dynamical instability of the MTP fitted including the 0 K structure arises purely from a deficiency of the MTP and is not a feature of DFT. In Supplementary Figure 11 (a) and (b), we plot the DFT energies of bcc Nb under shear along the two relevant transversal directions, at different electronic temperatures and volumes. As the electronic temperature increases, Nb becomes more bcc-stable, analogous with Ca as shown previously (cf. Fig. 5 in Ref. [33]). However, in contrast to Ca and in contrast to the MTP fitted with the 0 K structure, bcc Nb is always dynamically stable in DFT regardless of the electronic temperature and volume.

Additionally, we ensure that bcc Nb (group V element) is also stable with respect to the hexagonal $\omega$ phase observed in Nb-containing alloys[53] and known to be energetically the most favorable phase in DFT at 0 K for the group IV elements[54] (Ti, Zr, Hf). To this end, we analyze the respective energy-volume curves. Specifically, in Supplementary Figure 11 (c), the DFT $E$-$V$ curve of bcc Nb over a large volume range is compared to the $\omega$ phase ($\omega$') with $c/a = \sqrt{3/8} = 0.612$ ($c/a$ corresponding to bcc), to the $\omega$ phase with a relaxed $c/a = 0.549$, and additionally to the relaxed hcp structure with $c/a = 1.817$. The observed energy dependencies verify the stability of the bcc structure in Nb.

### E. Temperature integration with quantum correction

The free energy difference from TI can be corroborated by an alternative method, i.e., temperature integration. This can serve as a cross-check calculation (as utilized in the present study) or can become essential for cases where TI is unstable,

e.g., for systems with defects which allow for nearby atoms to migrate at higher temperatures.[55,56] In light of the high-accuracy requirements on the free energies, the usual temperature integration requires a correction, specifically a quantum correction, to yield exactly the same free energy differences as TI.

Within the temperature integration approach, the difference between two free energies $F_A$ and $F_B$ corresponding to temperatures $T_A$ and $T_B$ (at a fixed volume) can be calculated by integration of the internal energy $U_{cor}$ over the inverse temperature as[34]

$$\frac{F_B}{T_B} - \frac{F_A}{T_A} = \int_{1/T_A}^{1/T_B} U_{cor}(T)\, d\left(\frac{1}{T}\right). \quad (26)$$

As compared to the standard implementation,[34] the internal energy requires a correction (indicated by the subscript 'cor'), in order to match the free energy difference with the one obtained from TI. The correction is obtained as

$$U_{cor}(T) = U(T) + \Delta U_{qm}(T), \quad (27)$$

where $U = \langle E \rangle$ is the usual internal energy obtained by averaging total energies $E$ (here MTP energies) during a classical molecular dynamics run and $\Delta U_{qm}$ the quantum correction calculated as

$$\Delta U_{qm}(T) = U_{harm}^{qm}(T) - U_{harm}^{cl}(T). \quad (28)$$

Here, $U_{harm}^{qm}(T)$ and $U_{harm}^{cl}(T)$ are harmonic internal energies at temperature $T$ obtained from the usual analytical formulas in the quantum mechanical and classical version, respectively. For the classical part this is simply $U_{harm}^{cl}(T) = 3k_BT$. The harmonic frequencies entering the quantum-mechanical formula correspond to the effective QH dynamical matrix. In general, the same dynamical matrix has to be used as the one that functions as the reference for the complementary TI.

Only with this quantum correction, $F^{vib}$ calculated using temperature integration agrees very precisely with $F^{vib}$ from TI. The reason is that the TI based calculation implicitly includes $\Delta U_{qm}(T)$, since $F^{qh}$ in Eq. (2) is computed with the

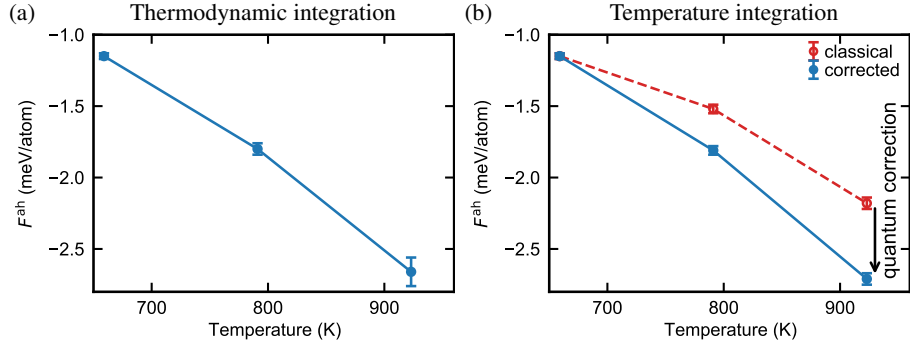

Supplementary Figure 12. Comparison of $F^{\mathrm{ah}}$ (specifically $\Delta F^{\mathrm{qh}\rightarrow\mathrm{MTP}}$) calculated by (a) thermodynamic integration and (b) temperature integration for Mg in a large-size supercell at the largest volume in the grid. (b) shows both the classical and quantum-corrected temperature integration.

quantum-mechanical formula, Eq. (3). Taking the quantum correction into account, either method can, in principle, be used to calculate the full vibrational free energy. Due to the significantly faster statistical convergence, we prefer TI within the general procedure.

Supplementary Figure 12 illustrates the quantum correction needed to obtain an accurate $F^{\mathrm{ah}}$ with temperature integration. When the quantum correction is included, $F^{\mathrm{ah}}$ agrees accurately with that of thermodynamic integration (both evaluated up to the MTP level). Although the correction can be less than 1 meV, it can be essential for second-derivative properties of the free-energy surface at high temperatures near the melting point.

## F. Finite size effects: Shifts of atomic planes in small hcp cells

Large supercell sizes become crucial in some systems while performing TI calculations due to finite-size effects. This is the case for hcp systems, where whole atomic layers can shift cooperatively in small supercells to change the stacking sequence. Such a behavior was observed previously for hcp Ni within stacking fault energy calculations.[21] Here, we observe it for our hcp prototype system, i.e., hcp Mg.

When running a TI from an effective QH to a high-MTP for hcp Mg in a small supercell of 36 atoms, we notice singular jumps in the QH energies for $\lambda = 1$ (cf. Supplementary Figure 13 (a)). A coupling of $\lambda = 1$ means that the forces are determined solely by the high-MTP. The trajectories of the atoms reveal that whole atomic layers (close packed planes) are able to perform correlated shifts to neighboring stacking positions as exemplified in Supplementary Figure 13 (c) and (d). The shifts are responsible for the jumps in the QH energies, because these QH energies are calculated with respect to a fixed lattice. The fixed lattice corresponds to the initial stacking layer sequence. The further the planes shift away from the initial order, the larger the divergence of the QH energies. Such shifts are more probable at higher temperatures and larger volumes as shown by the contour lines in Supplementary Figure 13 (b).

The observed atomic plane shifts are facilitated by the peri-

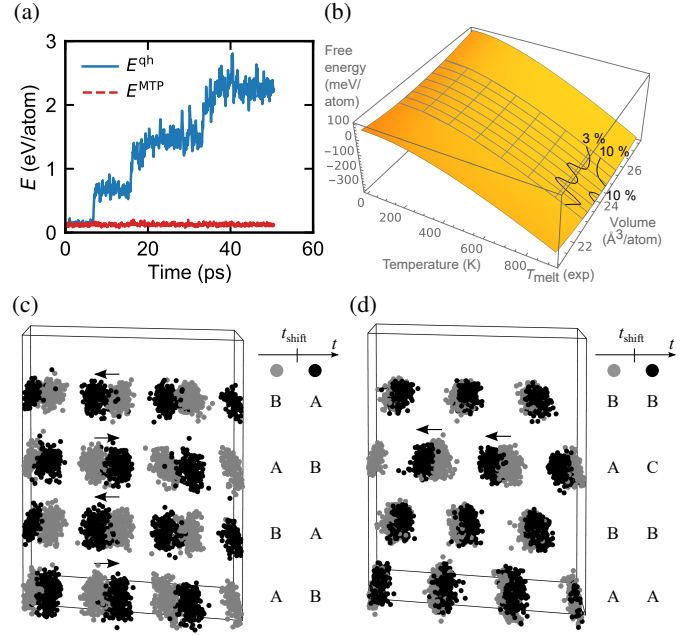

Supplementary Figure 13. Finite size effect: shifts of close-packed planes of atoms in hcp Mg in a small-size supercell during TI with MTP MD. (a) Energy from the effective harmonic reference and MTP during TI at 923 K. (b) Contour plot for the percentage of TI runs where instabilities occurred. (c, d) Thermal positions of atoms before and after the shift at (c) 923 K and (d) 791 K. Gray and black colors indicate positions at times before and after the shift (the time of the shift is labelled $t_{\mathrm{shift}}$).

odic boundary conditions in the small 36 atom supercell. The probability of a shift event decreases exponentially as the simulation cell size increases (in any of the plane directions). For macroscopic systems such events would be fully suppressed even on experimental time scales. As for the simulations, we are able to circumvent the problem by increasing the supercell size to $\geq 96$ atoms, resulting in negligible probabilities for the shifts on the available time scales. Increasing the size of supercells is facilitated by the MTPs, allowing us to simulate considerably larger systems compared to what is possible

with high accuracy DFT.

It should be stressed that the shifts are not a shortcoming of the MTPs. The shifts would be likewise observed (for small cells) for trajectories driven by DFT forces, if only the simulation time scale was long enough (tens of ps, cf. Supplementary Figure 13 (a)). Further, it is worth pointing out that employing temperature integration as just discussed in Supplementary Discussion 1 E would not solve the intrinsic problem of the shifts, even though the singular jumps in the QH energies could be avoided (because there is no QH reference, except for the quantum correction). It should be clear that the final free energy calculated by temperature integration would not correspond to pure hcp, but rather to an average over different stacking orders.

### SUPPLEMENTARY DISCUSSION 2. TOTAL CPU TIME

CPU time estimates for the four prototype systems are given in Supplementary Table 13. The computational load depends to a large extent on the system, and clearly on the number of valence electrons. The upsampling stage takes the most time (the darkest gray column in Supplementary Table 13) which involves two sets of DFT calculations on 42 snapshots at each $(V, T)$. The error of the upsampling from MTP to DFT (see Supplementary Methods 2 F) decreases as $1/\sqrt{N}$, where $N$ is the number of snapshots. Considering this, an improvement in the MTP energy RMSE by 10% will decrease the number of configurations needed to achieve the same accuracy by around 20%, and vice versa. It is therefore crucial for the efficiency of the workflow to have a high-MTP that reproduces the high-DFT energies and forces as accurately as possible.

### SUPPLEMENTARY DISCUSSION 3. ADDITIONAL RESULTS: Al AND Mg LDA FREE ENERGIES

In addition to the PBE XC results shown in Fig. 5 in the main text, we also present the results using the LDA XC functional for Al and Mg in Supplementary Figure 14. The overbinding tendency of the LDA XC and the resulting Gibbs energy curves at ambient pressure have been discussed already in the main text (cf. Results). The effect of the electronic and anharmonic contribution to the free energy at the melting point (second column) is similar to that observed for the PBE XC functional. In the third column, the anharmonic free energy of Al agrees within 1.5 meV to previous calculations.[21] The difference arises from using the effective QH reference (instead of the 0 K QH reference). In general, the anharmonic free energies are very close (< 1 meV/atom) to the PBE counterparts (check Fig. 5).

### SUPPLEMENTARY NOTE 1. ABBREVIATIONS IN TABLE 1 OF THE MAIN TEXT

AE: all-electron method

ASW: augmented spherical wave method

DAMA: decoupled anharmonic mode approximation

EMTO: exact muffin-tin orbital theory

GW: GW approximation

LDA: local density approximation

MD: molecular dynamics simulation

NC: norm-conserving pseudopotential formalism

PAW: projector augmented-wave method

PBE: generalized gradient approximation by Perdew *et al.*

PW91: generalized gradient approximation by Perdew *et al.*

SCAILD: self-consistent *ab initio* lattice dynamics

TI: thermodynamic integration

TBMD: tight-binding molecular dynamics

USP: ultrasoft pseudopotential formalism

UP-TILD: upsampled thermodynamic integration using Langevin dynamics

### SUPPLEMENTARY DISCUSSION 4. THERMODYNAMIC DATABASE

Thermodynamic properties at ambient pressure up to the meting point for all four prototype systems are tabulated on the following pages. The results are calculated including all relevant contributions—the 0 K $E$-$V$ curve, vibrational contribution including anharmonicity, electronic contribution including coupling, and magnetic contribution for Ni. The data shown in the tables are also available online on a mesh with 1 kelvin steps:

https://doi.org/10.18419/darus-3239

Contents:

Symbols in tables:

$T$: temperature

$G$: Gibbs energy (Eq. (16))

Supplementary Table 13. CPU time estimates in core hours for the prototype systems with the PBE XC functional. The given number of valence electrons are for the PAW potential used. The columns shaded in light gray correspond to low-DFT calculations, that in gray to high-DFT calculations, while those in white to MTP. The CPU time is given for an Intel Xeon E6252 Gold processor. A medium-size cell is used for generating snapshots for the MTP training and a large-size cell is used for the final TI.

| | Number of valence electrons | $(V, T)$ points | Low-DFT MD Small-size cell $4V$, $1T$ 1000 steps | High-DFT[a] Medium-size cell $9/11$[b] $V$, $2T$ 30 snapshots | TI Medium/large-size cell $9/11$[b] $V$, $10/13$[c] $T$ 22 $\lambda$ | Upsample Medium-size cell $9/11$[b] $V$, $10/13$[c] $T$ 42 snapshots | Sum |
|---|---|---|---|---|---|---|---|
| | | | | | CPU time (core hr) | | |
| Nb | 11 | 143 | 340 | 20,000 | 7700 | 410,000 | 440,000 |
| Ni | 10 | 110 | 400 | 33,000 | 8900 | 380,000 | 420,000 |
| Al | 3 | 90 | 25 | 2400 | 6700 | 18,000 | 28,000 |
| Mg | 2 | 90 | 73 | 1100 | 5600 | 16,000 | 22,000 |

[a] Used for MTP training and fitting of the effective QH potential.
[b] 9 $V$ for Al and Mg; and 11 $V$ for Nb and Ni.
[c] 10 $T$ for Al, Ni and Mg; 13 $T$ for Nb (excluding 0 K in the count).

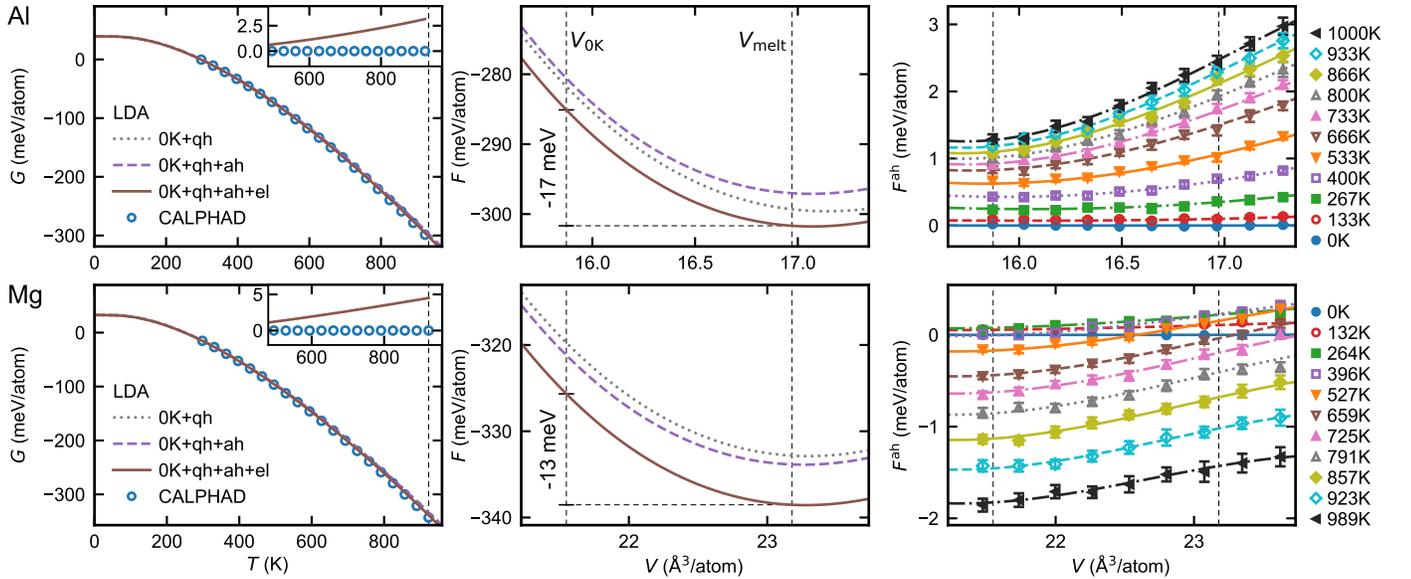

Supplementary Figure 14. *Ab initio* calculated Gibbs energy $G(T)$ at 100 kPa, free energy $F(V)$ at $T_{melt}$, and anharmonic free energy $F^{ah}(V,T)$, for Al and Mg using the LDA XC functional. The $G(T)$ values are referenced to the minimum energy of the static lattice at 0 K. Calculations using the CALPHAD method[57] (aligned to the *ab initio* values at room temperature) are shown in blue dots for comparison. The insets contain the full *ab initio* Gibbs energy at high temperatures with respect to the CALPHAD values. For $G(T)$ and $F(V, T_{melt})$, curves including different excitation mechanisms are shown. The vertical dashed line in the first column is the melting temperature and the lines in the second and third column are equilibrium volumes corresponding to 0 K and the melting point.

$H$: enthalpy (Eq. (17))

$S$: entropy (Eq. (18))

$C_p$: isobaric heat capacity (Eq. (19))

$C_V$: isochoric heat capacity (Eq. (20))

$k_B$: Boltzmann constant

$V^{eq}$: volume (Eq. (21))

$\alpha$: linear thermal expansion coefficient (Eq. (22))

$B_T$: isothermal bulk modulus (Eq. (23))

$B_S$: adiabatic bulk modulus (Eq. (24))

## A. Nb PBE

Supplementary Table 14. Thermodynamic properties of Nb by the PBE functional.

| T (K) | G (meV/atom) | H (meV/atom) | S ($k_B$) | $C_p$ ($k_B$) | $C_V$ ($k_B$) | $V^{eq}$ (Å³/atom) | α ($10^{-5}/K$) | $B_T$ (GPa) | $B_S$ (GPa) |
|---|---|---|---|---|---|---|---|---|---|
| 1 | 26.3 | 26.3 | 0.000 | 0.000 | 0.000 | 18.289 | 0.000 | 170.9 | 170.9 |
| 40 | 26.1 | 27.0 | 0.258 | 0.612 | 0.612 | 18.290 | 0.160 | 170.9 | 170.9 |
| 80 | 23.8 | 31.3 | 1.085 | 1.808 | 1.804 | 18.297 | 0.498 | 170.4 | 170.8 |
| 120 | 18.5 | 38.6 | 1.944 | 2.396 | 2.385 | 18.311 | 0.671 | 169.9 | 170.7 |
| 160 | 10.5 | 47.3 | 2.674 | 2.671 | 2.653 | 18.326 | 0.752 | 169.3 | 170.5 |
| 200 | 0.1 | 56.8 | 3.286 | 2.818 | 2.793 | 18.343 | 0.795 | 168.7 | 170.3 |
| 240 | -12.1 | 66.6 | 3.807 | 2.907 | 2.875 | 18.361 | 0.820 | 168.1 | 170.0 |
| 280 | -26.1 | 76.7 | 4.259 | 2.967 | 2.927 | 18.380 | 0.836 | 167.6 | 169.8 |
| 298.15 | -32.9 | 81.4 | 4.446 | 2.988 | 2.945 | 18.388 | 0.842 | 167.3 | 169.7 |
| 300 | -33.6 | 81.8 | 4.464 | 2.990 | 2.947 | 18.389 | 0.842 | 167.3 | 169.7 |
| 350 | -53.9 | 94.8 | 4.928 | 3.035 | 2.984 | 18.412 | 0.854 | 166.5 | 169.3 |
| 400 | -76.0 | 107.9 | 5.335 | 3.069 | 3.010 | 18.436 | 0.862 | 165.7 | 169.0 |
| 450 | -99.8 | 121.1 | 5.698 | 3.096 | 3.029 | 18.460 | 0.868 | 165.0 | 168.6 |
| 500 | -125.1 | 134.5 | 6.025 | 3.120 | 3.045 | 18.484 | 0.873 | 164.2 | 168.2 |
| 550 | -151.7 | 148.0 | 6.323 | 3.141 | 3.058 | 18.508 | 0.877 | 163.4 | 167.8 |
| 600 | -179.6 | 161.5 | 6.597 | 3.160 | 3.069 | 18.533 | 0.880 | 162.6 | 167.4 |
| 650 | -208.6 | 175.2 | 6.851 | 3.179 | 3.080 | 18.557 | 0.884 | 161.8 | 167.0 |
| 700 | -238.6 | 188.9 | 7.087 | 3.197 | 3.090 | 18.582 | 0.887 | 160.9 | 166.5 |
| 750 | -269.6 | 202.7 | 7.308 | 3.215 | 3.100 | 18.607 | 0.890 | 160.1 | 166.1 |
| 800 | -301.6 | 216.6 | 7.516 | 3.234 | 3.110 | 18.632 | 0.894 | 159.3 | 165.6 |
| 850 | -334.4 | 230.5 | 7.712 | 3.253 | 3.121 | 18.657 | 0.897 | 158.4 | 165.1 |
| 900 | -368.0 | 244.6 | 7.899 | 3.272 | 3.132 | 18.682 | 0.901 | 157.5 | 164.6 |
| 950 | -402.5 | 258.7 | 8.076 | 3.293 | 3.144 | 18.707 | 0.905 | 156.7 | 164.1 |
| 1000 | -437.6 | 272.9 | 8.246 | 3.314 | 3.157 | 18.733 | 0.909 | 155.8 | 163.5 |
| 1050 | -473.5 | 287.2 | 8.408 | 3.337 | 3.171 | 18.758 | 0.913 | 154.9 | 163.0 |
| 1100 | -510.1 | 301.6 | 8.563 | 3.360 | 3.186 | 18.784 | 0.918 | 153.9 | 162.4 |
| 1150 | -547.3 | 316.2 | 8.713 | 3.385 | 3.202 | 18.810 | 0.923 | 153.0 | 161.8 |
| 1200 | -585.2 | 330.8 | 8.858 | 3.412 | 3.219 | 18.836 | 0.929 | 152.1 | 161.2 |
| 1250 | -623.7 | 345.5 | 8.998 | 3.440 | 3.237 | 18.862 | 0.935 | 151.1 | 160.6 |
| 1300 | -662.7 | 360.4 | 9.133 | 3.470 | 3.257 | 18.889 | 0.942 | 150.1 | 159.9 |
| 1350 | -702.4 | 375.4 | 9.265 | 3.501 | 3.278 | 18.916 | 0.949 | 149.1 | 159.3 |
| 1400 | -742.6 | 390.6 | 9.393 | 3.535 | 3.300 | 18.943 | 0.957 | 148.1 | 158.6 |
| 1450 | -783.3 | 405.9 | 9.517 | 3.570 | 3.324 | 18.970 | 0.966 | 147.1 | 158.0 |
| 1500 | -824.6 | 421.3 | 9.639 | 3.608 | 3.350 | 18.998 | 0.975 | 146.0 | 157.3 |
| 1550 | -866.4 | 436.9 | 9.758 | 3.647 | 3.377 | 19.026 | 0.985 | 145.0 | 156.6 |
| 1600 | -908.7 | 452.7 | 9.874 | 3.689 | 3.405 | 19.054 | 0.996 | 143.9 | 155.9 |
| 1650 | -951.5 | 468.7 | 9.988 | 3.733 | 3.435 | 19.083 | 1.008 | 142.8 | 155.2 |
| 1700 | -994.8 | 484.9 | 10.100 | 3.779 | 3.467 | 19.112 | 1.020 | 141.7 | 154.4 |
| 1750 | -1038.5 | 501.3 | 10.211 | 3.828 | 3.500 | 19.141 | 1.034 | 140.5 | 153.7 |
| 1800 | -1082.8 | 517.9 | 10.319 | 3.880 | 3.535 | 19.171 | 1.049 | 139.3 | 152.9 |
| 1850 | -1127.5 | 534.7 | 10.426 | 3.934 | 3.571 | 19.202 | 1.065 | 138.1 | 152.2 |
| 1900 | -1172.6 | 551.8 | 10.532 | 3.991 | 3.610 | 19.233 | 1.082 | 136.9 | 151.4 |
| 1950 | -1218.2 | 569.1 | 10.636 | 4.052 | 3.649 | 19.264 | 1.101 | 135.6 | 150.6 |
| 2000 | -1264.3 | 586.7 | 10.740 | 4.115 | 3.691 | 19.296 | 1.120 | 134.3 | 149.8 |
| 2050 | -1310.8 | 604.5 | 10.842 | 4.182 | 3.734 | 19.329 | 1.142 | 133.0 | 149.0 |
| 2100 | -1357.7 | 622.7 | 10.944 | 4.252 | 3.778 | 19.362 | 1.165 | 131.6 | 148.1 |
| 2150 | -1405.1 | 641.2 | 11.045 | 4.326 | 3.825 | 19.397 | 1.190 | 130.2 | 147.3 |
| 2200 | -1452.9 | 660.0 | 11.145 | 4.404 | 3.873 | 19.432 | 1.217 | 128.8 | 146.5 |
| 2250 | -1501.1 | 679.1 | 11.245 | 4.487 | 3.922 | 19.468 | 1.246 | 127.3 | 145.6 |
| 2300 | -1549.8 | 698.6 | 11.344 | 4.574 | 3.973 | 19.505 | 1.277 | 125.8 | 144.8 |
| 2350 | -1598.9 | 718.5 | 11.444 | 4.666 | 4.026 | 19.542 | 1.311 | 124.2 | 143.9 |
| 2400 | -1648.4 | 738.8 | 11.543 | 4.763 | 4.081 | 19.582 | 1.348 | 122.6 | 143.1 |
| 2450 | -1698.4 | 759.6 | 11.642 | 4.867 | 4.137 | 19.622 | 1.388 | 120.9 | 142.2 |
| 2500 | -1748.8 | 780.8 | 11.742 | 4.977 | 4.195 | 19.663 | 1.431 | 119.1 | 141.4 |
| 2550 | -1799.6 | 802.5 | 11.841 | 5.095 | 4.255 | 19.706 | 1.478 | 117.3 | 140.5 |
| 2600 | -1850.8 | 824.7 | 11.942 | 5.220 | 4.316 | 19.751 | 1.529 | 115.5 | 139.6 |
| 2650 | -1902.5 | 847.5 | 12.042 | 5.355 | 4.380 | 19.797 | 1.585 | 113.5 | 138.8 |
| 2700 | -1954.6 | 870.8 | 12.144 | 5.501 | 4.445 | 19.845 | 1.646 | 111.5 | 137.9 |
| 2750 | -2007.1 | 894.9 | 12.246 | 5.658 | 4.513 | 19.895 | 1.713 | 109.3 | 137.1 |

## B. Ni PBE

Supplementary Table 15. Thermodynamic properties of Ni by the PBE functional.

| $T$ (K) | $G$ (meV/atom) | $H$ (meV/atom) | $S$ ($k_B$) | $C_p$ ($k_B$) | $C_V$ ($k_B$) | $V^{eq}$ (Å³/atom) | $\alpha$ ($10^{-5}/K$) | $B_T$ (GPa) | $B_S$ (GPa) |
|---|---|---|---|---|---|---|---|---|---|
| 1 | 37.6 | 37.6 | 0.000 | 0.000 | 0.000 | 10.845 | 0.000 | 190.4 | 190.4 |
| 40 | 37.5 | 37.8 | 0.096 | 0.264 | 0.264 | 10.845 | 0.095 | 190.4 | 190.4 |
| 80 | 36.4 | 40.4 | 0.582 | 1.261 | 1.258 | 10.849 | 0.506 | 189.8 | 190.2 |
| 120 | 33.2 | 46.1 | 1.248 | 2.009 | 1.998 | 10.858 | 0.829 | 188.8 | 189.8 |
| 160 | 27.7 | 53.8 | 1.890 | 2.443 | 2.421 | 10.870 | 1.018 | 187.6 | 189.3 |
| 200 | 20.2 | 62.7 | 2.465 | 2.709 | 2.675 | 10.884 | 1.133 | 186.2 | 188.6 |
| 240 | 10.8 | 72.3 | 2.975 | 2.892 | 2.846 | 10.899 | 1.208 | 184.9 | 187.9 |
| 280 | -0.3 | 82.5 | 3.432 | 3.036 | 2.978 | 10.915 | 1.262 | 183.5 | 187.0 |
| 298.15 | -5.8 | 87.3 | 3.624 | 3.095 | 3.031 | 10.923 | 1.282 | 182.8 | 186.7 |
| 320 | -12.9 | 93.2 | 3.845 | 3.164 | 3.093 | 10.932 | 1.304 | 182.0 | 186.2 |
| 360 | -26.8 | 104.3 | 4.225 | 3.290 | 3.207 | 10.950 | 1.338 | 180.6 | 185.2 |
| 400 | -42.0 | 115.8 | 4.578 | 3.428 | 3.332 | 10.967 | 1.369 | 179.1 | 184.2 |
| 440 | -58.4 | 127.9 | 4.912 | 3.588 | 3.479 | 10.986 | 1.397 | 177.6 | 183.1 |
| 480 | -75.9 | 140.5 | 5.232 | 3.784 | 3.662 | 11.004 | 1.423 | 176.0 | 181.9 |
| 520 | -94.5 | 154.0 | 5.544 | 4.033 | 3.896 | 11.023 | 1.450 | 174.5 | 180.6 |
| 560 | -114.1 | 168.4 | 5.854 | 4.353 | 4.201 | 11.043 | 1.477 | 172.9 | 179.2 |
| 600 | -134.8 | 184.1 | 6.168 | 4.772 | 4.604 | 11.062 | 1.505 | 171.3 | 177.6 |
| 640 | -156.7 | 200.8 | 6.482 | 4.160 | 3.975 | 11.083 | 1.534 | 169.7 | 177.6 |
| 680 | -179.4 | 214.3 | 6.720 | 3.759 | 3.558 | 11.103 | 1.557 | 168.1 | 177.6 |
| 720 | -203.0 | 227.0 | 6.930 | 3.627 | 3.411 | 11.124 | 1.579 | 166.4 | 177.0 |
| 760 | -227.2 | 239.4 | 7.125 | 3.600 | 3.367 | 11.145 | 1.601 | 164.8 | 176.2 |
| 800 | -252.1 | 251.8 | 7.310 | 3.614 | 3.364 | 11.167 | 1.623 | 163.0 | 175.2 |
| 840 | -277.6 | 264.3 | 7.487 | 3.644 | 3.376 | 11.189 | 1.646 | 161.3 | 174.1 |
| 880 | -303.7 | 276.9 | 7.657 | 3.681 | 3.395 | 11.211 | 1.670 | 159.6 | 173.0 |
| 920 | -330.4 | 289.7 | 7.821 | 3.721 | 3.416 | 11.234 | 1.694 | 157.8 | 171.9 |
| 960 | -357.7 | 302.5 | 7.981 | 3.763 | 3.438 | 11.257 | 1.719 | 156.0 | 170.7 |
| 1000 | -385.4 | 315.6 | 8.135 | 3.806 | 3.461 | 11.280 | 1.745 | 154.1 | 169.5 |
| 1040 | -413.7 | 328.8 | 8.285 | 3.851 | 3.485 | 11.304 | 1.772 | 152.3 | 168.3 |
| 1080 | -442.6 | 342.1 | 8.431 | 3.899 | 3.510 | 11.328 | 1.801 | 150.4 | 167.0 |
| 1120 | -471.9 | 355.6 | 8.574 | 3.948 | 3.536 | 11.353 | 1.831 | 148.4 | 165.8 |
| 1160 | -501.7 | 369.3 | 8.713 | 4.000 | 3.563 | 11.378 | 1.862 | 146.5 | 164.5 |
| 1200 | -532.0 | 383.2 | 8.850 | 4.054 | 3.591 | 11.404 | 1.895 | 144.5 | 163.1 |
| 1240 | -562.7 | 397.2 | 8.984 | 4.112 | 3.621 | 11.430 | 1.931 | 142.4 | 161.7 |
| 1280 | -593.9 | 411.5 | 9.115 | 4.173 | 3.653 | 11.457 | 1.968 | 140.4 | 160.3 |
| 1320 | -625.5 | 426.0 | 9.244 | 4.238 | 3.687 | 11.484 | 2.007 | 138.3 | 158.9 |
| 1360 | -657.6 | 440.7 | 9.372 | 4.307 | 3.724 | 11.512 | 2.050 | 136.1 | 157.4 |
| 1400 | -690.2 | 455.7 | 9.498 | 4.381 | 3.762 | 11.541 | 2.095 | 133.9 | 155.9 |
| 1440 | -723.1 | 470.9 | 9.622 | 4.460 | 3.804 | 11.570 | 2.143 | 131.6 | 154.3 |
| 1480 | -756.5 | 486.4 | 9.746 | 4.545 | 3.848 | 11.601 | 2.195 | 129.3 | 152.7 |
| 1520 | -790.3 | 502.2 | 9.868 | 4.637 | 3.896 | 11.632 | 2.251 | 126.9 | 151.1 |
| 1560 | -824.5 | 518.4 | 9.990 | 4.736 | 3.948 | 11.663 | 2.312 | 124.5 | 149.3 |
| 1600 | -859.2 | 534.9 | 10.111 | 4.844 | 4.003 | 11.696 | 2.377 | 122.0 | 147.6 |
| 1640 | -894.3 | 551.8 | 10.232 | 4.961 | 4.063 | 11.730 | 2.449 | 119.4 | 145.8 |
| 1680 | -929.7 | 569.1 | 10.353 | 5.089 | 4.128 | 11.765 | 2.527 | 116.7 | 143.9 |
| 1720 | -965.6 | 586.9 | 10.474 | 5.229 | 4.198 | 11.802 | 2.614 | 114.0 | 141.9 |
| 1728 | -972.9 | 590.5 | 10.499 | 5.258 | 4.213 | 11.809 | 2.632 | 113.4 | 141.5 |

## C.  Al PBE

Supplementary Table 16. Thermodynamic properties of Al by the PBE functional.

| $T$ (K) | $G$ (meV/atom) | $H$ (meV/atom) | $S$ ($k_B$) | $C_p$ ($k_B$) | $C_V$ ($k_B$) | $V^{eq}$ (Å³/atom) | $\alpha$ ($10^{-5}/K$) | $B_T$ (GPa) | $B_S$ (GPa) |
|---|---|---|---|---|---|---|---|---|---|
| 1 | 38.0 | 38.0 | 0.000 | 0.000 | 0.000 | 16.553 | 0.000 | 74.7 | 74.7 |
| 20 | 38.0 | 38.0 | 0.011 | 0.026 | 0.026 | 16.553 | 0.019 | 74.7 | 74.7 |
| 40 | 38.0 | 38.2 | 0.083 | 0.254 | 0.253 | 16.554 | 0.205 | 74.6 | 74.6 |
| 60 | 37.7 | 39.1 | 0.272 | 0.714 | 0.713 | 16.558 | 0.570 | 74.4 | 74.6 |
| 80 | 37.0 | 40.7 | 0.547 | 1.202 | 1.197 | 16.566 | 0.950 | 74.1 | 74.4 |
| 100 | 35.7 | 43.2 | 0.862 | 1.615 | 1.602 | 16.577 | 1.270 | 73.7 | 74.2 |
| 120 | 33.9 | 46.2 | 1.186 | 1.935 | 1.913 | 16.591 | 1.522 | 73.2 | 74.0 |
| 140 | 31.6 | 49.7 | 1.503 | 2.179 | 2.147 | 16.607 | 1.717 | 72.7 | 73.8 |
| 160 | 28.7 | 53.7 | 1.806 | 2.365 | 2.321 | 16.625 | 1.869 | 72.1 | 73.5 |
| 180 | 25.4 | 57.8 | 2.093 | 2.508 | 2.452 | 16.644 | 1.990 | 71.6 | 73.2 |
| 200 | 21.5 | 62.2 | 2.363 | 2.619 | 2.552 | 16.665 | 2.089 | 71.0 | 72.9 |
| 220 | 17.2 | 66.8 | 2.616 | 2.708 | 2.629 | 16.686 | 2.171 | 70.5 | 72.6 |
| 240 | 12.5 | 71.5 | 2.855 | 2.781 | 2.689 | 16.708 | 2.241 | 69.9 | 72.3 |
| 260 | 7.4 | 76.4 | 3.079 | 2.840 | 2.736 | 16.731 | 2.302 | 69.3 | 72.0 |
| 280 | 1.9 | 81.3 | 3.291 | 2.891 | 2.774 | 16.754 | 2.355 | 68.8 | 71.7 |
| 298.15 | -3.4 | 85.8 | 3.474 | 2.931 | 2.803 | 16.776 | 2.400 | 68.2 | 71.4 |
| 300 | -4.0 | 86.3 | 3.492 | 2.934 | 2.805 | 16.778 | 2.404 | 68.2 | 71.3 |
| 320 | -10.2 | 91.4 | 3.683 | 2.973 | 2.831 | 16.803 | 2.449 | 67.6 | 71.0 |
| 340 | -16.7 | 96.5 | 3.864 | 3.007 | 2.852 | 16.828 | 2.491 | 67.0 | 70.7 |
| 360 | -23.5 | 101.7 | 4.036 | 3.038 | 2.870 | 16.853 | 2.530 | 66.5 | 70.4 |
| 380 | -30.6 | 106.9 | 4.201 | 3.068 | 2.886 | 16.879 | 2.568 | 65.9 | 70.0 |
| 400 | -38.0 | 112.2 | 4.359 | 3.096 | 2.900 | 16.905 | 2.605 | 65.3 | 69.7 |
| 420 | -45.7 | 117.6 | 4.510 | 3.123 | 2.913 | 16.932 | 2.641 | 64.7 | 69.4 |
| 440 | -53.6 | 123.0 | 4.656 | 3.149 | 2.926 | 16.959 | 2.676 | 64.1 | 69.0 |
| 460 | -61.7 | 128.4 | 4.797 | 3.175 | 2.937 | 16.986 | 2.712 | 63.5 | 68.7 |
| 480 | -70.1 | 133.9 | 4.932 | 3.201 | 2.948 | 17.014 | 2.747 | 63.0 | 68.4 |
| 500 | -78.7 | 139.4 | 5.063 | 3.228 | 2.960 | 17.042 | 2.783 | 62.3 | 68.0 |
| 520 | -87.6 | 145.0 | 5.190 | 3.255 | 2.971 | 17.071 | 2.820 | 61.7 | 67.6 |
| 540 | -96.6 | 150.6 | 5.314 | 3.282 | 2.982 | 17.100 | 2.857 | 61.1 | 67.3 |
| 560 | -105.9 | 156.3 | 5.433 | 3.310 | 2.993 | 17.130 | 2.895 | 60.5 | 66.9 |
| 580 | -115.4 | 162.0 | 5.550 | 3.339 | 3.005 | 17.160 | 2.935 | 59.9 | 66.5 |
| 600 | -125.0 | 167.8 | 5.664 | 3.369 | 3.017 | 17.190 | 2.976 | 59.2 | 66.2 |
| 620 | -134.9 | 173.6 | 5.775 | 3.400 | 3.028 | 17.221 | 3.018 | 58.6 | 65.8 |
| 640 | -144.9 | 179.5 | 5.883 | 3.431 | 3.040 | 17.253 | 3.063 | 57.9 | 65.4 |
| 660 | -155.2 | 185.4 | 5.989 | 3.464 | 3.052 | 17.285 | 3.109 | 57.2 | 65.0 |
| 680 | -165.6 | 191.4 | 6.093 | 3.497 | 3.064 | 17.317 | 3.158 | 56.6 | 64.5 |
| 700 | -176.2 | 197.5 | 6.195 | 3.531 | 3.076 | 17.350 | 3.209 | 55.9 | 64.1 |
| 720 | -186.9 | 203.6 | 6.295 | 3.566 | 3.087 | 17.384 | 3.264 | 55.1 | 63.7 |
| 740 | -197.9 | 209.8 | 6.393 | 3.602 | 3.098 | 17.418 | 3.321 | 54.4 | 63.2 |
| 760 | -209.0 | 216.0 | 6.489 | 3.638 | 3.108 | 17.453 | 3.382 | 53.6 | 62.8 |
| 780 | -220.3 | 222.3 | 6.584 | 3.676 | 3.117 | 17.489 | 3.447 | 52.9 | 62.3 |
| 800 | -231.7 | 228.7 | 6.678 | 3.714 | 3.125 | 17.526 | 3.517 | 52.1 | 61.9 |
| 820 | -243.3 | 235.1 | 6.770 | 3.752 | 3.132 | 17.563 | 3.591 | 51.2 | 61.4 |
| 840 | -255.0 | 241.6 | 6.861 | 3.791 | 3.137 | 17.602 | 3.671 | 50.4 | 60.9 |
| 860 | -266.9 | 248.1 | 6.950 | 3.831 | 3.140 | 17.641 | 3.757 | 49.5 | 60.4 |
| 880 | -279.0 | 254.8 | 7.039 | 3.871 | 3.141 | 17.681 | 3.850 | 48.6 | 59.9 |
| 900 | -291.2 | 261.5 | 7.126 | 3.912 | 3.139 | 17.723 | 3.951 | 47.6 | 59.4 |
| 920 | -303.6 | 268.2 | 7.213 | 3.954 | 3.134 | 17.765 | 4.061 | 46.6 | 58.8 |
| 933 | -311.7 | 272.7 | 7.268 | 3.982 | 3.130 | 17.794 | 4.138 | 46.0 | 58.5 |

## D. Al LDA

Supplementary Table 17. Thermodynamic properties of Al by the LDA functional.

| $T$ (K) | $G$ (meV/atom) | $H$ (meV/atom) | $S$ ($k_B$) | $C_p$ ($k_B$) | $C_V$ ($k_B$) | $V^{eq}$ (Å$^3$/atom) | $\alpha$ ($10^{-5}/K$) | $B_T$ (GPa) | $B_S$ (GPa) |
|---|---|---|---|---|---|---|---|---|---|
| 1 | 39.4 | 39.4 | 0.000 | 0.000 | 0.000 | 15.874 | 0.000 | 81.1 | 81.1 |
| 20 | 39.4 | 39.4 | 0.010 | 0.023 | 0.023 | 15.874 | 0.015 | 81.1 | 81.1 |
| 40 | 39.3 | 39.6 | 0.075 | 0.228 | 0.228 | 15.874 | 0.173 | 81.0 | 81.0 |
| 60 | 39.0 | 40.3 | 0.248 | 0.661 | 0.660 | 15.878 | 0.511 | 80.8 | 81.0 |
| 80 | 38.4 | 41.9 | 0.505 | 1.136 | 1.131 | 15.884 | 0.880 | 80.5 | 80.9 |
| 100 | 37.2 | 44.2 | 0.805 | 1.546 | 1.534 | 15.894 | 1.197 | 80.1 | 80.7 |
| 120 | 35.6 | 47.1 | 1.117 | 1.871 | 1.851 | 15.907 | 1.448 | 79.6 | 80.5 |
| 140 | 33.4 | 50.6 | 1.425 | 2.122 | 2.091 | 15.922 | 1.642 | 79.1 | 80.3 |
| 160 | 30.6 | 54.4 | 1.721 | 2.314 | 2.272 | 15.938 | 1.792 | 78.6 | 80.0 |
| 180 | 27.4 | 58.5 | 2.002 | 2.463 | 2.410 | 15.956 | 1.910 | 78.0 | 79.7 |
| 200 | 23.7 | 62.8 | 2.268 | 2.581 | 2.516 | 15.975 | 2.004 | 77.5 | 79.5 |
| 220 | 19.6 | 67.3 | 2.518 | 2.675 | 2.598 | 15.995 | 2.082 | 76.9 | 79.2 |
| 240 | 15.0 | 72.0 | 2.753 | 2.751 | 2.663 | 16.015 | 2.146 | 76.3 | 78.9 |
| 260 | 10.1 | 76.8 | 2.976 | 2.814 | 2.715 | 16.036 | 2.201 | 75.8 | 78.6 |
| 280 | 4.8 | 81.6 | 3.186 | 2.868 | 2.756 | 16.057 | 2.248 | 75.2 | 78.3 |
| 298.15 | -0.4 | 86.2 | 3.367 | 2.910 | 2.787 | 16.077 | 2.287 | 74.7 | 78.0 |
| 300 | -0.9 | 86.6 | 3.385 | 2.914 | 2.790 | 16.079 | 2.291 | 74.6 | 77.9 |
| 320 | -6.9 | 91.7 | 3.574 | 2.953 | 2.818 | 16.101 | 2.329 | 74.1 | 77.6 |
| 340 | -13.2 | 96.8 | 3.754 | 2.989 | 2.842 | 16.124 | 2.364 | 73.5 | 77.3 |
| 360 | -19.9 | 101.9 | 3.926 | 3.021 | 2.862 | 16.147 | 2.396 | 72.9 | 77.0 |
| 380 | -26.8 | 107.1 | 4.090 | 3.050 | 2.879 | 16.171 | 2.427 | 72.3 | 76.6 |
| 400 | -34.0 | 112.4 | 4.247 | 3.077 | 2.894 | 16.194 | 2.457 | 71.8 | 76.3 |
| 420 | -41.4 | 117.7 | 4.398 | 3.103 | 2.908 | 16.218 | 2.486 | 71.2 | 76.0 |
| 440 | -49.1 | 123.1 | 4.542 | 3.128 | 2.920 | 16.243 | 2.514 | 70.6 | 75.6 |
| 460 | -57.1 | 128.5 | 4.682 | 3.152 | 2.932 | 16.267 | 2.543 | 70.0 | 75.3 |
| 480 | -65.3 | 133.9 | 4.816 | 3.176 | 2.943 | 16.292 | 2.571 | 69.4 | 74.9 |
| 500 | -73.7 | 139.4 | 4.946 | 3.200 | 2.953 | 16.318 | 2.599 | 68.8 | 74.6 |
| 520 | -82.3 | 144.9 | 5.072 | 3.224 | 2.963 | 16.343 | 2.628 | 68.2 | 74.2 |
| 540 | -91.2 | 150.5 | 5.194 | 3.248 | 2.973 | 16.369 | 2.658 | 67.6 | 73.8 |
| 560 | -100.3 | 156.1 | 5.313 | 3.273 | 2.983 | 16.396 | 2.689 | 67.0 | 73.5 |
| 580 | -109.5 | 161.8 | 5.428 | 3.298 | 2.993 | 16.422 | 2.720 | 66.3 | 73.1 |
| 600 | -119.0 | 167.5 | 5.540 | 3.324 | 3.003 | 16.449 | 2.753 | 65.7 | 72.7 |
| 620 | -128.6 | 173.2 | 5.649 | 3.350 | 3.014 | 16.477 | 2.787 | 65.1 | 72.3 |
| 640 | -138.5 | 179.0 | 5.756 | 3.378 | 3.024 | 16.504 | 2.823 | 64.4 | 71.9 |
| 660 | -148.5 | 184.8 | 5.861 | 3.406 | 3.035 | 16.532 | 2.860 | 63.7 | 71.5 |
| 680 | -158.7 | 190.7 | 5.963 | 3.435 | 3.046 | 16.561 | 2.899 | 63.1 | 71.1 |
| 700 | -169.0 | 196.7 | 6.063 | 3.466 | 3.058 | 16.590 | 2.940 | 62.4 | 70.7 |
| 720 | -179.6 | 202.7 | 6.161 | 3.498 | 3.070 | 16.620 | 2.984 | 61.7 | 70.3 |
| 740 | -190.3 | 208.7 | 6.257 | 3.531 | 3.081 | 16.650 | 3.029 | 61.0 | 69.8 |
| 760 | -201.1 | 214.8 | 6.351 | 3.565 | 3.093 | 16.680 | 3.078 | 60.2 | 69.4 |
| 780 | -212.2 | 221.0 | 6.444 | 3.600 | 3.106 | 16.711 | 3.130 | 59.5 | 69.0 |
| 800 | -223.4 | 227.2 | 6.536 | 3.637 | 3.118 | 16.743 | 3.184 | 58.7 | 68.5 |
| 820 | -234.7 | 233.5 | 6.626 | 3.676 | 3.130 | 16.775 | 3.242 | 57.9 | 68.0 |
| 840 | -246.2 | 239.9 | 6.715 | 3.716 | 3.142 | 16.808 | 3.304 | 57.1 | 67.6 |
| 860 | -257.9 | 246.3 | 6.803 | 3.758 | 3.154 | 16.842 | 3.371 | 56.3 | 67.1 |
| 880 | -269.7 | 252.8 | 6.890 | 3.801 | 3.165 | 16.876 | 3.442 | 55.5 | 66.6 |
| 900 | -281.6 | 259.4 | 6.976 | 3.847 | 3.177 | 16.912 | 3.518 | 54.6 | 66.1 |
| 920 | -293.7 | 266.1 | 7.061 | 3.894 | 3.187 | 16.948 | 3.600 | 53.7 | 65.6 |
| 933 | -301.7 | 270.4 | 7.116 | 3.926 | 3.194 | 16.972 | 3.656 | 53.1 | 65.3 |

## E.  Mg PBE

Supplementary Table 18. Thermodynamic properties of Mg by the PBE functional.

| $T$ (K) | $G$ (meV/atom) | $H$ (meV/atom) | $S$ ($k_B$) | $C_p$ ($k_B$) | $C_V$ ($k_B$) | $V^{eq}$ (Å³/atom) | $\alpha$ ($10^{-5}/K$) | $B_T$ (GPa) | $B_S$ (GPa) |
|---|---|---|---|---|---|---|---|---|---|
| 1 | 31.2 | 31.2 | 0.000 | 0.000 | 0.000 | 22.992 | 0.000 | 35.3 | 35.3 |
| 20 | 31.1 | 31.2 | 0.016 | 0.044 | 0.044 | 22.992 | 0.034 | 35.3 | 35.3 |
| 40 | 31.0 | 31.5 | 0.139 | 0.409 | 0.409 | 22.994 | 0.334 | 35.3 | 35.3 |
| 60 | 30.6 | 32.7 | 0.419 | 1.005 | 1.002 | 23.002 | 0.845 | 35.2 | 35.3 |
| 80 | 29.5 | 34.9 | 0.786 | 1.538 | 1.531 | 23.018 | 1.314 | 35.1 | 35.3 |
| 100 | 27.8 | 37.9 | 1.174 | 1.935 | 1.921 | 23.038 | 1.669 | 34.9 | 35.2 |
| 120 | 25.4 | 41.5 | 1.553 | 2.219 | 2.196 | 23.063 | 1.924 | 34.7 | 35.1 |
| 140 | 22.4 | 45.5 | 1.910 | 2.422 | 2.390 | 23.091 | 2.110 | 34.5 | 35.0 |
| 160 | 18.8 | 49.8 | 2.243 | 2.571 | 2.529 | 23.122 | 2.248 | 34.3 | 34.9 |
| 180 | 14.7 | 54.3 | 2.552 | 2.684 | 2.632 | 23.154 | 2.354 | 34.1 | 34.8 |
| 200 | 10.0 | 59.0 | 2.839 | 2.771 | 2.710 | 23.187 | 2.438 | 33.9 | 34.7 |
| 220 | 4.9 | 63.8 | 3.106 | 2.840 | 2.769 | 23.222 | 2.507 | 33.7 | 34.6 |
| 240 | -0.7 | 68.7 | 3.356 | 2.896 | 2.816 | 23.257 | 2.565 | 33.5 | 34.4 |
| 260 | -6.7 | 73.7 | 3.589 | 2.943 | 2.854 | 23.293 | 2.615 | 33.2 | 34.3 |
| 280 | -13.1 | 78.8 | 3.808 | 2.984 | 2.884 | 23.330 | 2.660 | 33.0 | 34.1 |
| 298.15 | -19.2 | 83.5 | 3.996 | 3.016 | 2.908 | 23.364 | 2.698 | 32.8 | 34.0 |
| 300 | -19.8 | 84.0 | 4.015 | 3.019 | 2.910 | 23.368 | 2.702 | 32.8 | 34.0 |
| 320 | -26.9 | 89.2 | 4.211 | 3.051 | 2.931 | 23.406 | 2.740 | 32.5 | 33.9 |
| 340 | -34.4 | 94.4 | 4.396 | 3.079 | 2.950 | 23.445 | 2.777 | 32.3 | 33.7 |
| 360 | -42.1 | 99.8 | 4.573 | 3.106 | 2.966 | 23.484 | 2.812 | 32.0 | 33.6 |
| 380 | -50.1 | 105.1 | 4.741 | 3.131 | 2.981 | 23.524 | 2.847 | 31.8 | 33.4 |
| 400 | -58.4 | 110.5 | 4.902 | 3.156 | 2.995 | 23.564 | 2.881 | 31.5 | 33.2 |
| 420 | -67.0 | 116.0 | 5.057 | 3.179 | 3.007 | 23.605 | 2.916 | 31.3 | 33.1 |
| 440 | -75.9 | 121.5 | 5.205 | 3.202 | 3.019 | 23.647 | 2.951 | 31.0 | 32.9 |
| 460 | -85.0 | 127.0 | 5.348 | 3.225 | 3.030 | 23.689 | 2.987 | 30.7 | 32.7 |
| 480 | -94.3 | 132.6 | 5.485 | 3.248 | 3.041 | 23.732 | 3.023 | 30.5 | 32.6 |
| 500 | -103.9 | 138.2 | 5.618 | 3.272 | 3.052 | 23.775 | 3.061 | 30.2 | 32.4 |
| 520 | -113.7 | 143.8 | 5.747 | 3.295 | 3.063 | 23.819 | 3.099 | 29.9 | 32.2 |
| 540 | -123.7 | 149.5 | 5.872 | 3.320 | 3.074 | 23.864 | 3.139 | 29.6 | 32.0 |
| 560 | -133.9 | 155.2 | 5.993 | 3.344 | 3.085 | 23.909 | 3.180 | 29.4 | 31.8 |
| 580 | -144.4 | 161.0 | 6.110 | 3.370 | 3.096 | 23.955 | 3.223 | 29.1 | 31.6 |
| 600 | -155.0 | 166.8 | 6.225 | 3.396 | 3.108 | 24.002 | 3.267 | 28.8 | 31.5 |
| 620 | -165.8 | 172.7 | 6.337 | 3.424 | 3.120 | 24.049 | 3.314 | 28.5 | 31.3 |
| 640 | -176.9 | 178.6 | 6.446 | 3.452 | 3.132 | 24.098 | 3.362 | 28.2 | 31.1 |
| 660 | -188.1 | 184.6 | 6.552 | 3.481 | 3.144 | 24.147 | 3.412 | 27.9 | 30.9 |
| 680 | -199.5 | 190.6 | 6.657 | 3.512 | 3.157 | 24.197 | 3.464 | 27.6 | 30.7 |
| 700 | -211.0 | 196.7 | 6.759 | 3.543 | 3.170 | 24.247 | 3.519 | 27.3 | 30.5 |
| 720 | -222.8 | 202.8 | 6.859 | 3.576 | 3.183 | 24.299 | 3.575 | 27.0 | 30.3 |
| 740 | -234.7 | 209.0 | 6.958 | 3.610 | 3.196 | 24.352 | 3.635 | 26.6 | 30.1 |
| 760 | -246.8 | 215.2 | 7.054 | 3.645 | 3.210 | 24.405 | 3.696 | 26.3 | 29.9 |
| 780 | -259.0 | 221.5 | 7.149 | 3.681 | 3.224 | 24.460 | 3.761 | 26.0 | 29.7 |
| 800 | -271.4 | 227.9 | 7.243 | 3.719 | 3.238 | 24.516 | 3.828 | 25.7 | 29.5 |
| 820 | -284.0 | 234.3 | 7.335 | 3.757 | 3.252 | 24.573 | 3.899 | 25.3 | 29.3 |
| 840 | -296.7 | 240.9 | 7.426 | 3.798 | 3.266 | 24.631 | 3.972 | 25.0 | 29.1 |
| 860 | -309.6 | 247.4 | 7.516 | 3.839 | 3.280 | 24.690 | 4.049 | 24.7 | 28.9 |
| 880 | -322.6 | 254.1 | 7.605 | 3.882 | 3.294 | 24.751 | 4.128 | 24.3 | 28.7 |
| 900 | -335.8 | 260.8 | 7.692 | 3.927 | 3.307 | 24.813 | 4.212 | 24.0 | 28.5 |
| 920 | -349.1 | 267.6 | 7.779 | 3.972 | 3.321 | 24.876 | 4.299 | 23.6 | 28.3 |
| 923 | -351.1 | 268.6 | 7.792 | 3.979 | 3.323 | 24.886 | 4.312 | 23.6 | 28.2 |

## F. Mg LDA

Supplementary Table 19. Thermodynamic properties of Mg by the LDA functional.

| $T$ (K) | $G$ (meV/atom) | $H$ (meV/atom) | $S$ ($k_B$) | $C_p$ ($k_B$) | $C_V$ ($k_B$) | $V^{eq}$ (Å³/atom) | $\alpha$ ($10^{-5}/K$) | $B_T$ (GPa) | $B_S$ (GPa) |
|---|---|---|---|---|---|---|---|---|---|
| 1 | 32.7 | 32.7 | 0.000 | 0.000 | 0.000 | 21.548 | 0.000 | 39.6 | 39.6 |
| 20 | 32.7 | 32.7 | 0.014 | 0.038 | 0.038 | 21.548 | 0.028 | 39.6 | 39.6 |
| 40 | 32.6 | 33.0 | 0.121 | 0.361 | 0.361 | 21.550 | 0.275 | 39.6 | 39.6 |
| 60 | 32.2 | 34.1 | 0.375 | 0.924 | 0.923 | 21.557 | 0.719 | 39.5 | 39.5 |
| 80 | 31.2 | 36.2 | 0.717 | 1.452 | 1.446 | 21.569 | 1.149 | 39.3 | 39.5 |
| 100 | 29.7 | 39.0 | 1.087 | 1.857 | 1.845 | 21.586 | 1.487 | 39.1 | 39.4 |
| 120 | 27.5 | 42.5 | 1.453 | 2.151 | 2.131 | 21.607 | 1.739 | 38.9 | 39.3 |
| 140 | 24.6 | 46.4 | 1.801 | 2.365 | 2.336 | 21.631 | 1.926 | 38.7 | 39.2 |
| 160 | 21.2 | 50.6 | 2.126 | 2.522 | 2.485 | 21.657 | 2.068 | 38.5 | 39.0 |
| 180 | 17.3 | 55.0 | 2.430 | 2.641 | 2.595 | 21.685 | 2.178 | 38.2 | 38.9 |
| 200 | 12.8 | 59.6 | 2.713 | 2.733 | 2.678 | 21.714 | 2.266 | 38.0 | 38.8 |
| 220 | 7.9 | 64.4 | 2.976 | 2.806 | 2.742 | 21.744 | 2.339 | 37.8 | 38.6 |
| 240 | 2.6 | 69.2 | 3.223 | 2.866 | 2.792 | 21.775 | 2.400 | 37.5 | 38.5 |
| 260 | -3.2 | 74.2 | 3.454 | 2.915 | 2.832 | 21.807 | 2.453 | 37.3 | 38.4 |
| 280 | -9.3 | 79.2 | 3.671 | 2.957 | 2.865 | 21.839 | 2.499 | 37.0 | 38.2 |
| 298.15 | -15.2 | 83.9 | 3.858 | 2.990 | 2.889 | 21.869 | 2.537 | 36.8 | 38.1 |
| 300 | -15.9 | 84.3 | 3.876 | 2.993 | 2.892 | 21.872 | 2.541 | 36.8 | 38.1 |
| 320 | -22.7 | 89.5 | 4.070 | 3.026 | 2.915 | 21.906 | 2.579 | 36.5 | 37.9 |
| 340 | -29.9 | 94.7 | 4.254 | 3.055 | 2.934 | 21.940 | 2.615 | 36.3 | 37.8 |
| 360 | -37.4 | 100.0 | 4.429 | 3.081 | 2.951 | 21.975 | 2.649 | 36.0 | 37.6 |
| 380 | -45.2 | 105.3 | 4.596 | 3.106 | 2.966 | 22.010 | 2.681 | 35.8 | 37.4 |
| 400 | -53.2 | 110.7 | 4.756 | 3.130 | 2.980 | 22.046 | 2.712 | 35.5 | 37.3 |
| 420 | -61.6 | 116.1 | 4.909 | 3.152 | 2.992 | 22.082 | 2.742 | 35.2 | 37.1 |
| 440 | -70.2 | 121.5 | 5.056 | 3.175 | 3.004 | 22.118 | 2.772 | 35.0 | 37.0 |
| 460 | -79.0 | 127.0 | 5.198 | 3.196 | 3.015 | 22.155 | 2.802 | 34.7 | 36.8 |
| 480 | -88.1 | 132.5 | 5.334 | 3.218 | 3.026 | 22.193 | 2.832 | 34.5 | 36.6 |
| 500 | -97.4 | 138.1 | 5.466 | 3.240 | 3.037 | 22.231 | 2.862 | 34.2 | 36.5 |
| 520 | -107.0 | 143.7 | 5.593 | 3.262 | 3.047 | 22.269 | 2.893 | 33.9 | 36.3 |
| 540 | -116.7 | 149.3 | 5.717 | 3.284 | 3.058 | 22.308 | 2.924 | 33.6 | 36.1 |
| 560 | -126.7 | 155.0 | 5.836 | 3.307 | 3.069 | 22.347 | 2.956 | 33.4 | 36.0 |
| 580 | -136.8 | 160.7 | 5.953 | 3.330 | 3.080 | 22.387 | 2.989 | 33.1 | 35.8 |
| 600 | -147.2 | 166.4 | 6.066 | 3.354 | 3.091 | 22.428 | 3.024 | 32.8 | 35.6 |
| 620 | -157.7 | 172.2 | 6.176 | 3.378 | 3.102 | 22.469 | 3.059 | 32.5 | 35.4 |
| 640 | -168.5 | 178.1 | 6.284 | 3.404 | 3.113 | 22.510 | 3.096 | 32.2 | 35.2 |
| 660 | -179.4 | 184.0 | 6.389 | 3.430 | 3.125 | 22.552 | 3.135 | 32.0 | 35.1 |
| 680 | -190.5 | 189.9 | 6.492 | 3.457 | 3.137 | 22.595 | 3.176 | 31.7 | 34.9 |
| 700 | -201.8 | 195.9 | 6.592 | 3.485 | 3.149 | 22.639 | 3.219 | 31.4 | 34.7 |
| 720 | -213.2 | 201.9 | 6.691 | 3.514 | 3.162 | 22.683 | 3.264 | 31.0 | 34.5 |
| 740 | -224.9 | 207.9 | 6.787 | 3.544 | 3.174 | 22.727 | 3.311 | 30.7 | 34.3 |
| 760 | -236.7 | 214.1 | 6.882 | 3.575 | 3.187 | 22.773 | 3.361 | 30.4 | 34.1 |
| 780 | -248.6 | 220.3 | 6.975 | 3.607 | 3.200 | 22.819 | 3.415 | 30.1 | 33.9 |
| 800 | -260.7 | 226.5 | 7.067 | 3.641 | 3.213 | 22.867 | 3.471 | 29.8 | 33.7 |
| 820 | -273.0 | 232.8 | 7.157 | 3.676 | 3.226 | 22.915 | 3.531 | 29.4 | 33.5 |
| 840 | -285.4 | 239.2 | 7.246 | 3.712 | 3.239 | 22.964 | 3.595 | 29.1 | 33.3 |
| 860 | -297.9 | 245.6 | 7.334 | 3.750 | 3.252 | 23.014 | 3.664 | 28.7 | 33.1 |
| 880 | -310.7 | 252.1 | 7.421 | 3.789 | 3.265 | 23.065 | 3.737 | 28.4 | 32.9 |
| 900 | -323.5 | 258.6 | 7.506 | 3.830 | 3.277 | 23.117 | 3.815 | 28.0 | 32.7 |
| 920 | -336.5 | 265.3 | 7.591 | 3.873 | 3.289 | 23.171 | 3.899 | 27.6 | 32.5 |
| 923 | -338.5 | 266.3 | 7.604 | 3.880 | 3.291 | 23.179 | 3.912 | 27.6 | 32.5 |

# SUPPLEMENTARY REFERENCES

[1] Kresse, G. & Furthmüller, J. Efficient iterative schemes for *ab initio* total-energy calculations using a plane-wave basis set. *Phys. Rev. B* **54**, 11169–11186 (1996).

[2] Blöchl, P. E. Projector augmented-wave method. *Phys. Rev. B* **50**, 17953–17979 (1994).

[3] Kresse, G. & Joubert, D. From ultrasoft pseudopotentials to the projector augmented-wave method. *Phys. Rev. B* **59**, 1758–1775 (1999).

[4] Perdew, J. P., Burke, K. & Ernzerhof, M. Generalized gradient approximation made simple. *Phys. Rev. Lett.* **77**, 3865–3868 (1996).

[5] Ceperley, D. M. & Alder, B. J. Ground state of the electron gas by a stochastic method. *Phys. Rev. Lett.* **45**, 566–569 (1980).

[6] Perdew, J. *et al.* Restoring the density-gradient expansion for exchange in solids and surfaces. *Phys. Rev. Lett.* **100**, 136406 (2008).

[7] Methfessel, M. & Paxton, A. T. High-precision sampling for Brillouin-zone integration in metals. *Phys. Rev. B* **40**, 3616–3621 (1989).

[8] Mermin, N. D. Thermal properties of the inhomogeneous electron gas. *Phys. Rev.* **137**, A1441–A1443 (1965).

[9] Shapeev, A. V. Moment tensor potentials: A class of systematically improvable interatomic potentials. *Multiscale Model. Simul.* **14**, 1153–1173 (2016).

[10] Novikov, I. S., Gubaev, K., Podryabinkin, E. V. & Shapeev, A. V. The MLIP package: moment tensor potentials with MPI and active learning. *Mach. Learn.: Sci. Technol.* **2**, 025002 (2021).

[11] Plimpton, S. Fast parallel algorithms for short-range molecular dynamics. *J. Comput. Phys.* **117**, 1–19 (1995).

[12] Schneider, T. & Stoll, E. Molecular-dynamics study of a three-dimensional one-component model for distortive phase transitions. *Phys. Rev. B* **17**, 1302–1322 (1978).

[13] Grabowski, B., Ismer, L., Hickel, T. & Neugebauer, J. Ab initio up to the melting point: Anharmonicity and vacancies in aluminum. *Phys. Rev. B* **79**, 134106 (2009).

[14] Hoover, W. G. Canonical dynamics: Equilibrium phase-space distributions. *Phys. Rev. A* **31**, 1695–1697 (1985).

[15] Glensk, A., Grabowski, B., Hickel, T. & Neugebauer, J. Break-down of the arrhenius law in describing vacancy formation energies: The importance of local anharmonicity revealed by ab initio thermodynamics. *Phys. Rev. X* **4**, 011018 (2014).

[16] Hellman, O., Abrikosov, I. A. & Simak, S. I. Lattice dynamics of anharmonic solids from first principles. *Phys. Rev. B* **84**, 180301 (2011).

[17] Eriksson, F., Fransson, E. & Erhart, P. The Hiphive package for the extraction of high-order force constants by machine learning. *Adv. Theory Simul.* **2**, 1800184 (2019).

[18] Togo, A. & Tanaka, I. First principles phonon calculations in materials science. *Scr. Mater.* **108**, 1–5 (2015).

[19] Momma, K. & Izumi, F. *VESTA 3* for three-dimensional visualization of crystal, volumetric and morphology data. *J. Appl. Crystallogr.* **44**, 1272–1276 (2011).

[20] Vinet, P., Ferrante, J., Rose, J. H. & Smith, J. R. Compressibility of solids. *J. Geophys. Res.: Solid Earth* **92**, 9319–9325 (1987).

[21] Zhang, X. *et al.* Temperature dependence of the stacking-fault Gibbs energy for Al, Cu, and Ni. *Phys. Rev. B* **98**, 224106 (2018).

[22] Rogal, L. *et al.* Computationally-driven engineering of sublattice ordering in a hexagonal AlHfScTiZr high entropy alloy. *Sci. Rep.* **7**, 2209 (2017).

[23] Mehta, S., Price, G. D. & Alfè, D. Ab initio thermodynamics and phase diagram of solid magnesium: A comparison of the LDA and GGA. *J. Chem. Phys.* **125**, 194507 (2006).

[24] Moruzzi, V. L., Janak, J. F. & Schwarz, K. Calculated thermal properties of metals. *Phys. Rev. B* **37**, 790–799 (1988).

[25] van de Walle, A. & Ceder, G. The effect of lattice vibrations on substitutional alloy thermodynamics. *Rev. Mod. Phys.* **74**, 11–45 (2002).

[26] Haynes, W. M. (ed.) *CRC Handbook of Chemistry and Physics* (CRC Press, Boca Raton, Fla., 2007).

[27] Kaufman, L. & Bernstein, H. *Computer Calculation of Phase Diagrams. With Special Reference to Refractory Metals* (Academic Press, NY, 1970).

[28] Chang, Y. *et al.* Phase diagram calculation: past, present and future. *Prog. Mater. Sci.* **49**, 313–345 (2004).

[29] Zhou, Y. *et al.* Thermodynamics up to the melting point in a TaVCrW high entropy alloy: Systematic *ab initio* study aided by machine learning potentials. *Phys. Rev. B* **105**, 214302 (2022).

[30] de Wijs, G. A., Kresse, G. & Gillan, M. J. First-order phase transitions by first-principles free-energy calculations: The melting of Al. *Phys. Rev. B* **57**, 8223–8234 (1998).

[31] Körmann, F., Ikeda, Y., Grabowski, B. & Sluiter, M. H. F. Phonon broadening in high entropy alloys. *npj Comput. Mater.* **3**, 36 (2017).

[32] Vočadlo, L. & Alfè, D. Ab initio melting curve of the fcc phase of aluminum. *Phys. Rev. B* **65**, 214105 (2002).

[33] Grabowski, B., Söderlind, P., Hickel, T. & Neugebauer, J. Temperature-driven phase transitions from first principles including all relevant excitations: The fcc-to-bcc transition in Ca. *Phys. Rev. B* **84**, 214107 (2011).

[34] van Gunsteren, W. F., Daura, X. & Mark, A. E. Computation of free energy. *Helv. Chim. Acta* **85**, 3113–3129 (2002).

[35] Zwanzig, R. W. High-temperature equation of state by a perturbation method. I. nonpolar gases. *J. Chem. Phys.* **22**, 1420–1426 (1954).

[36] These energies can be total energies, or should be from the same reference energy (which will then cancel).

[37] Kresse, G. & Furthmüller, J. Efficiency of ab-initio total energy calculations for metals and semiconductors using a plane-wave basis set. *Comput. Mater. Sci.* **6**, 15–50 (1996).

[38] In Ref. [33], when Eq. (12) is inserted in Eq. (10) and Eq. (10) is integrated over the energy $\varepsilon$, $S^{el}(T)$ becomes a polynomial in $T$ with the lowest order term of $T$. When $S^{el}(T)$ is inserted in Eq. (9), $F^{el}$ becomes a polynomial in $T$ with the lowest order term of $T^2$, analogous to our current parametrization.

[39] Zhang, X., Grabowski, B., Körmann, F., Freysoldt, C. & Neugebauer, J. Accurate electronic free energies of the 3d, 4d, and 5d transition metals at high temperatures. *Phys. Rev. B* **95**, 165126 (2017).

[40] Chuang, Y.-Y., Schmid, R. & Chang, Y. Magnetic contributions to the thermodynamic functions of pure Ni, Co, and Fe. *Metall. Trans. A* **16**, 153–165 (1985).

[41] Straatsma, T. P. & McCammon, J. A. Computational alchemy. *Annu. Rev. Phys. Chem.* **43**, 407–435 (1992).

[42] Flyvbjerg, H. & Petersen, H. G. Error estimates on averages of correlated data. *J. Chem. Phys.* **91**, 461–466 (1989).

[43] Pohorille, A., Jarzynski, C. & Chipot, C. Good practices in

free-energy calculations. *J. Phys. Chem. B* **114**, 10235–10253 (2010).

[44] Wang, K. & Reeber, R. R. The role of defects on thermophysical properties: Thermal expansion of V, Nb, Ta, Mo and W. *Mater. Sci. Eng., R* **23**, 101–137 (1998).

[45] Bujard, P., Sanjines, R., Walker, E., Ashkenazi, J. & Peter, M. Elastic constants in Nb-Mo alloys from zero temperature to the melting point: experiment and theory. *J. Phys. F: Met. Phys.* **11**, 775–786 (1981).

[46] Abdullaev, R. N., Kozlovskii, Y. M., Khairulin, R. A. & Stankus, S. V. Density and thermal expansion of high purity nickel over the temperature range from 150 K to 2030 K. *Int. J. Thermophys.* **36**, 603–619 (2015).

[47] Prikhodko, S. V. *et al.* Elastic constants of face-centered cubic and L1$_2$ Ni-Si alloys: Composition and temperature dependence. *Metall. Mater. Trans. A* **34**, 1863–1868 (2003).

[48] Touloukian, Y. S., Kirby, R., Taylor, R. & Desai, P. *Thermal Expansion Metallic Elements and Alloys* (Thermophysical Properties of Matter - the TPRC Data Series vol. 12, IFI/Plenum, 1975).

[49] Wang, K. & Reeber, R. R. The perfect crystal, thermal vacancies and the thermal expansion coefficient of aluminium. *Phi-los. Mag. A* **80**, 1629–1643 (2000).

[50] Slutsky, L. J. & Garland, C. W. Elastic constants of magnesium from 4.2°K to 300°K. *Phys. Rev.* **107**, 972–976 (1957).

[51] Tidholm, J. *et al.* Temperature dependence of the Kohn anomaly in bcc Nb from first-principles self-consistent phonon calculations. *Phys. Rev. B* **101**, 115119 (2020).

[52] Guthoff, F. *et al.* Lattice dynamics and self-diffusion in niobium at elevated temperatures. *J. Phys.: Condens. Matter* **6**, 6211–6220 (1994).

[53] Lai, M. *et al.* Origin of shear induced $\beta$ to $\omega$ transition in Ti-Nb-based alloys. *Acta Mater.* **92**, 55–63 (2015).

[54] Huang, L.-F. *et al.* From electronic structure to phase diagrams: A bottom-up approach to understand the stability of titanium–transition metal alloys. *Acta Mater.* **113**, 311–319 (2016).

[55] Gong, Y. *et al.* Temperature dependence of the gibbs energy of vacancy formation of fcc ni. *Phys. Rev. B* **97**, 214106 (2018).

[56] Forslund, A. & Ruban, A. Ab initio surface free energies of tungsten with full account of thermal excitations. *Phys. Rev. B* **105**, 045403 (2022).

[57] Dinsdale, A. SGTE data for pure elements. *Calphad* **15**, 317–425 (1991).

  
\fi

\end{document}